\documentclass[ aps,prd,onecolumn,eqsecnum,nofootinbib,superscriptaddress]{revtex4}

\usepackage{graphicx}% Include figure files
\usepackage{dcolumn}% Align table columns on decimal point
\usepackage{bm}% bold math
\usepackage[a4paper,top=3.5cm,bottom=3.2cm,left=2.3cm,right=2.3cm,bindingoffset=5mm]{geometry}
\usepackage{amsfonts}
\usepackage{latexsym}
\usepackage{amsmath}
\usepackage[english]{babel}
\usepackage[latin1]{inputenc}
\usepackage[T1]{fontenc}
\usepackage{slashed}
\usepackage{float}
\usepackage{subfigure}

\newcommand{\be}{\begin{equation}}
\newcommand{\ee}{\end{equation}}
\newcommand{\ba}{\begin{eqnarray}}
\newcommand{\ea}{\end{eqnarray}}
\newcommand{\eite}{\end{itemize}}
\newcommand{\bite}{\begin{itemize}}
\newcommand{\betabeta}{\mbox{$(\beta \beta)_{0 \nu}  $}}
\def\ltap{\ \raisebox{-.4ex}{\rlap{$\sim$}} \raisebox{.4ex}{$<$}\ }
\def\gtap{\ \raisebox{-.4ex}{\rlap{$\sim$}} \raisebox{.4ex}{$>$}\ }
\newcommand{\meff}{\mbox{$\left|  < \!  m \!  > \right| \ $}}
\newcommand{\mefff}{\mbox{$ < \! m \! > $}}

\newcommand{\eq}{\begin{eqnarray}}
\newcommand{\en}{\end{eqnarray}}

\def\rp{$R_p\hspace{-1em}/\ \ $}

\newcommand{\Mn}[1]{   \ensuremath{{M}_{#1}}  }
\newcommand{\Ml}[1]{   \ensuremath{{\Lambda}_{#1}}      }

\begin{document}

\hfill{{\small Ref. SISSA 11/2011/EP}}

\hfill{{\small Ref. IPMU11--0042}}

\hfill{{\small arXiv:1103.2434 [hep-ph]}}

\title{Uncovering Multiple CP-Nonconserving  Mechanisms of $\betabeta$-Decay}

\author{Amand  Faessler}
\affiliation{Institute of Theoretical Physics, University of Tuebingen,
 72076 Tuebingen, Germany.}

\author{A. Meroni}
\affiliation{SISSA, Via Bonomea 265, 34136 Trieste, Italy.}
\affiliation{Istituto Nazionale di Fisica Nucleare, Sezione  di Trieste, Via Valerio 2, 34126 Trieste, Italy.}

\author{S. T. Petcov\footnote{Also at: Institute of Nuclear Research and Nuclear Energy, Bulgarian Academy of Sciences, 1784 Sofia, Bulgaria}}
\affiliation{SISSA, Via Bonomea 265, 34136 Trieste, Italy.}
\affiliation{Istituto Nazionale di Fisica Nucleare, Sezione  di Trieste, Via Valerio 2, 34126 Trieste, Italy.}
\affiliation{ IPMU, University of Tokyo, Tokyo, Japan.}

\author{F. \v Simkovic}
\affiliation{Department of Nuclear Physics and Biophysics, Comenius  University, Mlynska dolina F1, SK-842 15 Bratislava, Slovakia}
\affiliation{Bogoliubov Laboratory of Theoretical Physics, JINR,  141980 Dubna, Moscow region, Russia.}

\author{J. Vergados}
\affiliation{Theoretical Physics  Devision, University of Ioannina, GR-451 10 Ioannina, Greece.}

% \begin{center}
% {\bf{\large Uncovering Multiple CP-Nonconserving 
% Mechanisms of $\betabeta$-Decay}}
% 
% \vspace{0.4cm}
% Amand  Faessler$\mbox{}^{a)}$,
% A. Meroni$\mbox{}^{b,c)}$,
% S. T. Petcov$\mbox{}^{b,c,d)}$
% \footnote{Also at: Institute of Nuclear Research and
% Nuclear Energy, Bulgarian Academy of Sciences, 1784 Sofia, Bulgaria},
% F. \v Simkovic$\mbox{}^{e,f)}$ and 
% J. Vergados$\mbox{}^{g)}$
% 
% 
% \vspace{0.2cm}
% $\mbox{}^{a)}${\em Institute of Theoretical Physics, University of Tuebingen,
% 72076 Tuebingen, Germany.\\}
% 
% \vspace{0.1cm}
% $\mbox{}^{b)}${\em  SISSA, Via Bonomea 265, 34136 Trieste, Italy.\\}
% 
% 
% \vspace{0.1cm}
% $\mbox{}^{c)}${\em  Istituto Nazionale di Fisica Nucleare, Sezione 
% di Trieste, Via Valerio 2, 34126 Trieste, Italy.\\}
% 
% 
% \vspace{0.1cm}
% $\mbox{}^{d)}${\em IPMU, University of Tokyo, Tokyo, Japan.\\
% }
% 
% \vspace{0.1cm}
% $\mbox{}^{e)}${\em Department of Nuclear Physics and Biophysics, Comenius 
% University, Mlynska dolina F1, SK-842 15 Bratislava, Slovakia\\
% }
% 
% \vspace{0.1cm}
% $\mbox{}^{f)}${\em Bogoliubov Laboratory of Theoretical Physics, JINR, 
% 141980 Dubna, Moscow region, Russia.\\
% }
% 
% \vspace{0.1cm}
% $\mbox{}^{g)}${\em Theoretical Physics  Devision, University of Ioannina, 
% GR-451 10 Ioannina, Greece.\\
% }
% 
% \end{center}

\begin{abstract}
We consider the possibility of several different 
mechanisms contributing to the
$\betabeta$-decay amplitude in the general case
of CP nonconservation: light Majorana
neutrino exchange, heavy left-handed (LH) and heavy
right-handed (RH) Majorana neutrino exchanges,
lepton charge non-conserving couplings in SUSY theories
with $R$-parity breaking. If the $\betabeta$-decay
is induced by, e.g., two ``non-interfering'' mechanisms
(light Majorana neutrino and heavy RH Majorana neutrino
exchanges), one can determine $|\eta_i|^2$ and  $|\eta_j|^2$, 
$\eta_i$ and  $\eta_j$ being the
two fundamental parameters characterising these
mechanisms, from data on the half-lives
of two nuclear isotopes. In the case when two
``interfering'' mechanisms are responsible
for the $\betabeta$-decay, $|\eta_i|^2$ and  $|\eta_j|^2$ 
and the interference term
can be uniquely determined, in principle,
from data on the half-lives of three nuclei.  
Given the half-life of one isotope,
the ``positivity conditions'' 
 $|\eta_i|^2\geq 0$ and  $|\eta_j|^2\geq 0$
lead to stringent constraints on the half-lives of  
the other $\betabeta$-decaying  isotopes.
These conditions,  as well as the conditions for 
constructive (destructive) interference 
are derived and their implications  are 
analysed in two specific cases.
The experimental limits on neutrino masses obtained 
in the $^3H$ $\beta$-decay experiments 
can constrain further the multiple 
mechanisms of $\betabeta$-decay 
if one of the mechanisms involved is the light 
Majorana neutrino exchange. The measurements of the 
half-lives with rather high precision and the 
knowledge of the relevant nuclear matrix elements with 
relatively small uncertainties is crucial for establishing 
that more than one mechanisms are operative in 
$\betabeta$-decay. The method considered by us
can be generalised to the case of more than two
$\betabeta$-decay mechanisms. It allows to treat 
the cases of CP conserving and CP nonconserving 
couplings generating the $\betabeta$-decay
in a unique way.

\end{abstract}

\vspace{0.6cm}
\maketitle

%%%%%%%%%%%%%%%%%%%%%%%%%%%%
%
\section{Introduction}
%
%%%%%%%%%%%%%%%%%%%%%%%%%%%%
%

If neutrinoless double beta ($\betabeta$-) decay
will be observed, it will be of fundamental
importance to determine the mechanism which
induces the decay. We know that neutrinos 
have mass and mix, and if they are Majorana 
particles they should trigger the decay 
at some probability level. 
The fundamental parameter which controls the 
$\betabeta$-decay rate in this case
is the effective Majorana mass: 
%%%%%%%%%%%%%%%%%%%%%%
\begin{equation}
% m_{\beta\beta}  
\mefff = \sum_j^{light} \left (U_{ej} \right)^2 m_j ~,
~({\rm all~} m_j \ge 0)~,
\label{meff}
\end{equation}
%%%%%%%%%%%%%%%%%%%
%
where $U$ is the Pontecorvo, Maki, Nakagawa and
Sakata (PMNS) neutrino mixing matrix 
\cite{BPont57,MNS62,BPont67}
and $m_j$ are the light Majorana neutrino masses, 
$m_j \ltap 1$ eV.
The $\betabeta$-decay rate depends  
on the type of neutrino mass spectrum 
which can be hierarchical, with partial hierarchy 
or quasi-degenerate (see, e.g., \cite{BPP1}).
Using the data on the neutrino oscillation 
parameters it is possible to show 
(see, e.g., \cite{PDG10}) that
in the case of normal hierarchical spectrum
one has $\meff \ltap 0.005$ eV, 
while if the spectrum is with inverted 
hierarchy, $0.01~{\rm eV}\ltap \meff \ltap 0.05$ eV.
A larger value of $\meff$ is possible 
if the light neutrino mass spectrum is 
with partial hierarchy or of quasi-degenerate 
type. In the latter case $\meff$ can be close 
to the existing upper limits. 

   The most stringent 
upper limits on $\meff$ were set by the 
IGEX \cite{IGEX00}, CUORICINO  \cite{CUORI} 
and NEMO3 \cite{NEMO3} experiments with  $^{76}$Ge, 
 $^{130}$Te and $^{100}$Mo, respectively 
\footnote{The NEMO3 collaboration has searched for 
$\betabeta$-decay of  $^{82}$Se and other isotopes as well.}.   
The IGEX collaboration has obtained for the half-life of 
$^{76}$Ge,  $T_{1/2}^{0\nu} > 1.57\times 10^{25}~\text{yr}$ (90\%~C.L.),
from which the limit $\meff < (0.33 - 1.35)$~eV was derived
\cite{IGEX00}. 
%{\bf 
Using the recent more advanced calculations 
of the corresponding nuclear matrix elements (including the relevant 
uncertainties) \cite{SFMRS09} 
one finds:  $\meff < (0.22 - 0.35)$~eV.
% }
The NEMO3 and CUORICINO experiments,
designed to reach a sensitivity to $\meff\sim (0.2-0.3)$ eV, 
set the limits: $\meff < (0.61\,\text{-}\,1.26)$~eV~\cite{NEMO3} and 
$\meff < (0.19 - 0.68)$~eV~\cite{CUORI} (90\% C.L.), 
where estimated uncertainties in the NME are accounted for. 
The two upper limits were derived from the
experimental lower limits on the half-lives of $^{100}$Mo and
$^{130}$Te, $T_{1/2}^{0\nu} > 5.8\times 10^{23}~\text{yr}$ 
(90\%C.L.)~\cite{NEMO3} and 
$T_{1/2}^{0\nu} > 3.0\times 10^{24}~\text{yr}$
(90\%C.L.)~\cite{CUORI}. 
% {\bf 
With the NMEs and their uncertainties calculated 
in \cite{SFMRS09}, the NEMO3 and CUORICINO upper limits read, 
respectively:  $\meff < (0.50 - 0.96)$~eV and 
$\meff < (0.25 - 0.43)$~eV.
% }
The best lower limit on the half-life of $^{76}$Ge,
$T_{1/2}^{0\nu} > 1.9\times 10^{25}~\text{yr}$ (90\%~C.L.),
was found in the Heidelberg-Moscow $^{76}$Ge experiment~\cite{HMGe76}. 
% {\bf 
It corresponds to the upper limit \cite{SFMRS09}
$\meff < (0.20 - 0.35)$~eV.
% }
A positive $\betabeta$-decay signal at $> 3\sigma$,
corresponding to $T_{1/2}^{0\nu} = (0.69 - 4.18)\times
10^{25}~\text{yr}$ (99.73\%~C.L.) and implying $\meff = (0.1 -
0.9)~{\rm eV}$, is claimed to have been observed in 
\cite{KlapdorMPLA}, while a later analysis reports 
evidence for $\betabeta$-decay at 6$\sigma$ 
% with $T_{1/2}^{0\nu} =  2.23^{+0.44}_{-0.31}\times 10^{25}~\text{yr}$,
corresponding to $\meff = 0.32 \pm 0.03$~eV~\cite{Klap04}. 

  Most importantly, a large number of projects aim at a
sensitivity to $\meff \sim (0.01 - 0.05)$ eV \cite{bb0nu}: CUORE
($^{130}$Te), GERDA ($^{76}$Ge), SuperNEMO, EXO ($^{136}$Xe), MAJORANA
($^{76}$Ge), MOON ($^{100}$Mo), COBRA ($^{116}$Cd), XMASS
($^{136}$Xe), CANDLES ($^{48}$Ca), KamLAND-Zen ($^{136}$Xe), 
SNO+ ($^{150}Nd$), etc.  These experiments, in
particular, will test the positive result claimed 
in \cite{Klap04}.
Let us note that the measurement 
of $\meff$ can provide 
unique information on the absolute scale 
of neutrino masses, the type of neutrino 
mass spectrum and the Majorana phases in the 
PMNS matrix \cite{meff}.

  The light Majorana neutrino exchange
can be called the ``standard'' 
mechanism of the $\betabeta$-decay.
The observation of 
$\betabeta$-decay would imply that the 
total lepton charge $L$ is not conserved.
This would also imply that the massive 
neutrinos get a Majorana mass \cite{SchVal82} 
and therefore are Majorana particles 
(see, e.g. \cite{BiPet87}). However, 
the latter does not guarantee that 
the dominant mechanism inducing the 
$\betabeta$-decay is the light Majorana 
neutrino exchange since the Majorana mass 
thus generated is exceedingly small.
The $\betabeta$-decay can well be due 
to the  existence of interactions which 
do not conserve the total lepton charge $L$,
$\Delta L = \pm 2$.
A number of such interactions have been 
proposed in the literature: 
heavy Majorana neutrinos
coupled to the electron 
in the $V-A$ charged current weak 
interaction Lagrangian, 
supersymmetric (SUSY) theories with 
$R$-parity breaking terms which do not 
conserve the total lepton charge $L$, 
$L$-nonconserving couplings in the Left-Right 
symmetric theories, etc. At present we do 
not have evidence for the existence of 
$\Delta L \neq 0$ terms in the Lagrangian 
describing the particle interactions.
Nevertheless, such terms can exist 
and they can be operative in the $\betabeta$-decay.
Moreover, it is impossible to exclude the 
hypothesis that, if observed, the 
$\betabeta$-decay is triggered by more 
than one competing mechanisms.

   The possibility of several different 
mechanisms contributing to the $\betabeta$-decay 
amplitude was considered recently in 
\cite{FSV10MM} assuming that the 
corresponding $\Delta L = \pm 2$ couplings 
are CP conserving. By exploiting  
the dependence of the nuclear
matrix elements on the decaying nucleus, 
it was shown that, given the experimental 
observation of the $\betabeta$-decay 
of sufficient number of nuclei, one can 
determine and/or sufficiently constrain 
the fundamental parameters associated 
with the lepton charge nonconserving couplings
generating the $\betabeta$-decay.

  The present work is a natural continuation of the 
study performed in \cite{FSV10MM}.
We consider the possibility of several different 
mechanisms contributing to the
$\betabeta$-decay amplitude in the general case
of CP nonconservation: light Majorana
neutrino exchange, heavy left-handed (LH) and heavy
right-handed (RH) Majorana neutrino exchanges,
lepton charge non-conserving couplings in SUSY theories
with $R$-parity breaking. If the $\betabeta$-decay
is induced by, e.g., two ``non-interfering'' mechanisms
(light Majorana neutrino and heavy RH Majorana neutrino
exchanges), one can determine the absolute values of the
two fundamental parameters, characterising these
mechanisms, from data on the half-lives
of two nuclear isotopes. In the case when two
``interfering'' mechanisms are responsible
for the $\betabeta$-decay, the absolute values of the
two relevant parameters and the interference term
can be uniquely determined from data on the
half-lives of three nuclei.
In the  specific examples considered  
of two ``noninterfering'' and two ``interfering'' 
mechanisms, 
% involved in the $\betabeta$-decay,   
namely, the light Majorana neutrino and 
the heavy RH Majorana neutrino exchanges, and 
the light Majorana neutrino and the dominant gluino
exchanges, we present 
illustrative examples of determination 
of the relevant fundamental parameters 
and of possible tests of the hypothesis 
that more than one mechanism is responsible 
for the $\betabeta$-decay, using as  
input hypothetical half-lives of  $^{76}$Ge, 
$^{130}$Te and $^{100}$Mo.
The effects of the uncertainties in the values 
of the nuclear matrix elements (NMEs) on the results of the 
indicated analyzes are also discussed and
illustrated.

  The method considered by us
can be generalised to the case of more than two
$\betabeta$-decay mechanisms. It has also the advantage
that it allows to treat the cases of CP conserving and
CP nonconserving couplings generating the $\betabeta$-decay
in a unique way.

%%%%%%%%%%%%%%%%%%%%%%%%
%
\section{Different Mechanisms of $\betabeta$-Decay}
%
%%%%%%%%%%%%%%%%%%%%%%%

 We will consider in the present article the following
mechanisms of  $\betabeta$-decay: the exchange of light 
Majorana neutrinos; the exchange of heavy ``left-handed'' 
(LH) Majorana neutrinos; the exchange of
heavy ``right-handed'' (RH) Majorana neutrinos;
and two mechanisms associated with possible 
$R$-parity breaking in SUSY theories.
Below we discuss briefly the lepton number 
violating (LNV) parameters and the 
nuclear matrix elements associated with 
each of the indicated mechanisms.

 Assuming the dominance of a single LNV 
mechanism characterised by a
parameter $\eta_{\kappa}^{LNV}$, where the 
index $\kappa$ denotes the mechanism,    
the inverse value of the $\betabeta$-decay
half-life for a given isotope $(A,Z)$ 
can be written as
%%%%%%%%%%%%%%%%%%%%%%%%%%%%%%%%%%
\begin{eqnarray}
\frac{1}{T^{0\nu}_{1/2}} &=& |\eta_{\kappa}^{LNV}|^2~
G^{0\nu}(E_0,Z) |{M'}_{\kappa}^{0\nu}|^2\,,
\label{eq.1}
\end{eqnarray}
%%%%%%%%%%%%%%%%%%%%%%%%%%%%%%%%%%%%%
%
where $G^{0\nu}(E_0,Z)$ and ${M'}_{\kappa}^{0\nu}$ are, 
respectively, the known phase-space factor 
($E_0$ is the energy release) and the nuclear matrix 
element of the decay. The latter depends on the 
mechanism generating the decay and  
on the nuclear structure of the specific 
isotopes $(A,Z)$, $(A,Z+1)$ and $(A,Z+2)$ under study. 

The phase space factors $G^{0\nu}(E_0,Z)$, which include 
the fourth power of the ``standard'' value 
of the  axial-coupling constant $g_A = 1.25$, 
are tabulated in ref. \cite{Sim99}; for $^{76}Ge$, $^{82}Se$,  
$^{100}Mo$ and $^{130}Te$ 
% and $g_A = 1.25$ they
are given in Table I. 
For a given isotope $(A,Z)$, $G^{0\nu}(E_0,Z)$ 
contains the inverse square of the nuclear 
radius $R(A)$ of the isotope, $R^{-2}(A)$,
compensated by the
factor $R(A)$ in ${M'}_{\kappa}^{0\nu}$. 
The assumed value of the nuclear radius is 
$R(A) = r_0 A^{1/3}$ with $r_0 = 1.1~fm$.

The nuclear matrix element ${M'}_{\kappa}^{0\nu}$ is defined as 
%%%%%%%%%%%%%%%%%%%%%%%%%%%%%%%%%%
\begin{equation}
{M'}_{\kappa}^{0\nu} =  \left(\frac{g_A}{1.25}\right)^2 {M}_{\kappa}^{0\nu}.
\label{Mprime}
\end{equation}
%%%%%%%%%%%%%%%%%%%%%%%%%%%%%%%%%
%
This definition of ${M'}_{\kappa}^{0\nu}$ \cite{Rodin} allows
to display the effects of uncertainties in $g_A$ and to use
the same phase factor $G^{0\nu}(E_0,Z)$ when calculating 
%the $0\nu\beta\beta$-decay rate.
the $\betabeta$-decay rate.
%

%%%%%%%%%%%%%%%%%%%%%%%%%%%%
%
\subsection{Light Majorana Neutrino Exchange}
%
%%%%%%%%%%%%%%%%%%%%%%%%%%%
%
In the case of the light Majorana neutrino 
exchange mechanism of $\betabeta$-decay,
% $0\nu\beta\beta$-decay 
the LNV parameter is given by:
%%%%%%%%%%%%%%%%%%%%%%%%%%
\begin{eqnarray}
% \eta_{\nu} = \frac{m_{\beta\beta}}{m_e},
\eta_{\nu} = \frac{\mefff}{m_e},
\label{etanu}
\end{eqnarray}
%%%%%%%%%%%%%%%%%%%%%%%%%%%
%
where 
% $m_{\beta\beta}$ 
$\mefff$ is the effective Majorana mass 
(see, e.g., \cite{BiPet87}).
Under the assumption of $n$ 
light massive Majorana neutrinos coupled to 
the electron in the weak charged lepton current, 
the effective Majorana mass  
% $\langle m_{\beta\beta} \rangle$ 
is given in eq. (\ref{meff}).
Thus, $\mefff$ depends on the elements of 
first row of the PMNS neutrino mixing matrix,
$U_{ej}$, $j=1,2.3,..$. 
The PMNS matrix $U$ is not assumed 
to be CP conserving and 
at least two of the elements $U_{ej}$
contain physical CP violating phases 
\cite{BHP80,SchValle80} 
(see also, e.g., \cite{PDG10}).
In the case of 3 light neutrinos
and the ``standard'' parametrisation of 
$U$ \cite{PDG10}, the elements 
$U_{e2}$ and $U_{e3}$ contain the two physical 
CP violating Majorana phases 
\cite{BHP80} and $U_{e3}$
contains the Dirac phase as well.

The expression for $\mefff$, eq. (\ref{meff}),
corresponds to the contribution from 
the standard $(V-A)$ charged current (CC) 
weak interaction. 
The nuclear matrix element ${M}^{0\nu}_{\nu}$
for different isotopes $(A,Z)$
is given in \cite{Sim99,anatomy} 
(see also Table I).

%%%%%%%%%%%%%%%%%%%%%%%%%%
%
\subsection{Heavy Majorana Neutrino Exchange Mechanisms}
%
%%%%%%%%%%%%%
%
We assume that the neutrino mass spectrum includes, 
in addition to the three light Majorana neutrinos, 
heavy Majorana states $N_k$ with masses 
$M_k$ much larger than the typical energy scale of the 
% $0\nu\beta\beta$-decay, 
$\betabeta$-decay, $M_k \gg 100$ MeV; 
we will consider the case of $M_k \gtap 10$ GeV. 
Such a possibility arises 
if the weak interaction Lagrangian includes 
right-handed (RH) sterile neutrino fields which
couple to the LH flavour neutrino fields 
via the neutrino Yukawa coupling and possess a 
Majorana mass term.
The heavy Majorana neutrinos $N_k$
can mediate the $\betabeta$-decay \cite{HMPR76}
similar to the light Majorana neutrinos
via the $V-A$ charged current weak interaction.  
The difference between the two mechanisms is that,
unlike the light Majorana neutrino exchange
which leads to a long range inter-nucleon interactions,
in the case of $M_k \gtap 10$ GeV 
of interest the momentum dependence 
of the heavy Majorana neutrino propagators 
can be neglected (i.e., the $N_k$ propagators 
can be contracted to points) 
and, as a consequence, the corresponding 
effective nucleon transition operators are local.
The LNV parameter in the case when the  
$\betabeta$-decay is generated by the $(V-A)$ CC
weak interaction due to the 
exchange of $N_k$ can be written as:
%%%%%%%%%%%%%%%%%%%%%%%%%%%%%%
\begin{eqnarray}
\eta^{L}_{_N}
~&=& ~ \sum^{heavy}_k~ U_{ek}^2
\frac{m_p}{M_k}\,, 
\label{etaL}
\end{eqnarray}
%%%%%%%%%%%%%%%%%%%%%%%%%%%%%%%%
%
%%%%%%%%%%%%%%%%%%%%%%%%%%%%%%%%%%%%%%%%%%%%%%
%%%%%%%%%%%%%%%%%%%%%%%%%%%%%%%%%%%%%%%%%%%%%%
 \begin{figure}[h!]
  \begin{center}
 {\includegraphics[width=10cm]{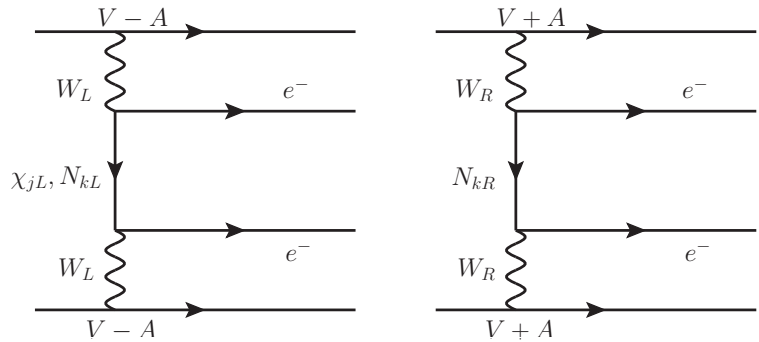}}
  \end{center}
\caption{\label{Feyn1} Feynman diagrams for the
$\betabeta$-decay, generated by the 
light and heavy LH Majorana neutrino exchange 
(left panel) and the heavy (RH) 
Majorana neutrino exchange (right panel).}
\end{figure}
%%%%%%%%%%%%%%%%%%%%%%%%%%%%%%%%%%%%%%%%%%%%%%
%%%%%%%%%%%%%%%%%%%%%%%%%%%%%%%%%%%%%%%%%%%%%%
%
\noindent where $m_p$ is the proton mass and  $U_{ek}$ 
is the element of the neutrino mixing matrix 
through which $N_k$ couples to the electron in the 
weak charged lepton current. We note that 
$|\eta^{L}_{_N}|$ is suppressed by both 
the ratio $m_p/M_k$ and the magnitude of 
$U_{ek}$ (see, e.g., \cite{Antusch:2008tz}).

 If the weak interaction Lagrangian contains also
$(V+A)$ (i.e., right-handed (RH)) charged currents 
coupled to a RH charged weak boson $W_R$, 
as, e.g., in the 
% $SU(2)_L\times SU(2)_R\times U(1)$ theories,
Left-Right Symmetric theories, 
we can have also a contribution to the 
% $0\nu\beta\beta$-decay 
$\betabeta$-decay amplitude generated 
by the exchange of virtual $N_k$
coupled to the electron in 
the hypothetical $(V+A)$ CC
part of the weak interaction 
Lagrangian \cite{WR1}.
In this case the corresponding 
LNV parameter can be written as:
%%%%%%%%%%%%%%%%%%%%%%%%%%%%%%
\begin{eqnarray}
\eta^{R}_{_N}
~&=& ~ \left (\frac{M_W}{M_{WR}}\right )^{4}\,\sum^{heavy}_k~ V_{e k}^2  
% {\xi'}_k ~
\frac{m_p}{M_k}.
\label{etaR}
\end{eqnarray}
%%%%%%%%%%%%%%%%%%%%%%%%%%%%%%%%
%
Here $V_{e k}$ are the elements of a mixing matrix 
by which $N_k$ couple to the electron  
in the $(V+A)$ charged lepton current
\footnote{We have neglected the 
contributions to 
$\eta^{R}_{_N}$, and, more generally,
to the $\betabeta$-decay amplitude 
due to the possible but small mixing 
between $W$ and $W_R$ bosons.}, 
$M_W$ is the mass of the 
Standard Model charged weak boson,  
$M_W \cong 80$ GeV, and $M_{WR}$ is the mass of $W_R$. 
It follows from the existing data that \cite{WR2}
$M_{WR} \gtap 2.5$ TeV. Thus, $|\eta^{R}_{_N}|$ is 
suppressed by the factor  $(M_W/M_{WR})^{4}$.

If CP invariance does not hold, which 
we will assume to be the case in what 
follows, $U_{ek}$ and $V_{ek}$ 
will contain physical CP violating 
phases at least for some $k$ and thus 
the parameters $\eta^{L}_{_N}$ 
and $\eta^{R}_{_N}$ will not be real.

 As can be shown, 
the nuclear matrix elements 
corresponding to the two mechanisms of 
$\betabeta$-decay with exchange of heavy 
Majorana neutrinos $N_k$, described in the 
present subsection, are the same
and are  given in \cite{Sim99}.
We will denote them by ${M}^{0\nu}_{_N}$ 
(and ${M'}^{0\nu}_{_N}$).

Finally, it is important to note that 
the current factor in the  
$\betabeta$-decay amplitude describing 
the two final state electrons, 
has different forms in the cases of 
$\betabeta$-decay mediated by  $(V-A)$ and 
by $(V+A)$ CC weak interactions, namely,
$\bar{e}(1 + \gamma_5)e^c \equiv 
2\bar{e_L}\, (e^c)_R$ 
and  $\bar{e}(1 - \gamma_5)e^c \equiv 2\bar{e_R}\,(e^c)_L $, 
respectively, where $e^c = C(\bar{e})^{T}$, 
$C$ being the charge conjugation  matrix 
(see, e.g., \cite{BiPet87}).
The difference in the chiral structure
of the two currents leads to a 
specific phase space factor
of the interference term
in the rate of $\betabeta$-decay,
triggered by two mechanisms 
whose respective contributions to the 
$\betabeta$-decay amplitude 
involve the two different 
electron current factors.
The phase space factor of the 
interference term under discussion
is significantly smaller than 
the phase space factors 
of the contributions 
to the $\betabeta$-decay rate
due to each of the two mechanisms, 
which leads to a relatively 
strong suppression of the 
interference term \cite{HPR83}
(see further).

%%%%%%%%%%%%%%%%%%%%%%%%%%
%
\subsection{SUSY Models with R-Parity Non-conservation}
%
%%%%%%%%%%%%%%%%%%%%%%%%%%%
%
The SUSY models with R-parity non-conservation 
include LNV couplings which  can trigger the
% $0\nu\beta\beta$ decay.
$\betabeta$-decay.
Let us recall that the R-parity is a multiplicative
quantum number defined by $R=(-1)^{2S+3B+L}$, where $S$, 
$B$ and $L$ are the spin, the baryon
and lepton numbers of a given particle. 
The ordinary (Standard Model) particles have $R=+1$,
while their superpartners carry 

%%%%%%%%%%%%%%%%%%%%%%%%%%%%%%%%%%%%%%%%%%%%%%
%%%%%%%%%%%%%%%%%%%%%%%%%%%%%%%%%%%%%%%%%%%%%%
 \begin{figure}[h!]
  \begin{center}
 \subfigure
 {\includegraphics[width=4.2cm]{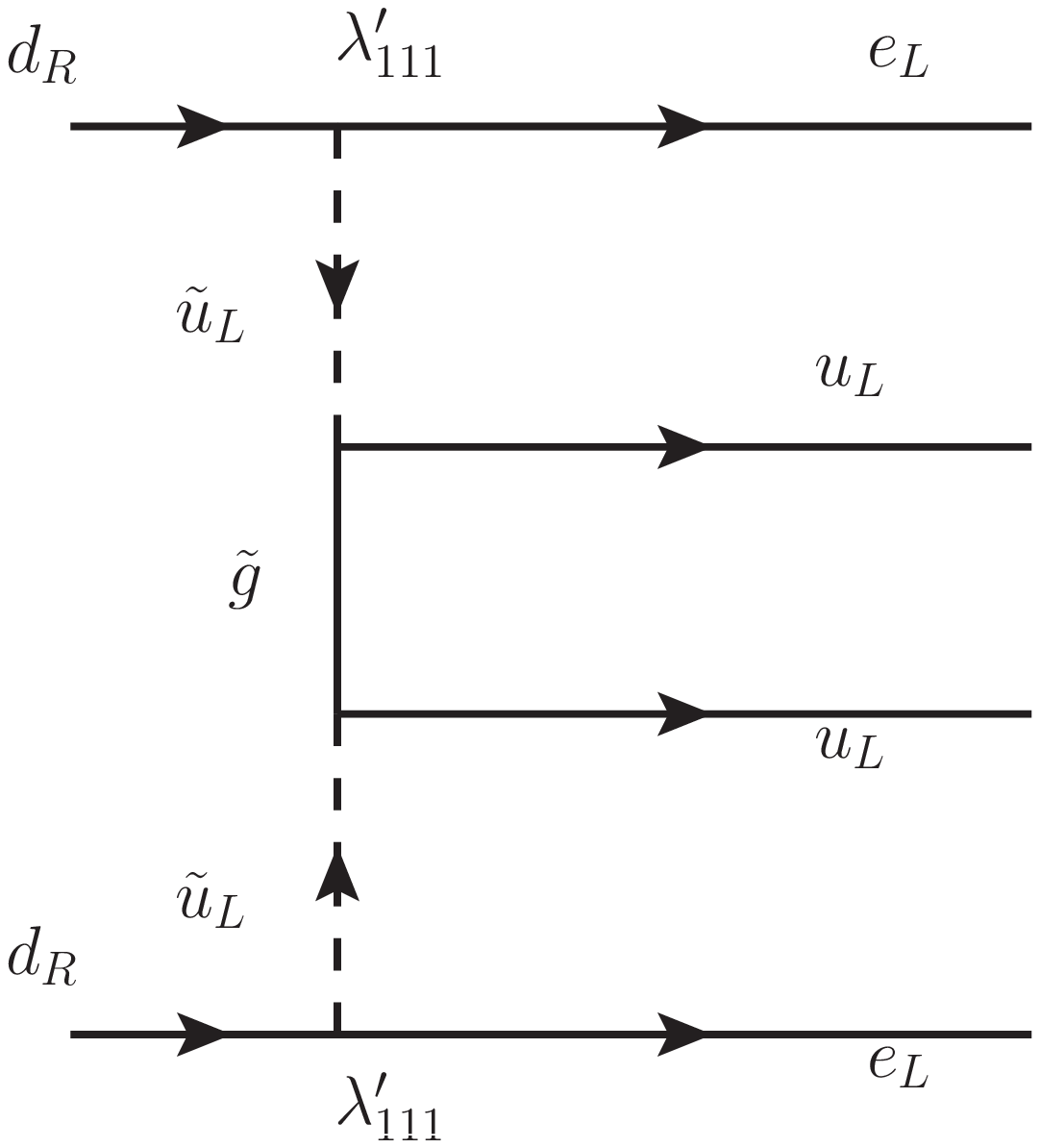}}
\hspace{20pt}
 \subfigure
   {\includegraphics[width=4.7cm]{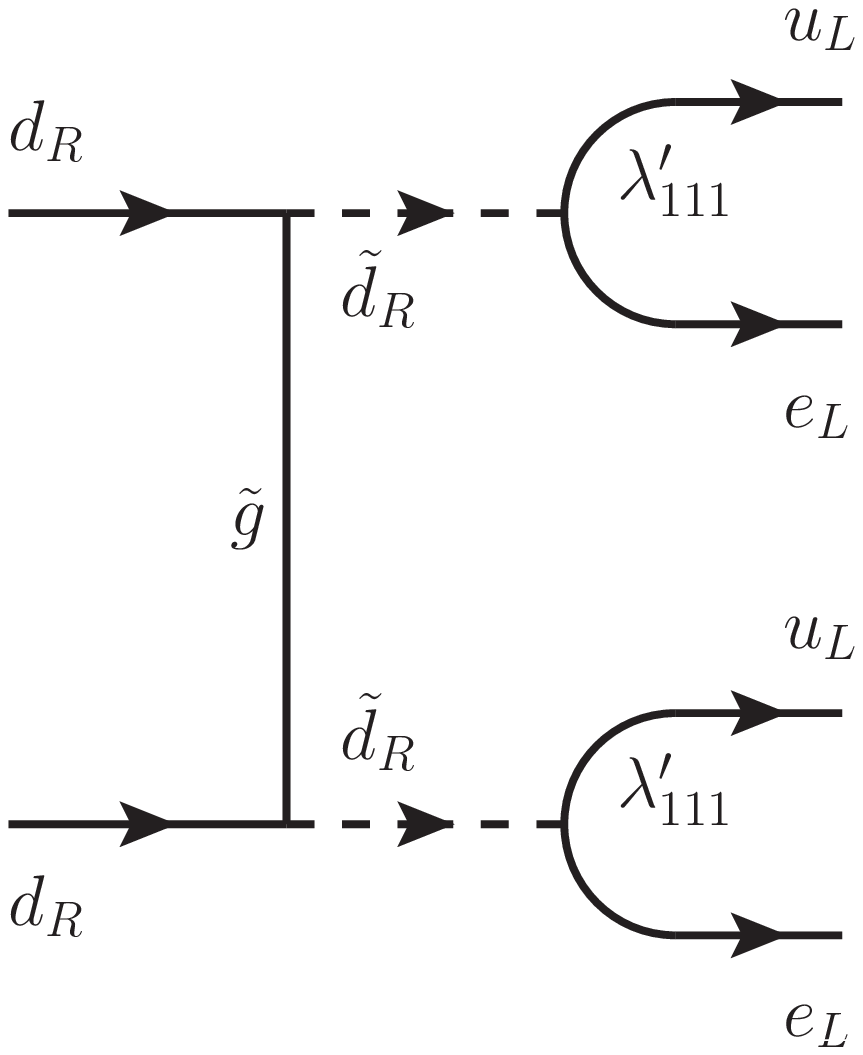}}
\hspace{20pt}
\subfigure
   {\includegraphics[width=4.2cm]{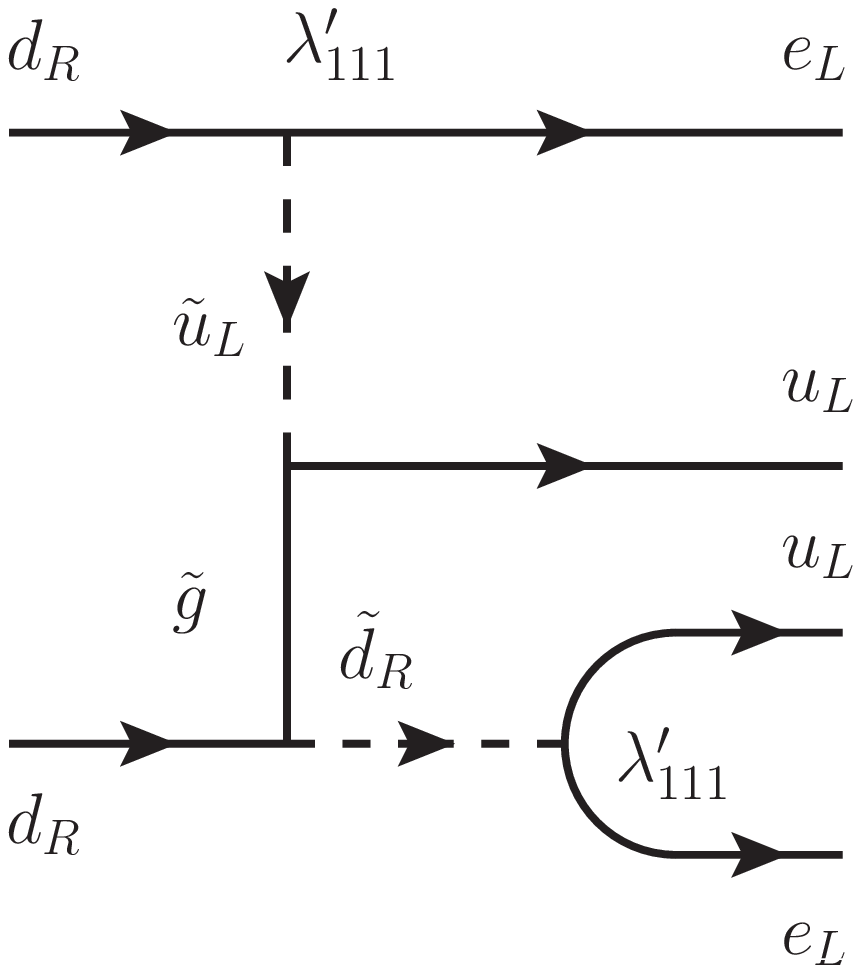}}
   \end{center}
     \caption{\label{Feyn2} Feynman diagrams 
for $\betabeta$-decay due to the gluino exchange mechanism. 
}
\end{figure}
%%%%%%%%%%%%%%%%%%%%%%%%%%%%%%%%%%%%%%%%%%%%%%
%%%%%%%%%%%%%%%%%%%%%%%%%%%%%%%%%%%%%%%%%%%%%%
%
\noindent $R=-1$.
The LNV couplings emerge in this class of SUSY models from
the R-parity breaking (\rp) part of the superpotential
%%%%%%%%%%%%%%%%%%%%%%%%%
\begin{equation}
W_{\slashed{R}_{p}}\, = \, \lambda _{ijk}L_{i}L_{j}E_{k}^{c}+\lambda
_{ijk}^{\prime }L_{i}Q_{j}D_{k}^{c} + \mu_i L_i H_2,  
\label{W-Rp}
\end{equation}
%%%%%%%%%%%%%%%%%%%%%%%%%%
%
where $L$, $Q$ stand for lepton and quark $SU(2)_{L}$
doublet left-handed superfields, while $E^{c},D^{c}$ for
lepton and down quark singlet superfields. Here
we concentrate only on the trilinear $\lambda'$-couplings.
The $\lambda'$-couplings of the first family of particles and 
sparticles relevant  for $\betabeta$-decay are given
in terms of the fields of the 
LH electron, electron neutrino $\nu_{eL}$, 
LH selectron $\tilde e_L$ and sneutrino $\tilde \nu_{eL}$, 
LH and RH $u$- and $d$-quarks,
$u_{L,R}$ and $d_{L,R}$, and LH and RH 
$u$- and $d$-squarks, 
$\tilde u_{L,R}$,  $\tilde d_{L,R}$, by:
%%%%%%%%%%%%%%%%%%%%%%%%%%%%%%%%
\be 
\mathcal{L}_{\slashed{R}_p} = \lambda'_{111}\left[ (\bar u_L\,\,
\bar d_L) \binom{e_R^c}{-\nu^c_{eR}}\tilde d_R + (\bar e_L \,\, \bar \nu_{eL})
d_R \binom{\tilde u^{\ast}_L}{-\tilde d^{\ast}_L}+ (\bar u_L \,\, \bar
d_L) d_R\binom{\tilde e_L^{\ast}}{-\tilde \nu^{\ast}_{eL}} \right]  
+ h.c.
\ee
%%%%%%%%%%%%%%%%%%%%%%%%%%%%%%%%%%%%%%%
%

At the quark-level there are basically 
two types of \rp SUSY mechanisms 
of $\betabeta$-decay:  
a short-range one with exchange of heavy 
Majorana and scalar SUSY particles 
(gluinos and squarks, and/or 
neutralinos and selectrons)
\cite{Moh86,Ver87,dbd-gluino-neutralino,dbd-gluino-neutralino1,Fedor-Wodecki,ramsey}, and a long-range
mechanism involving the exchange of 
both heavy squarks 
and light Majorana neutrinos
\cite{FKS98b,bivalle,HKK:96,Pes,squark}. We 
will call the latter 
the ``squark-neutrino'' mechanism. 

%%%%%%%%%%%%%%%%%%%%%%%%%%%%%%%%%%
%
\subsubsection*{The Case of Gluino Exchange Dominance}
%
%%%%%%%%%%%%%%%%%%%%%%%%%%%%%%%%%%%

 Assuming the dominance of the gluino exchange
in the short-range mechanism, 
one obtains  the following 
% simplified 
expression for the corresponding 
LNV parameter:
%%%%%%%%%%%%%%%%%%%%%%%%%%%%
\begin{equation}
\eta_{\lambda'} =
\frac{\pi \alpha_s}{6}
\frac{\lambda^{'2}_{111}}{G_F^2 m_{\tilde d_R}^4}
\frac{m_p}{m_{\tilde g}}\left[
1 + \left(\frac{m_{\tilde d_R}}{m_{\tilde u_L}}\right)^2\right]^2.
\label{etalambda}
\end{equation}
%%%%%%%%%%%%%%%%%%%%%%%%%%%%%
%
Here, $G_F$ is the Fermi constant,
$\alpha_s = g^2_3/(4\pi )$, $g_3$ being the $\rm SU(3)_c$ 
gauge coupling constant.
$m_{{\tilde u}_L}$, $m_{{\tilde d}_R}$ and $m_{\tilde g}$ are masses 
of the LH u-squark, RH d-squark and gluino, respectively.  

 The nuclear matrix element associated with 
the gluino exchange mechanism,
${M}^{0\nu}_{\lambda'}$,
was calculated in \cite{FKSS97,FKS98a}.
The electron current factor in the 
term of the $\betabeta$-decay
amplitude corresponding to the gluino 
exchange mechanism under discussion
has the form 
$\bar{e}(1 + \gamma_5)e^c \equiv 
2\bar{e_L}\, (e^c)_R$, 
i.e., it coincides with that 
of the light (or heavy LH) Majorana 
neutrino exchange. 
Thus,  when calculating the 
$\betabeta$-decay rate,
the interference between the two terms present 
in the  $\betabeta$-decay amplitude,
corresponding to the indicated 
two mechanisms, has the same 
phase space factor as 
the contributions due to each
of the two mechanisms.
As a consequence, the interference term
will not suffer from phase space suppression. 

% %%%%%%%%%%%%%%%%%%%%
% %
%
\subsubsection*{The Squark-Neutrino Mechanism}
%
% %
% %%%%%%%%%%%%%%%%%%%%%
% %
In the case of squark-neutrino exchange \cite{squark}, 
the $\betabeta$-decay amplitude 
does not vanish in the limit of zero 
Majorana neutrino mass,  
in contrast to the case of the ``standard'' light (LH)
Majorana neutrino exchange.
This is a consequence of 
the chiral structure of the corresponding 
\rp SUSY couplings.
The contribution due to the squark-neutrino exchange
to the $\betabeta$-decay amplitude is
roughly proportional to the momentum of the 
virtual neutrino, 
which is of the order of the Fermi momentum 
of the nucleons inside of nucleus, $p_F\approx 100$ MeV. 
The corresponding LNV parameter is defined as
\cite{squark}
%%%%%%%%%%%%%%%%%%%%%%%%%%%%%%%%
\begin{eqnarray}
\label{eta}
\eta_{\tilde q} &=& \sum_{k} \frac{\lambda'_{11k}\lambda'_{1k1}}{2
\sqrt{2} G_F}
\sin{2\theta^{d}_{(k)} }\left( \frac{1}{m^2_{\tilde d_1 (k)}} -
\frac{1}{m^2_{\tilde d_2 (k)}}\right)\,. 
\end{eqnarray}
%%%%%%%%%%%%%%%%%%%%%%%%%%%%%%%%%%%
%
Here we use the notations $d_{(k)} = d, s, b$ and 
assumed that there are 3 light Majorana neutrinos.
This LNV parameter vanishes in the absence 
of $\tilde d_{kL}-\tilde d_{kR}$ - mixing,
i.e.,  when $\theta^{d}=0$. The nuclear matrix 
element for the squark-neutrino mechanism,
${M}^{0\nu}_{\tilde q}$, is given in 
\cite{squark}.

%%%%%%%%%%%%%%%%%%%%%%%
%
\section{Nuclear Structure Calculations}
%
%%%%%%%%%%%%%%%%%%%%%%%%%%%%%%
%
  In what follows the 
% $0\nu\beta\beta$-decay 
$\betabeta$-decay nuclear matrix elements
${M'}^{0\nu}_\nu$, ${M'}^{0\nu}_N$, ${M'}^{0\nu}_{\lambda'}$
and ${M'}^{0\nu}_{\tilde q}$ are evaluated
for ${^{76}Ge}$, ${^{82}Se}$,  ${^{100}Mo}$  
and ${^{130}Te}$.  These nuclei are considered as most 
probable candidate sources for the next generation 
of the experiments searching for 
% $0\nu\beta\beta$-decay.
$\betabeta$-decay.

We used the Self-consistent Renormalized 
Quasiparticle Random Phase Approximation\\ 
\noindent (SRQRPA) \cite{srpa} to calculate the 
nuclear matrix elements (NMEs) of interest. 
The SRQRPA takes into account the Pauli 
exclusion principle and conserves the mean 
particle number in correlated ground state.

 For each of the four nuclei, two choices of single-particle 
basis are considered. The intermediate size model 
space has 12 levels (oscillator shells N=2-4)
for ${^{76}Ge}$ and ${^{82}Se}$, 16 levels 
(oscillator shells N=2-4 plus 
the f+h orbits from N=5) for ${^{100}Mo}$  and 18 levels 
(oscillator shells
%%%%%%%%%%%%%%%%%%%%%%%%%%%%%%%%
\begin{table}
\centering
% \begin{table*}[htb]
% \begin{center}
\caption{\label{table.1} 
The phase-space factor $G^{0\nu}(E_0,Z)$
and the nuclear matrix elements ${M'}^{0\nu}_\nu$ (light Majorana neutrino 
exchange mechanism),
${M'}^{0\nu}_N$ (heavy Majorana neutrino exchange mechanism), 
${M'}^{0\nu}_{\lambda'}$ (mechanism of gluino exchange dominance  
in SUSY with trilinear R-parity breaking term)  and 
${M'}^{0\nu}_{\tilde q}$ (squark-neutrino mechanism)
for the $\betabeta$-decays of $^{76}Ge$, $^{100}Se$, $^{100}Mo$ and
$^{130}Te$. The nuclear matrix elements were 
obtained within the Self-consistent
Renormalized Quasiparticle Random Phase 
Approximation (SRQRPA). See text for details. 
}
\renewcommand{\tabcolsep}{1.1mm}
 \renewcommand{\arraystretch}{1}
{\footnotesize
\begin{tabular}{lcccccccccccccc}
\hline\hline
&&&&&&&\\
Nuclear & $G^{0\nu}(E_0,Z)$ & & & \multicolumn{2}{c}{$|{M'}^{0\nu}_\nu|$} &  & 
\multicolumn{2}{c}{$|{M'}^{0\nu}_N|$} & & \multicolumn{2}{c}{$|{M'}^{0\nu}_{\lambda'}|$} & & \multicolumn{2}{c}{$|{M'}^{0\nu}_{\tilde q}|$} \\ 
\cline{5-6} \cline{8-9} \cline{11-12} \cline{14-15}
 transition & [$y^{-1}$] & &  & \multicolumn{2}{c}{$g_A =$}  & & 
\multicolumn{2}{c}{$g_A =$}  & & \multicolumn{2}{c}{$g_A =$} 
& & \multicolumn{2}{c}{$g_A =$} \\ 
 & & NN pot. & m.s.  & 1.0 & 1.25  & & 1.0 & 1.25 & & 1.0 & 1.25 & & 1.0 & 1.25 \\\hline
   & & &
    & & & & & & & & & & &    \\
$^{76}Ge\rightarrow {^{76}Se}$ & $7.98~10^{-15}$  & Argonne &
  intm.  & 3.85 & 4.75  & & 172.2 & 232.8 & & 387.3 & 587.2 & & 396.1 & 594.3 \\
   & & &
  large  & 4.39 & 5.44  & & 196.4 & 264.9 & & 461.1 & 699.6 & & 476.2 & 717.8 \\
   & & CD-Bonn &
  intm.  & 4.15 & 5.11  & & 269.4 & 351.1 & & 339.7 & 514.6 & & 408.1 & 611.7 \\
   & & &
  large  & 4.69 & 5.82  & & 317.3 & 411.5 & & 392.8 & 595.6 & & 482.7 & 727.6 \\
   & & &
     & & & & & & & & & & &    \\
$^{82}Se\rightarrow {^{82}Kr}$  & $3.53~10^{-14}$ & Argonne &
  intm.  & 3.59 &  4.54  & & 164.8 & 225.7 & & 374.5 & 574.2 & & 379.3 & 577.9 \\
   & & &
  large  & 4.18 &  5.29 & & 193.1 & 262.9 & & 454.9 & 697.7 & & 465.1 & 710.2 \\
   & & CD-Bonn &
  intm.  & 3.86 &  4.88 & & 258.7 & 340.4 & & 328.7 & 503.7 & & 390.4 & 594.5 \\
   & & &
  large  & 4.48 &  5.66 & & 312.4 & 408.4 & & 388.0 & 594.4 & & 471.8 & 719.9 \\
   & & &
     & & & & & & & & & & &    \\
$^{100}Mo\rightarrow {^{100}Ru}$  & $5.73~10^{-14}$ &  Argonne &
  intm.  & 3.62 & 4.39 & & 184.9 & 249.8 & & 412.0 & 629.4 & & 405.1 & 612.1 \\
   & & &
  large  & 3.91 & 4.79 & & 191.8 & 259.8 & & 450.4 & 690.3 & & 449.0 & 682.6 \\
   & & CD-Bonn &
  intm.  & 3.96 & 4.81 & & 298.6 & 388.4 & & 356.3 & 543.7 & & 415.9 & 627.9 \\
   & & &
  large  & 4.20 & 5.15 & & 310.5 & 404.3 & & 384.4 & 588.6 & & 454.8 & 690.5 \\
   & & &
     & & & & & & & & & & &    \\
$^{130}Te\rightarrow {^{130}Xe}$ & $5.54~10^{-14}$ &  Argonne &
  intm. &  3.29 & 4.16 & & 171.6 & 234.1 & & 385.1 & 595.2 & & 382.2 & 588.9 \\
   & & &
  large  & 3.34 & 4.18 & & 176.5 & 239.7 & & 405.5 & 626.0 & & 403.1 & 620.4 \\
   & & CD-Bonn &
  intm.  & 3.64 & 4.62 & & 276.8 & 364.3 & & 335.8 & 518.8 & & 396.8 & 611.1 \\
   & & &
  large  & 3.74 & 4.70 & & 293.8 & 384.5 & & 350.1 & 540.3 & & 416.3 & 640.7 \\
   & & &
     & & & & & & & & & & &    \\
\hline\hline
\end{tabular}}
% \end{center}
%\end{table*}
\end{table}
\noindent N=3,4 plus f+h+p orbits from N=5) 
for  ${^{130}Te}$.
The large size single particle space contains 21 levels 
(oscillator shells N=0-5)  for  ${^{76}Ge}$, ${^{82}Se}$ 
and ${^{100}Mo}$, 
and 23 levels for ${^{130}Te}$ (N=1-5 and $i$ orbits from N=6).  
In comparison with previous studies \cite{Rodin}, 
we omitted the small space model 
which is not sufficient to describe 
realistically the tensor part of the 
$\betabeta$-decay nuclear 
matrix elements.

The single particle energies were obtained by using
a  Coulomb--corrected Woods--Saxon potential. 
Two-body G-matrix elements we derived from 
the Argonne and the Charge Dependent Bonn (CD-Bonn) 
one-boson exchange potential within the Brueckner theory. 
The schematic pairing interactions have been
adjusted to fit the empirical pairing gaps 
\cite{cheo93}. The
particle-particle and particle-hole channels of the G-matrix
interaction of the nuclear Hamiltonian $H$ are renormalized by
introducing the parameters $g_{pp}$ and $g_{ph}$, respectively.
The calculations have been carried out for $g_{ph} = 1.0$.
The particle-particle strength parameter $g_{pp}$ 
of the SRQRPA is fixed by the data on the two-neutrino double 
beta decays \cite{Rodin,anatomy}.
In the calculation of the 
% $0\nu\beta\beta$-decay 
$\betabeta$-decay NMEs,  
the two-nucleon short-range correlations  
derived from same potential as residual interactions,
namely from the Argonne or CD-Bonn potentials, were considered 
\cite{src}.

The calculated NMEs ${M'}^{0\nu}_\nu$, ${M'}^{0\nu}_N$, 
${M'}^{0\nu}_{\lambda'}$ and ${M'}^{0\nu}_{\tilde q}$ are listed 
in Table \ref{table.1}. We see that a significant source 
of uncertainty is the value of the axial-vector coupling 
constant $g_A$. Further, the NMEs associated with 
heavy neutrino exchange 
are sensitive also to the choice of the NN interaction,
the CD-Bonn or Argonne potential. These types of realistic 
NN interaction differ mostly by the description of the 
short-range interactions.

% {\bf 
Finally, we notice that all NMEs given in Table \ref{table.1} are 
real and positive.
% }

% \newpage
%%%%%%%%%%%%%%%%%%%%%%%%%%%%%%
%
\section{Analysis}
%
%%%%%%%%%%%%%%%%%%%%%%%%%%%%%%

  We illustrate the possibility to get information 
about the different LNV parameters when 
two or more mechanisms are operative 
in  $\betabeta$-decay, analysing 
the following two cases. 
First we consider two competitive ``not-interfering''
mechanisms of $\betabeta$-decay: light left-handed Majorana
neutrino exchange and heavy right-handed 
Majorana neutrino exchange.
In this case the interference term 
arising in the $\betabeta$-decay half-life 
from the product of the 
contributions due to the two mechanisms  
in the $\betabeta$-decay amplitude,
is strongly suppressed \cite{HPR83} as a consequence of
the different chiral structure of the 
final state electron current 
in the two amplitudes.
The latter leads to a different 
phase-space factor for the 
interference term, which is typically 
by a factor of 10 smaller 
than the standard one
(corresponding to the contribution
to the $\betabeta$-decay half-life 
of each of the two mechanisms).
More specifically, the suppression factors 
for ${^{76}Ge}$, ${^{82}Se}$,  ${^{100}Mo}$  
and ${^{130}Te}$ read, respectively
\cite{HPR83}: 0.13; 0.08; 0.075 and 0.10.
It is particularly small for
$^{48}Ca$: 0.04.
In the analysis which follows we will neglect 
the contribution of the interference term 
in the  $\betabeta$-decay half-life. 
The effect of taking into account 
the interference term on the results thus 
obtained, as our numerical
calculations have shown, does not exceed
approximately 10\%.
 
  In the case of negligible 
interference term, the inverse value of the $\betabeta$-decay
half-life for a given isotope (A,Z) is given by:
%%%%%%%%%%%%%%%%%%%%%%%%%%%%%%%%%%
\be
\frac{1}{T^{0\nu}_{1/2,i}G^{0\nu}_i(E, Z)} \cong |\eta_\nu|^2
|{M'}^{0\nu}_{i, \nu}|^2 + |\eta_R|^2|{M'}^{0\nu}_{i,N }|^2\,,
\label{hl}
\ee
%%%%%%%%%%%%%%%%%%%%%%%%%%%%%%%%
%
where the index $i$ denotes the isotope. 
The values of the phase space factor $G^{0\nu}_i(E, Z)$ 
and of the NMEs ${M'}^{0\nu}_{i, \nu}$ and ${M'}^{0\nu}_{i,N }$
for $^{76}$Ge, $^{82}Se$, $^{100}$Mo and $^{130}$Te
are listed in Table \ref{table.1}.
The parameters  $|\eta_\nu|$ and $|\eta_R|$ are 
defined in eqs. (\ref{etanu}) and (\ref{etaR}).

  In the second illustrative 
case we consider $\betabeta$-decay triggered by
two active and ``interfering'' mechanisms: the 
light Majorana neutrino exchange and the 
gluino exchange. In this case, for a given nucleus,
the inverse of the $\betabeta$-decay half-life is given by:
%%%%%%%%%%%%%%%%%%%%%%%%%%%%%
\be
\frac{1}{T^{0\nu}_{1/2,i}G^{0\nu}_i(E, Z) }=|\eta_\nu|^2
|{M'}^{0\nu}_{i, \nu}|^2 + |\eta_{\lambda'}|^2|{M'}^{0\nu}_{i,\lambda'}|^2 +
2\cos\alpha |{M'}^{0\nu}_{i,\lambda'}||{M'}^{0\nu}_{i,\nu}||\eta_\nu||\eta_{\lambda'}|\,.
\label{hlint}
\ee
%%%%%%%%%%%%%%%%%%%%%%%%%%%%
%
Here $|\eta_{\lambda'}|$ 
is the basic parameter
of the gluino exchange mechanism
defined in eq. (\ref{etalambda})
and $\alpha$ is the relative phase of
$\eta_{\lambda'}$ and $ \eta_\nu$.
The values of the NMEs of the mechanisms considered
are listed in Table \ref{table.1}.

 In the illustrative examples of how one 
can extract information about
$ |\eta_\nu|$, $|\eta_R|$, etc. we use as input
hypothetical values of the $\betabeta$-decay
half-life of $^{76}$Ge satisfying 
the existing lower limits
and the value claimed in ref. \cite{Klap04}
\cite{HMGe76},
as well as the following hypothetical ranges
for $T^{0\nu}_{1/2}$($^{100}$Mo) and T$^{0\nu}_{1/2}$($^{130}$Te):
%%%%%%%%%%%%%%%%%%%%%%%%%%%%%%
\be
\begin{split} T^{0\nu}_{1/2}(^{76}Ge)\geq 1.9\times 10^{25} y
,&\quad
T^{0\nu}_{1/2}(^{76}Ge)= 2.23^{+0.44}_{-0.31}\times 10^{25} y\\
5.8\times 10^{23}y\leq T^{0\nu}_{1/2}(^{100}Mo)\leq5.8\times
10^{24}y,&\quad 3.0\times 10^{24}y\leq T^{0\nu}_{1/2}(^{130}Te)
\leq 3.0\times 10^{25}y
\end{split}
\label{limit}
\ee
%%%%%%%%%%%%%%%%%%%%%%%%%%%%%%
%
Let us note that $5.8\times 10^{23}$ y and $3.0\times 10^{24}$ y
are the existing lower bounds on the half-lives of 
$^{100}Mo$ and $^{130}Te$ \cite{NEMO3,CUORI}.

 In the analysis which follows we will present numerical 
results first for $g_A=1.25$ and using 
the NMEs calculated with 
the large size single particle basis 
(``large basis'')
and the Charge Dependent Bonn (CD-Bonn) 
potential. Later results for  
$g_A=1.0$, as well as for NMEs calculated with the 
Argonne potential, will also be reported.
 
 As we will see, in certain cases of 
at least one more mechanism being 
operative in $\betabeta$-decay  
beyond the light neutrino exchange, 
one has to take into account the 
upper limit on the absolute scale of neutrino masses 
set by the $^3H$ $\beta$-decay experiments 
\cite{MoscowH3,MainzKATRIN}: $m(\bar{\nu}_e) < 2.3$ eV.
In the case of $\betabeta$-decay, this limit implies 
a similar limit on the effective Majorana mass
\footnote{We remind the reader that for 
$m_{1,2,3} \gtap 0.1$ eV the neutrino mass spectrum 
is quasi-degenerate (QD), $m_1\cong m_2\cong m_3 \equiv m$,
$m^2_j >> \Delta m^2_{21},|\Delta m^2_{31}|$. 
In this case we have $m(\bar{\nu}_e) \cong m$ and 
$\meff \ltap m$.} 
$\meff < 2.3$ eV. The latter translates into 
the following limit on the conveniently 
rescaled parameter $|\eta_\nu|^2$:
%%%%%%%%%%%%%%%%%%%%%%%%%%%%%%%%%%%%%%%%
\be
|\eta_\nu|^2\times 10^{10} < 0.21\,.
\label{etanuH3MM}
\ee
%%%%%%%%%%%%%%%%%%%%%%%%%%%%%%%%%%%%%%%
% 

 A more stringent limit on the absolute neutrino mass scale
and therefore on $\meff$ is planned to be obtained in the 
KATRIN experiment \cite{MainzKATRIN}: 
$\meff < 0.2$ eV (90\% C.L.). This corresponds to the 
following prospective limit on $|\eta_\nu|^2$:
%%%%%%%%%%%%%%%%%%%%%%%%%%%%%%%%%%%%%%%%
\be
|\eta_\nu|^2\times 10^{10} < 1.6\times 10^{-3}\,.
\label{etanuKatrin}
\ee
%%%%%%%%%%%%%%%%%%%%%%%%%%%%%%%%%%%%%%%
% 
As the results presented in Section 6 indicate,
if the limit of 0.2 eV will be reached in the 
KATRIN experiment, this will lead to severe 
constraints on some of the solutions for 
$|\eta_\nu|^2$ obtained in the case 
of two ``interfering'' mechanisms, one of which 
is the light neutrino exchange.

%%%%%%%%%%%%%%%%%%%%%%%%%%%%%%%%%%%%%%%%
%
\section{\label{sec:anal} Two ``Non-Interfering'' Mechanisms}
%
%%%%%%%%%%%%%%%%%%%%%%%%%%%%%%%%%%%%%%%

  In the case of two
active non-interfering 
$\betabeta$-decay generating mechanisms, 
which we will assume to be
the light LH and heavy RH Majorana 
neutrino exchanges \cite{HPR83}, it is
possible to extract the absolute values of 
the corresponding two LNV fundamental 
parameters $|\eta_\nu|$ and $|\eta_R|$,
using the ``data'' on the half-lives
of two different
nuclei undergoing the $\betabeta$-decay.
Indeed, using eq. (\ref{hl}) we can set 
a system of two linear equations for two
unknowns using as input the two half-lives:
%%%%%%%%%%%%%%%%%%%%%%%%%%%%%%%%%%%%%%%%
\be
\frac{1}{T_{1} G_1}= |\eta_\nu|^2 |{M'}^{0\nu}_{1,\nu }|^2 +
|\eta_R|^2|{M'}^{0\nu}_{1,N }|^2,  \quad \frac{1}{T_{2} G_2}= |\eta_\nu|^2
|{M'}^{0\nu}_{2,\nu }|^2 + |\eta_R|^2|{M'}^{0\nu}_{2,N }|^2.
\label{2nonint}
\ee
%%%%%%%%%%%%%%%%%%%%%%%%%%%%%%%
%
The solutions read:
%%%%%%%%%%%%%%%%%%%%%%%%%%%%%%%%%
\be
\begin{split}
|\eta_\nu|^2 =\frac{|{M'}^{0\nu}_{2,N }|^2/T_1 G_1- 
|{M'}^{0\nu}_{1,N }|^2/T_2 G_2}
{|{M'}^{0\nu}_{1,\nu }|^2|{M'}^{0\nu}_{2,N }|^2-|{M'}^{0\nu}_{1,N }|^2
|{M'}^{0\nu}_{2,\nu }|^2},\quad
|\eta_R|^2=\frac{ |{M'}^{0\nu}_{1,\nu }|^2/T_2 G_2 - 
|{M'}^{0\nu}_{2,\nu }|^2/T_1 G_1}
{|{M'}^{0\nu}_{1,\nu }|^2|{M'}^{0\nu}_{2,N }|^2-
|{M'}^{0\nu}_{1,N }|^2|{M'}^{0\nu}_{2,\nu }|^2}.
\end{split}
\ee
%%%%%%%%%%%%%%%%%%%%%%%%%%
%
Obviously, solutions giving
$|\eta_\nu|^2 <0$ and/or $|\eta_R|^2<0$
are unphysical.
Given a pair of nuclei $(A_1,Z_1)$, $(A_2,Z_2)$
of the three $^{76}$Ge, $^{100}$Mo and $^{130}$Te
we will be considering, and $T_1$,
and choosing (for convenience) always
$A_1<A_2$, positive solutions
for $|\eta_\nu|^2$ and $|\eta_R|^2$
are possible for the following range of values of $T_2$:
%%%%%%%%%%%%%%%%%%%%%%%%%%%%%%%
\be
\frac{T_1G_1 |{M'}^{0\nu}_{1,N }|^2 }{G_2 |{M'}^{0\nu}_{2,N }|^2} \leq T_2 \leq
\frac{T_1G_1 |{M'}^{0\nu}_{1,\nu }|^2 }{G_2 |{M'}^{0\nu}_{2,\nu }|^2}\,,
\label{PosC}
\ee
%%%%%%%%%%%%%%%%%%%%%%%%%%%%%%%%%
%
where we have used the fact that
$|{M'}^{0\nu}_{1,\nu }|^2/|{M'}^{0\nu}_{2,\nu }|^2 > 
|{M'}^{0\nu}_{1,N }|^2 /|{M'}^{0\nu}_{2,N }|^2$
(see Table \ref{table.1})
\footnote{
 This condition will exhibit a 
relatively weak dependence
on the value of $g_A$ 
in the cases of mechanisms 
in which the Gamow-Teller 
term in the NMEs dominates
(as in the gluino exchange dominance 
and the squark-neutrino exchange mechanisms).
Indeed, the factor $(1.25)^4$
(corresponding to $g_A=1.25$) and
included in the definition of the 
phase space terms $G_{1,2}$,
cancels in the ratio $G_{1}/G_{2}$, and
$|{M'}_{1,\nu(N) }|^2/|{M'}_{2,\nu(N)}|^2 =
|M^{0\nu}_{1,\nu(N) }|^2/|M^{0\nu}_{2,\nu(N)}|^2$ 
(see eq. (\ref{Mprime})).}.
Using the values of the
phase-space factors and
the two relevant NME given in Table I 
in the columns ``CD-Bonn, large, $g_A=1.25$'',
we get the positivity
conditions for the ratio of the half-lives of the
different pairs of the three nuclei of interest:
%%%%%%%%%%%%%%%%%%%%%%%%%%%%%%%
\be
% 0.145  \leq \frac{T^{0\nu}_{1/2}(^{100}Mo)}{T^{0\nu}_{1/2}(^{76}Ge)} 
% \leq 0.178\,,
0.15  \leq \frac{T^{0\nu}_{1/2}(^{100}Mo)}{T^{0\nu}_{1/2}(^{76}Ge)} 
\leq 0.18\,,
% 0.165 \leq \frac{T^{0\nu}_{1/2}(^{130}Te)}{T^{0\nu}_{1/2}(^{76}Ge)} 
% \leq 0.221\,,
0.17 \leq \frac{T^{0\nu}_{1/2}(^{130}Te)}{T^{0\nu}_{1/2}(^{76}Ge)} 
\leq 0.22\,,
% 1.139 \leq \frac{T^{0\nu}_{1/2}(^{130}Te)}{T^{0\nu}_{1/2}(^{100}Mo)} 
% \leq 1.242\,.
1.14 \leq \frac{T^{0\nu}_{1/2}(^{130}Te)}{T^{0\nu}_{1/2}(^{100}Mo)} 
\leq 1.24\,.
\label{CDBonn125}
\ee
%%%%%%%%%%%%%%%%%%%%%%%%%%%%%%%%%
%
In  the case of $g_A=1.0$ we find:
%%%%%%%%%%%%%%%%%%%%%%%%%%%%%%%%%%%
\be
% 0.145  \leq \frac{T^{0\nu}_{1/2}(^{100}Mo)}{T^{0\nu}_{1/2}(^{76}Ge)} 
% \leq 0.174\,,
0.15  \leq \frac{T^{0\nu}_{1/2}(^{100}Mo)}{T^{0\nu}_{1/2}(^{76}Ge)} 
\leq 0.17\,,
% 0.167 \leq \frac{T^{0\nu}_{1/2}(^{130}Te)}{T^{0\nu}_{1/2}(^{76}Ge)} 
% \leq 0.227\,,
0.17 \leq \frac{T^{0\nu}_{1/2}(^{130}Te)}{T^{0\nu}_{1/2}(^{76}Ge)} 
\leq 0.23\,,
% 1.157 \leq \frac{T^{0\nu}_{1/2}(^{130}Te)}{T^{0\nu}_{1/2}(^{100}Mo)} 
% \leq 1.304\,.
1.16 \leq \frac{T^{0\nu}_{1/2}(^{130}Te)}{T^{0\nu}_{1/2}(^{100}Mo)} 
\leq 1.30\,.
\label{CDBonn10}
\ee
%%%%%%%%%%%%%%%%%%%%%%%%%%
%
It is quite remarkable that the physical solutions are possible only
if the ratio of the half-lives of all the pairs of
the three isotopes considered take values in very narrow
intervals. This result is a consequence of the
values of the phase space factors and of the
NME for the two mechanisms considered.
In the case of the Argonne potential, 
``large basis'' and $g_A=1.25~(1.0)$ 
(see Table \ref{table.1}) we get very 
similar results:
%%%%%%%%%%%%%%%%%%%%%%%%%%%%%%%
\be
% 0.145  \leq \frac{T^{0\nu}_{1/2}(^{100}Mo)}{T^{0\nu}_{1/2}(^{76}Ge)} 
% \leq 0.180\,,
% 0.176 \leq \frac{T^{0\nu}_{1/2}(^{130}Te)}{T^{0\nu}_{1/2}(^{76}Ge)} 
% \leq 0.244\,,
% 1.216 \leq \frac{T^{0\nu}_{1/2}(^{130}Te)}{T^{0\nu}_{1/2}(^{100}Mo)} 
% \leq 1.358\,.
0.15  \leq \frac{T^{0\nu}_{1/2}(^{100}Mo)}{T^{0\nu}_{1/2}(^{76}Ge)} 
\leq 0.18\,,
0.18 \leq \frac{T^{0\nu}_{1/2}(^{130}Te)}{T^{0\nu}_{1/2}(^{76}Ge)} 
\leq 0.24~(0.25)\,,
1.22 \leq \frac{T^{0\nu}_{1/2}(^{130}Te)}{T^{0\nu}_{1/2}(^{100}Mo)} 
\leq 1.36~(1.42)\,.
\label{Argonne12510}
\ee
%%%%%%%%%%%%%%%%%%%%%%%%%%%%%%%%%
%
If it is experimentally established that any of the three 
ratios of half-lives considered lies outside the interval of physical 
solutions  of $|\eta_\nu|^2$ and  $|\eta_R|^2$, obtained 
taking into account all relevant uncertainties, 
one would be led to conclude that the 
$\betabeta$-decay is not generated by the two mechanisms 
under discussion.
In order to show that the constraints 
given above are indeed satisfied, the 
relevant ratios of $\betabeta$-decay half-lives 
should be known with a remarkably small uncertainty 
(not  exceeding  approximately 5\% of the 
central values of the intervals). 

Obviously, given the half-life of one isotope, 
constraints similar to those described above 
can be derived on the half-life of any other isotope 
beyond those considered by us. 
Similar constraints can be obtained in all 
cases of two ``non-interfering'' mechanisms 
generating the $\betabeta$-decay. 
The predicted  intervals of half-lives 
of the various  isotopes will  
differ, in general, for the different 
pairs of  ``non-interfering''  mechanisms. 
However, these differences in the cases of the 
of the $\betabeta$-decay triggered by 
the exchange of heavy Majorana neutrinos 
coupled to (V+A) currents  and i) the gluino 
exchange mechanism, or ii)  the squark-neutrino exchange 
mechanism,  are extremely  small. One of the consequences of 
this feature of the different pairs of ``non-interfering'' 
mechanisms considered  by us is that if it will be possible 
to rule out one of them as the cause of  $\betabeta$-decay, 
most likely one will  be able to rule out all three of them. 
The set of constraints  under discussion will not be 
valid, in general,    if the  $\betabeta$-decay is 
triggered by two  ``interfering''  mechanisms 
with a non-negligible interference term, or 
by more than two mechanisms  
with significant contributions to the 
$\betabeta$-decay rates of the different 
nuclei.

  Evidently, if one of the two solutions is zero,
for example $|\eta_R|^2 = 0$, then
only one of the two $\betabeta$-decay mechanisms is active.
Since for the two mechanisms considered
we have $({M'}^{0\nu}_{1,\nu })^2({M'}^{0\nu}_{2,N })^2-({M'}^{0\nu}_{1,N })^2({M'}^{0\nu}_{2,\nu })^2\neq 0$,
the condition that $|\eta_R|^2 = 0$ reads:
%%%%%%%%%%%%%%%%%%%%%%%%%%%%%%%%
\be
 |{M'}^{0\nu}_{1,\nu }|^2\, T_1\,G_1 = |{M'}^{0\nu}_{2,\nu }|^2\,T_2\, G_2\,,~~|\eta_R|^2 = 0\,.
\label{etaR0}
\ee
%%%%%%%%%%%%%%%%%%%%%%%%%%%
%
The condition that $|\eta_\nu|^2 = 0$ has a similar form:
%%%%%%%%%%%%%%%%%%%%%%%%%%%%%%%%
\be
 |{M'}^{0\nu}_{1,N }|^2\, T_1\,G_1 = |{M'}^{0\nu}_{2,N }|^2\,T_2\, G_2\,,~~|\eta_{\nu}|^2 = 0\,.
\label{etanu0}
\ee
%%%%%%%%%%%%%%%%%%%%%%%%%%%
%
%%%%%%%%%%%%%%%%%%%%%%%%%%%%%%%%%%%%%%%%%%%%
\begin{table}
\centering \caption{\label{tab:table2} The predictions for the
half-life of a third nucleus $(A_3,Z_3)$, using as input in the
system of equations for $|\eta_\nu|^2$ and $|\eta_R|^2$, eq.
(\ref{2nonint}), the half-lives of two other nuclei $(A_1,Z_1)$ and
$(A_2,Z_2)$. The three nuclei used are $^{76}$Ge, $^{100}$Mo and
$^{130}$Te. The results shown are obtained for a fixed value of the
half-life of  $(A_1,Z_1)$ and assuming the half-life of  $(A_2,Z_2)$
to lie in a certain specific interval. The physical solutions for
$|\eta_\nu|^2$ and $|\eta_R|^2$ and then used to derive predictions
for the half-life of the third nucleus  $(A_3,Z_3)$. The latter are
compared with the lower limits given in eq.  (\ref{limit}).
The results quoted are obtained for NMEs given in
the columns  ``CD-Bonn, large, $g_A=1.25$'' 
in Table \ref{table.1}.
One star beside the isotope pair whose half-lives are used as input
for the system of equations (\ref{2nonint}), indicates predicted 
ranges of half-lives of the nucleus $(A_3,Z_3)$ 
that are not compatible with the lower bounds 
given in (\ref{limit}).}
%%%%%%%%%%%%%%%%%%%%%%%%%%%%%%%%%%%%%%%%%%%%
\renewcommand{\arraystretch}{0.8}
{\footnotesize\tt
\begin{tabular}{|l|l|c|c|}
\hline \hline
Pair  &  T$^{0\nu}_{1/2}(A_1,Z_1)$[yr] & T$^{0\nu}_{1/2}[A_2,Z_2]$[yr]   &  Prediction on $[A_3,Z_3]$[yr] \\
\hline
$^{76}Ge-^{100}Mo$  &  T$(Ge)= 2.23\cdot10^{25}$&  $3.23 \cdot10^{24}\leq T(Mo) \leq 3.97 \cdot10^{24}$   & $3.68 \cdot10^{24}\leq  T(Te) \leq 4.93 \cdot10^{24}$\\
$^{76}Ge-^{130}Te$  &   T$(Ge)=2.23\cdot10^{25}$&  $3.68 \cdot10^{24}\leq  T(Te) \leq 4.93 \cdot10^{24}$    & $3.23 \cdot10^{24}\leq T(Mo) \leq 3.97 \cdot10^{24}$\\
$^{76}Ge-^{100}Mo$  & T$(Ge)= 10^{26}$ & $1.45 \cdot10^{25}\leq
T(Mo)\leq 1.78 \cdot10^{25}$    & $1.65 \cdot10^{25}\leq  T(Te) \leq 2.21 \cdot10^{25}$  \\
$^{76}Ge-^{130}Te$  &  T$(Ge)= 10^{26}$ &  $1.65 \cdot10^{25}\leq
T(Te) \leq 2.21 \cdot10^{25}$   &$1.45 \cdot10^{25}\leq
T(Mo)\leq 1.78 \cdot10^{25}$ \\
$^{100}Mo-^{130}Te$ $\star$  &  T$(Mo)=5.8\cdot10^{23}$&  $6.61 \cdot10^{23}\leq T(Te)\leq 7.20 \cdot10^{23}$   &  $3.26 \cdot10^{24}\leq T(Ge) \leq4.00 \cdot10^{24}$\\
$^{100}Mo-^{130}Te$  &T$(Mo)=4\cdot10^{24}$& $ 4.56 \cdot10^{24}\leq T(Te) \leq 4.97 \cdot10^{24}$       & $ 2.25 \cdot10^{25}\leq T(Ge) \leq 2.76 \cdot10^{25}$\\
$^{100}Mo-^{130}Te$  & T$(Mo)=5.8\cdot10^{24}$&  $6.61\cdot10^{24}\leq T(Te) \leq 7.20 \cdot10^{24}$     &  $3.26\cdot10^{25}\leq T(Ge) \leq 4.00 \cdot10^{25}$\\
$^{100}Mo-^{130}Te$ $\star$ &  T$(Te)=3\cdot10^{24}$& $2.42
\cdot10^{24}\leq T(Mo) \leq 2.63 \cdot10^{24}$           & $1.36
\cdot10^{25}\leq T(Ge) \leq 1.82
\cdot10^{25}$\\
$^{100}Mo-^{130}Te$  &   T$(Te)=1.65\cdot10^{25}$&  $1.33 \cdot10^{25}\leq T(Mo) \leq 1.45 \cdot10^{25}$      &  $7.47 \cdot10^{25}\leq T(Ge) \leq 1.00 \cdot10^{26}$\\
$^{100}Mo-^{130}Te$  &   T$(Te)=3\cdot10^{25}$&  $2.42 \cdot10^{25}\leq T(Mo) \leq 2.63 \cdot10^{25}$         &  $1.36 \cdot10^{26}\leq T(Ge) \leq 1.82 \cdot10^{26}$\\
\hline\hline
\end{tabular}}
\end{table}
%%%%%%%%%%%%%%%%%%%%%%%%%%%%%%%%%%%%%%%%%
%

 If only the light neutrino exchange mechanism 
is present and the NME are correctly calculated, 
the $\betabeta$-decay effective Majorana
mass (and $|\eta_\nu|^2$) extracted from all three 
(or any number of) $\betabeta$-decay 
isotopes must be the same 
(see, e.g., \cite{PPW02,BiPet04}).
Similarly, if the heavy RH Majorana neutrino exchange
gives the dominant contribution, the
extracted value $|\eta_R|^2$ must be the same
for all three (or more) $\betabeta$-decay nuclei.

 We analyze next the possible solutions  for
different combinations of the
half-lives of the following isotopes:
$^{76}$Ge, $^{100}$Mo and $^{130}$Te. Assuming the half-lives of
two isotopes to be known and using the physical solutions
for $|\eta_\nu|^2$ and $|\eta_R|^2$ obtained using these half-lives,
one can obtain a prediction for the half-life of the
third isotope.
The predicted half-life should satisfy the existing lower
limits on it.
In the calculations the results of which are reported here,
we fixed the half-life of one of the two isotopes
and assumed the second half-life lies in an interval
compatible with the existing constraints.
We used the value of T$^{0\nu}_{1/2}(^{76}Ge)$
and values of  T$^{0\nu}_{1/2}(^{100}Mo)$ and T$^{0\nu}_{1/2}(^{130}Te)$
from the intervals given in (\ref{limit}).
The system of two equations is solved and
the values of  $|\eta_\nu|^2> 0$ and $|\eta_R|^2> 0$ thus
obtained were used to obtained predictions for the
half-life of the third isotope. The results 
for NMEs corresponding to the case 
 ``CD-Bonn, large, $g_A=1.25$'' 
(see Table \ref{table.1})
are given in Table \ref{tab:table2}.
We note that the experimental lower bounds quoted in
eq. (\ref{limit}) have to be taken into account
since they can further constrain the range of
allowed values of $|\eta_\nu|^2$ and $|\eta_R|^2$.
Indeed, an inspection of the values in
Table \ref{tab:table2} shows
that not all the ranges predicted for the third
half-life using the solutions obtained for
$|\eta_R|^2$ and $|\eta_{\nu}|^2$
are compatible with the lower bounds on the
half-live of the considered nuclear isotopes, given in
(\ref{limit}). In this case, some or all ``solution''
values of  $|\eta_R|^2$ and/or $|\eta_{\nu}|^2$
are ruled out. In Table \ref{tab:table2} 
these cases are marked by a star.

 The results reported in Table \ref{tab:table2}
are stable with respect to variations of the NMEs.
If we use the  NMEs corresponding to the case 
``CD-Bonn, large, $g_A=1.0$''
(see Table \ref{table.1}), the limits of the 

%%%%%%%%%%%%%%%%%%%%%%%%%%%%%%%%%%%%%%%%
\begin{figure}[h!]
   \begin{center}
 \subfigure
   {\includegraphics[width=7cm]{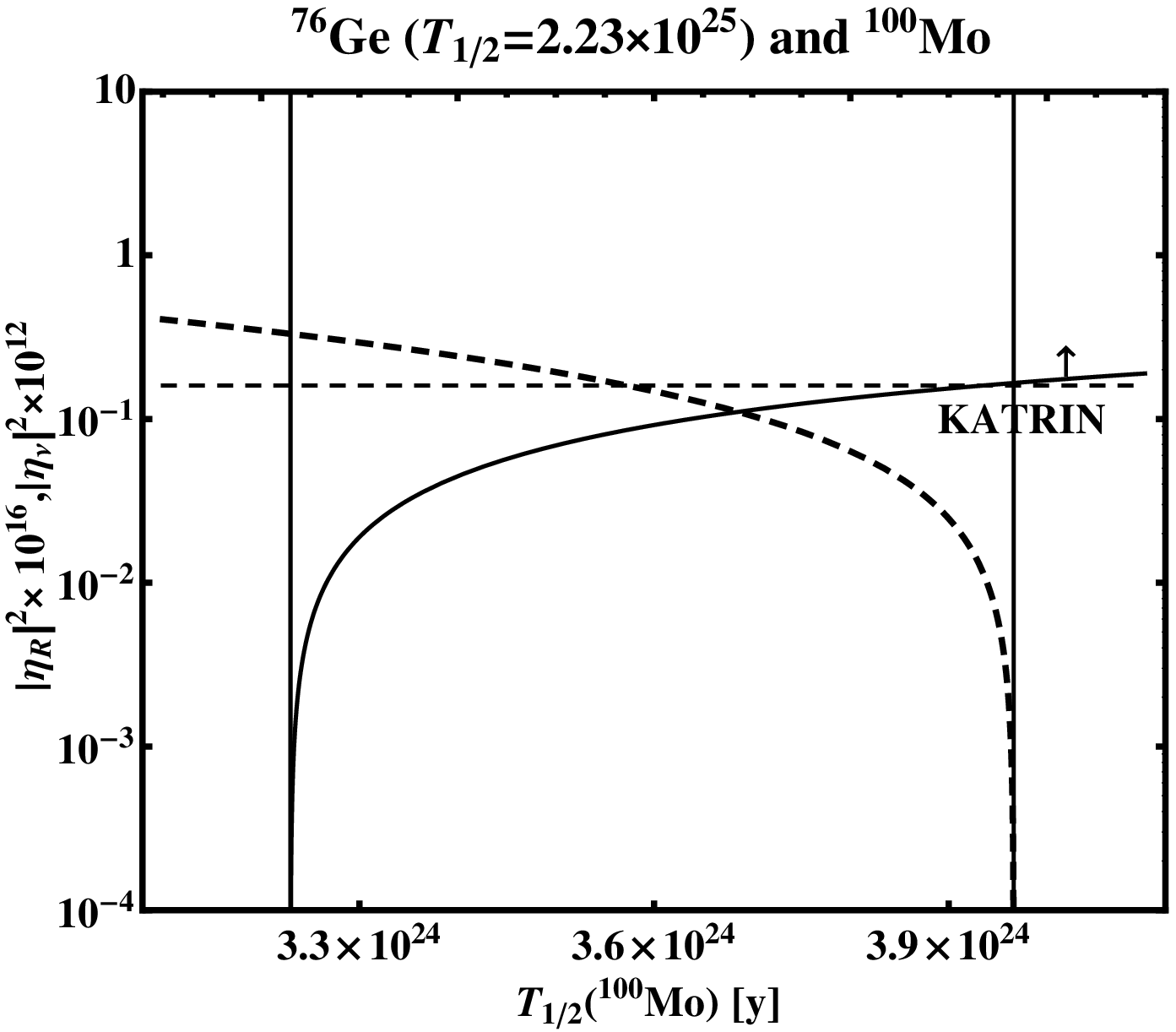}}
% \vspace{-0.8cm}
 \subfigure
   {\includegraphics[width=7cm]{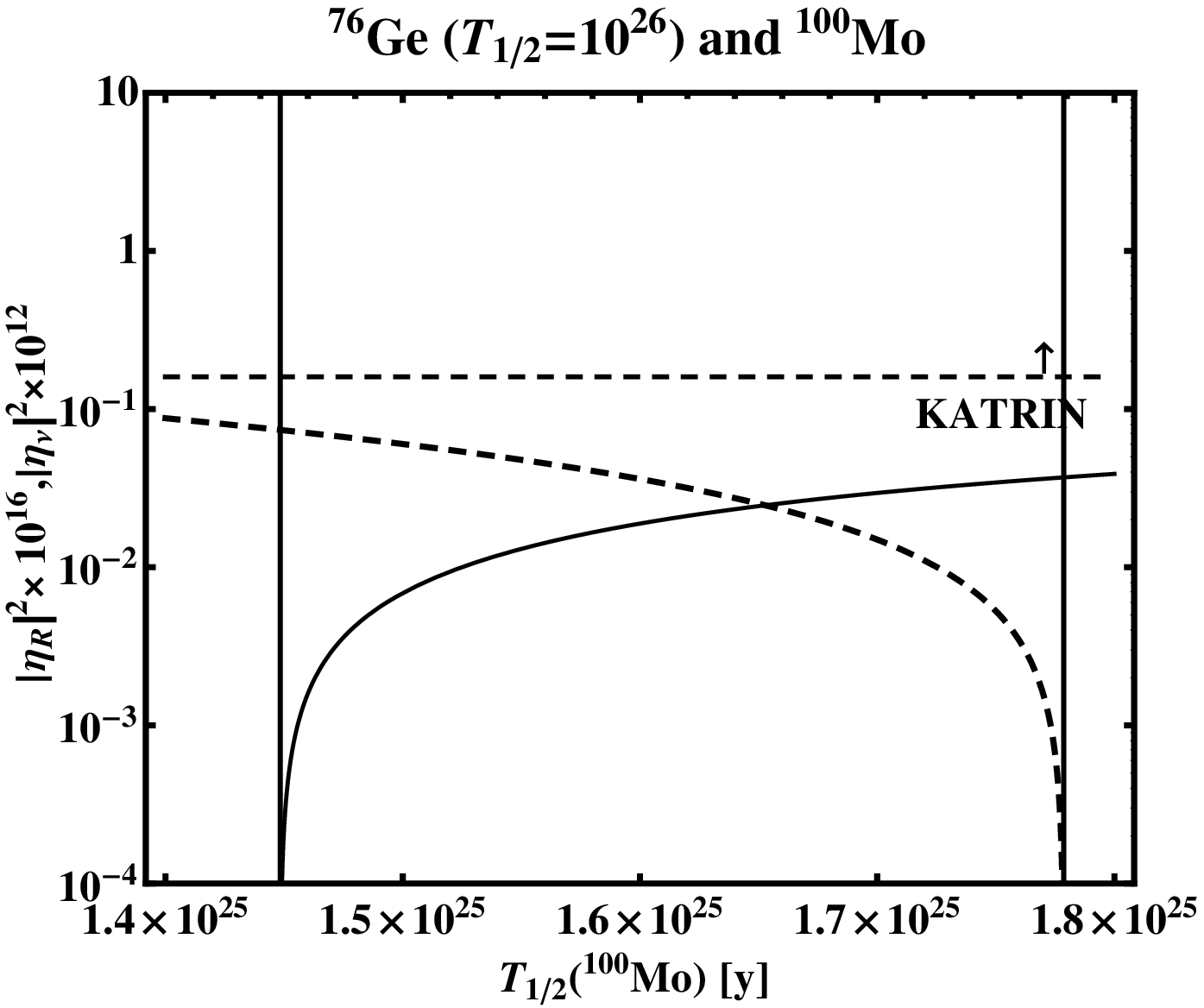}}
     \end{center}
 \vspace{-0.8cm}  
 \caption{ 
\label{fig:fig1}
 The values of the rescaled parameters
$|\eta_\nu|^2$ (solid line) and $|\eta_R|^2$ (dashed lined), obtained as
solutions of eq. (\ref{2nonint}) for two values of
$T^{0\nu}_{1/2}(^{76}Ge)$ and values
of $T^{0\nu}_{1/2}(^{100}Mo)$ lying in a specific
interval. The physical (positive) solutions
are delimited by the two  vertical lines.
The lower bound on  $T^{0\nu}_{1/2}(^{130}Te)$ given in
(\ref{limit}) does not lead to further 
constraints on $|\eta_\nu|^2$ and $|\eta_R|^2$. 
The horizontal dashed line corresponds to the prospective 
upper limit from the upcoming $^3H$ $\beta$-decay 
experiment KATRIN \cite{MainzKATRIN}. See text for further details.
}
\end{figure}
%%%%%%%%%%%%%%%%%%%%%%%%%%%%%%
% \vspace{-1.0cm}
%%%%%%%%%%%%%%%%%%%%%%%%%%%%%%%%%%%%%%%%
\begin{figure}[h!]
   \begin{center}
 \subfigure
   {\includegraphics[width=7cm]{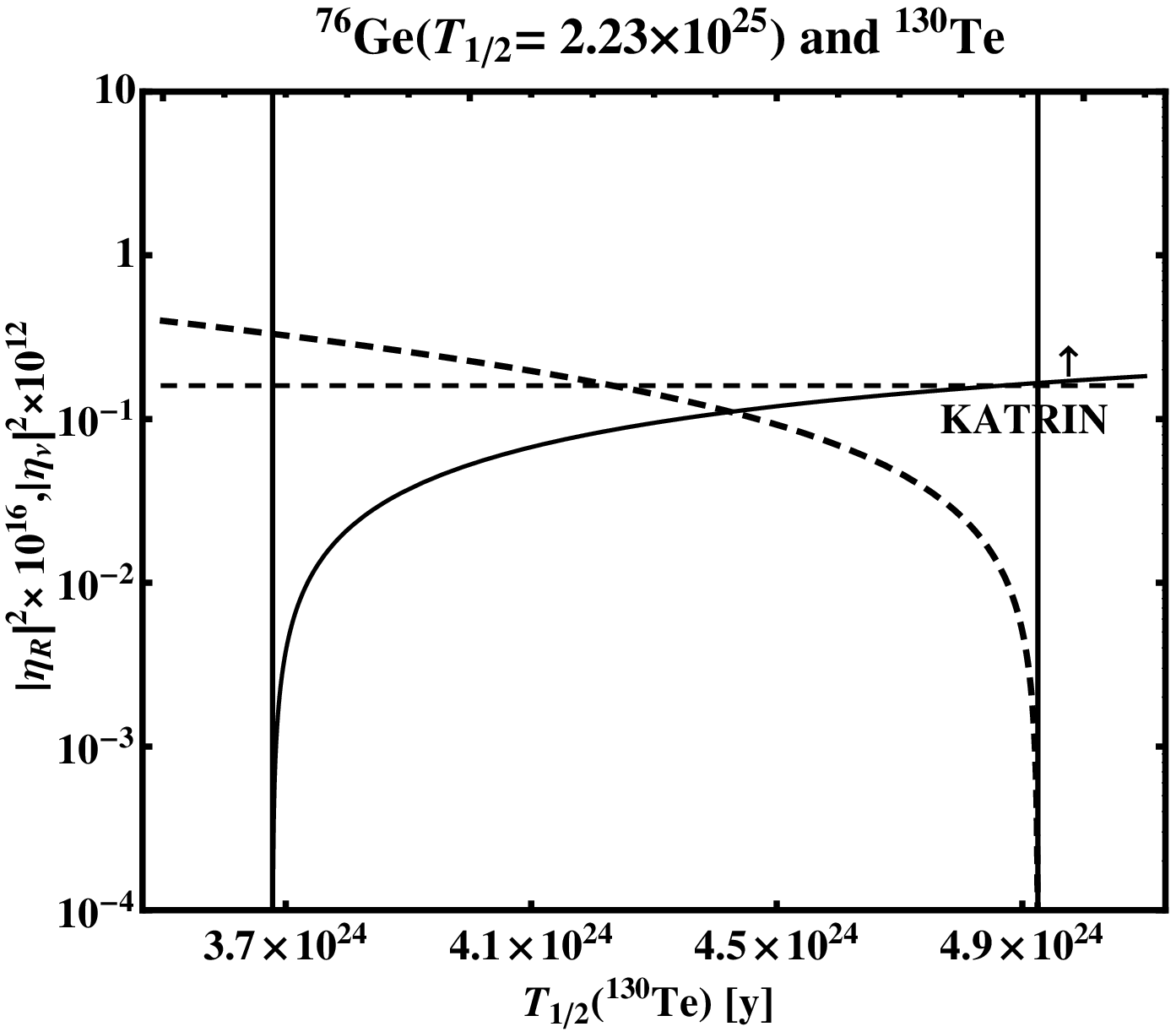}}
 \vspace{5mm}
 \subfigure
   {\includegraphics[width=7cm]{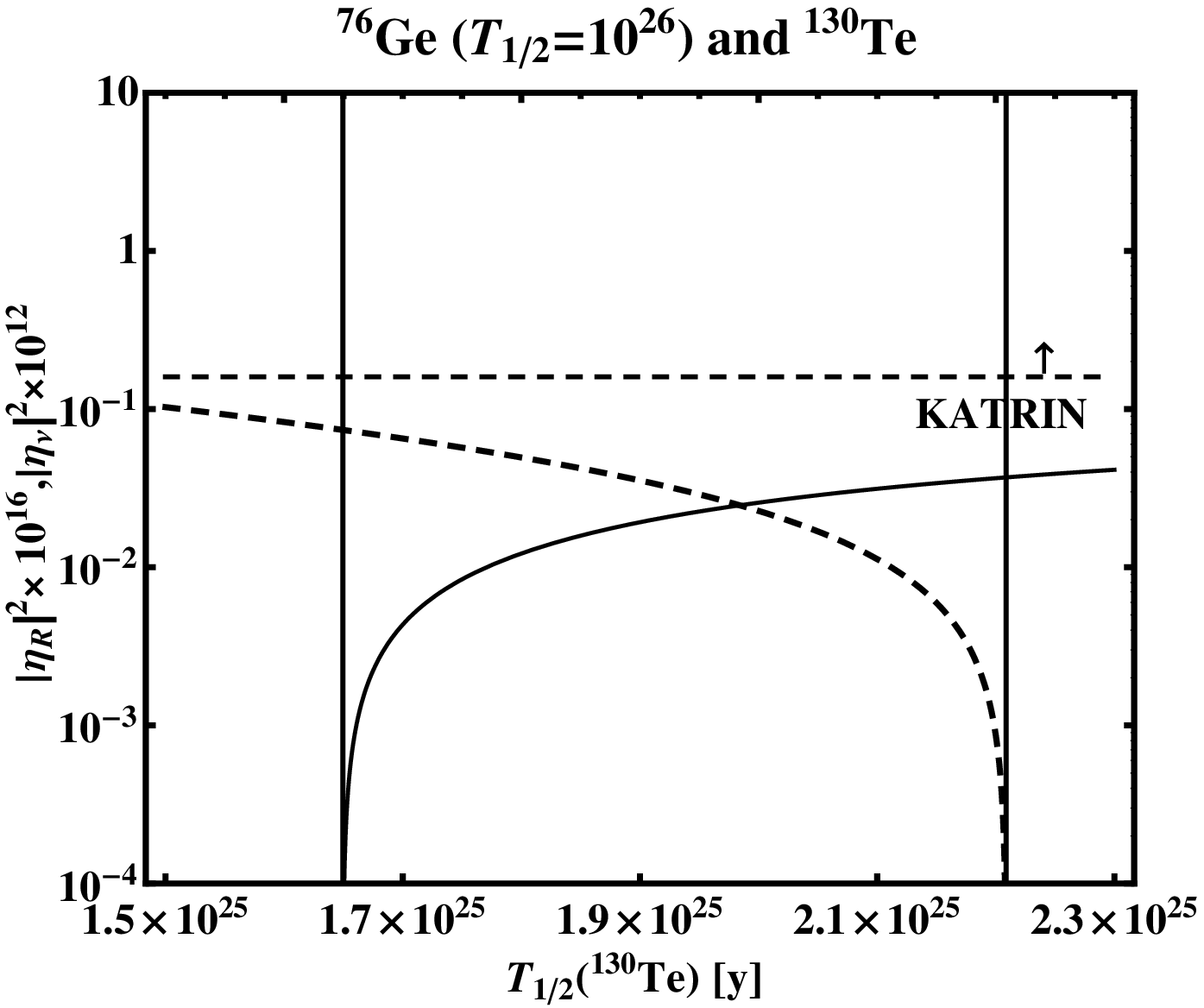}}
     \end{center}
\vspace{-0.8cm}
   \caption{ 
\label{fig:fig2}
The same as in Fig. \ref{fig:fig1}, but using as input 
hypothetical values of the half-lives of 
$^{76}Ge$ and $^{130}Te$, $T^{0\nu}_{1/2}(^{76}Ge)$ and 
 $T^{0\nu}_{1/2}(^{130}Te)$.
The physical (positive) solutions
are delimited by the two  vertical lines.
The lower bound on  $T^{0\nu}_{1/2}(^{100}Mo)$ given in
(\ref{limit}) does not lead to further constraints on $|\eta_{\nu,R}|^2$.
}
\end{figure}
%%%%%%%%%%%%%%%%%%%%%%%%%%%%%%
% \vspace{-0.8cm}
%

\noindent intervals quoted in  Table \ref{tab:table2}
change by $\pm 5\%$. If instead 
we use the NMEs corresponding 
to the Argonne potential,
``large basis''  
and $g_A=1.25$ ($g_A=1.0$),
the indicated limits change 
by $\pm 10\%$ ($\pm 14\%$).

\begin{figure}[h!]
  \begin{center}
 \subfigure
 %  {\includegraphics[scale=0.44]{MoFixedB.eps}}
{\includegraphics[width=7cm]{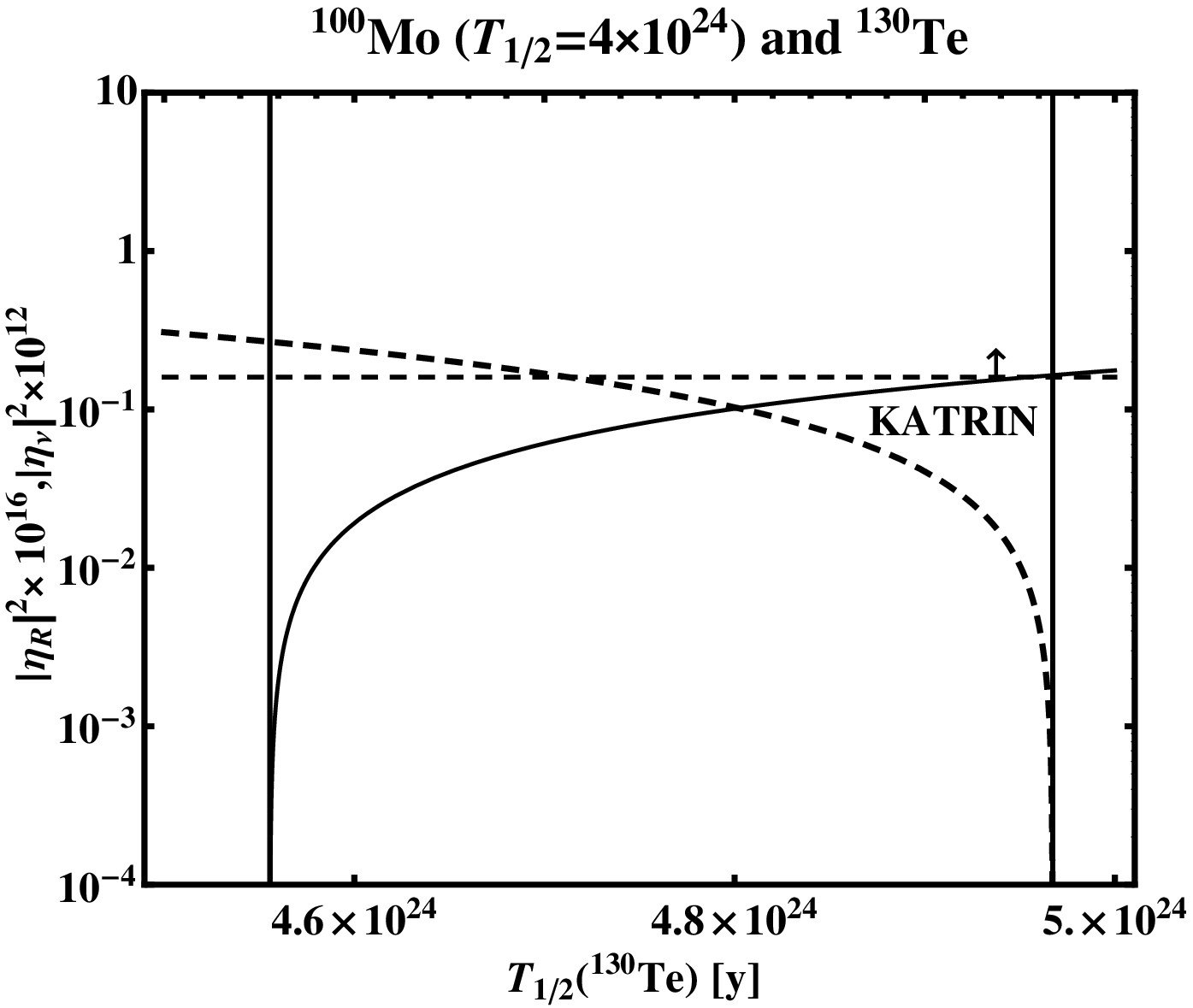}}
 % \vspace{5mm}
 \subfigure
%   {\includegraphics[scale=0.44]{MoFixedA.eps}}
{\includegraphics[width=7cm]{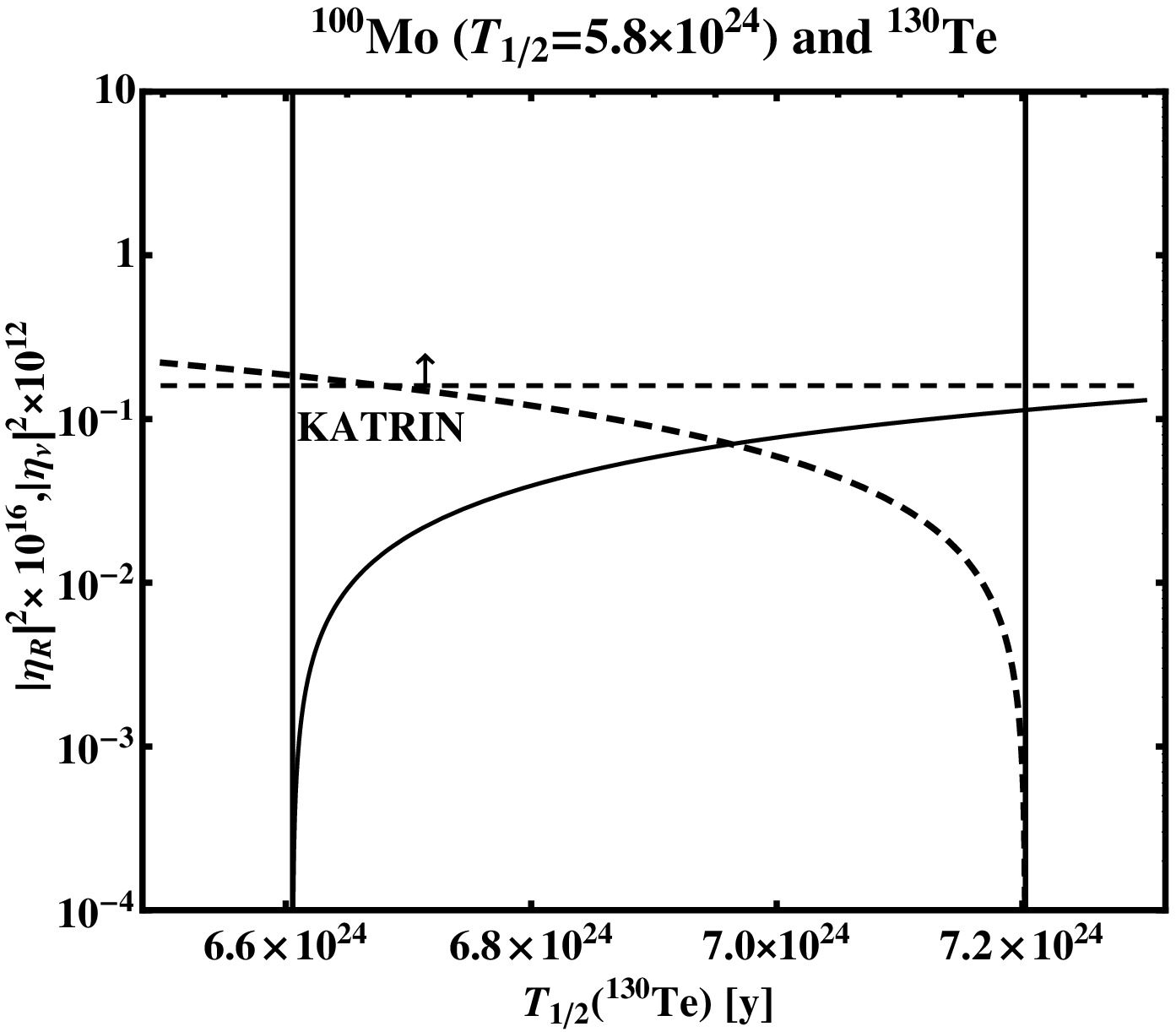}}
     \end{center}
\vspace{-0.8cm}   
 \caption{
\label{fig:fig3}
The same as in Fig. \ref{fig:fig1}, but using as input 
two values of the half-life of $^{100}Mo$ 
and values of the half-life of $^{130}Te$ 
lying in a specific interval.
% $^{130}Te$, $T^{0\nu}_{1/2}(^{100}Mo)$ and 
% $T^{0\nu}_{1/2}(^{130}Te)$.
% The values of the rescaled parameters $|\eta_\nu|^2$ (solid line)
% and $|\eta_R|^2$ (dashed lined), obtained as solutions of eq. (6)
% for two values of $T^{0\nu}_{1/2}(^{100}Mo)$ and values of
% $T^{0\nu}_{1/2}(^{130}Te)$ lying in a specific interval. 
The physical (positive) solutions are delimited by the two 
vertical lines. 
The lower bound on $T^{0\nu}_{1/2}(^{76}Ge)$ given in 
(\ref{limit}) does not lead to further constraints 
on $|\eta_\nu|^2$ and $|\eta_R|^2$.
}
\end{figure}

\begin{figure}[h!]
  \begin{center}
 \subfigure
   {\includegraphics[width=7cm]{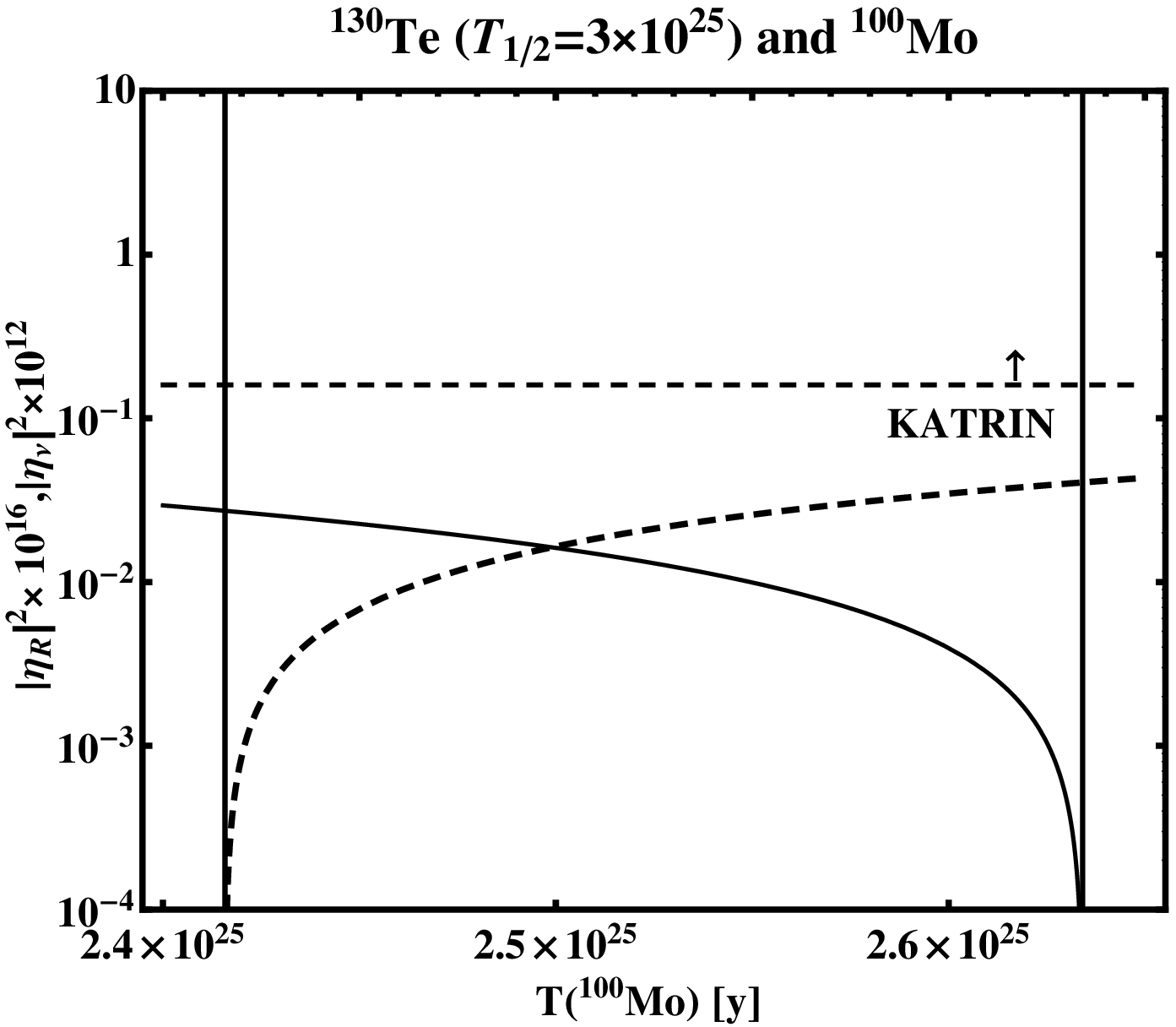}}
% \vspace{5mm}
 \subfigure
   {\includegraphics[width=7cm]{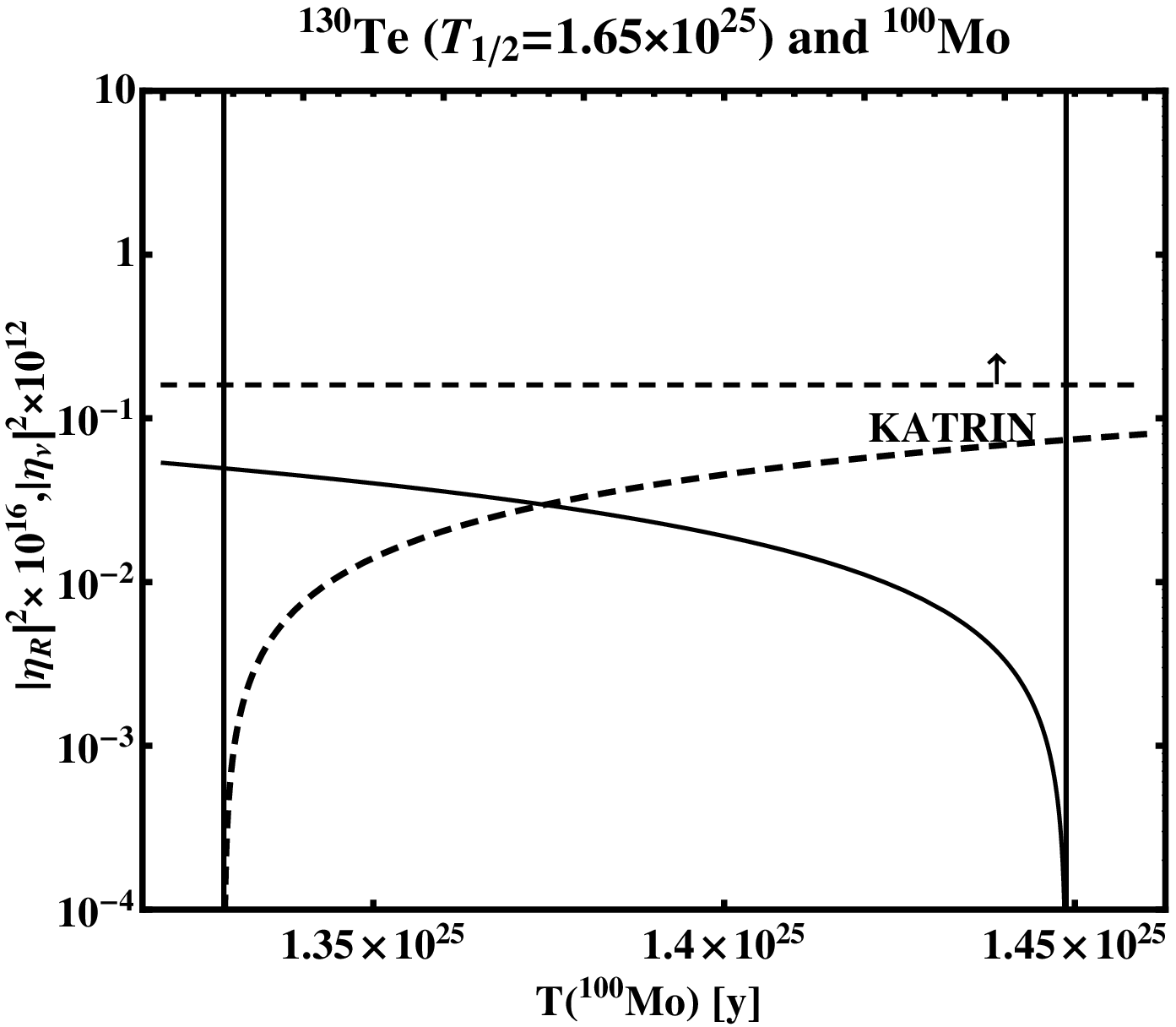}}
     \end{center}
\vspace{-0.8cm}   
\caption{ 
\label{fig:fig4}
The same as in Fig. \ref{fig:fig3}, but using as input 
two values of the half-life of $^{130}Te$ 
and values of the half-life of $^{100}Mo$ 
lying in a specific interval.
% The values of the rescaled parameters
% $|\eta_\nu|^2$ (thick line) and $|\eta_R|^2$ (dashed lined), obtained as
% solutions of eq. (6) for two values of
% $T^{0\nu}_{1/2}(^{130}Te)$ and values
% of $T^{0\nu}_{1/2}(^{100}Mo)$ lying in a specific
% interval. 
The physical (positive) solutions
are delimited by the two thick vertical lines.
The lower bound on $T^{0\nu}_{1/2}(^{76}Ge)$ 
% given in (\ref{limit}) 
does not lead to further constraints on
$|\eta_\nu|^2$ and $|\eta_R|^2$.
}
\end{figure}

\begin{figure}[h!]
  \begin{center}
   {\includegraphics[width=7cm]{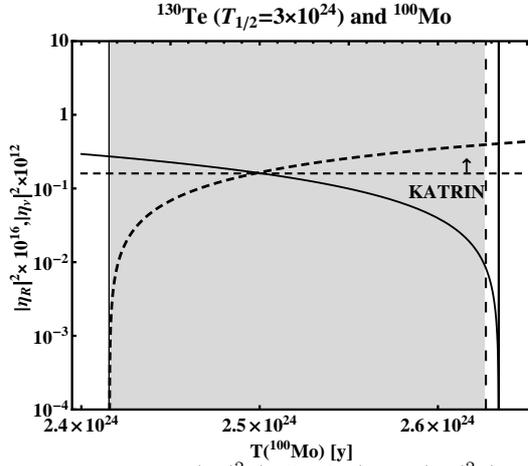}}
 \end{center}
\vspace{-0.8cm}
\caption{
\label{fig:figTeC} The values of the rescaled parameters
$|\eta_\nu|^2$ (solid line) and $|\eta_R|^2$ (dashed lined),
obtained as solutions of eq. (\ref{2nonint}) for the minimum  value of
$T^{0\nu}_{1/2}(^{130}Te)$ specified in eq. (\ref{limit}). The
physical (positive) solutions are delimited by the two 
vertical lines. The gray region is an excluded
due to the lower bound on $T^{0\nu}_{1/2}(^{76}Ge)$ 
quoted in (\ref{limit}).
}
\end{figure}

  These results and considerations are illustrated
in Figs. \ref{fig:fig1}-\ref{fig:figTeC}.
The  horizontal dashed line in these figures 
corresponds to the prospective limit 
planned to be obtained in the upcoming KATRIN 
experiment \cite{MainzKATRIN}.  
In figure \ref{fig:fig1} we show the solutions for
$|\eta_R|^2$ and/or $|\eta_{\nu}|^2$
(conveniently rescaled), obtained for two values of
$T^{0\nu}_{1/2}(^{76}Ge) = 2.23\times 10^{25}$ y and 10$^{26}$ y,
assuming $T^{0\nu}_{1/2}(^{100}Mo)$ has a value in
a certain interval.
In the case of $T^{0\nu}_{1/2}(^{76}Ge) = 2.23\times 10^{25}$ and
$T^{0\nu}_{1/2}(^{76}Ge) = 10^{26}$ y,
the derived physical values of 
$|\eta_R|^2$ and $|\eta_{\nu}|^2$ 
lead to predictions for $T^{0\nu}_{1/2}(^{100}Te)$ which are
compatible with the existing lower limit (Fig. \ref{fig:fig1}, 
left and right panel).  We get similar results  
using as input in the system of two equations for $|\eta_R|^2$
%
%%%%%%%%%%%%%%%%%%%%%%%%%%%%%%%%%%
%%%%%%%%%%%%%%%%%%%%%%%%%%%%%%%%%
% \vspace{-1.0cm}
%%%%%%%%%%%%%%%%%%%%%%%%%%%%%%%%%%%%%%%%%
%%%%%%%%%%%%%%%%%%%%%%%%%%%%%%%%%%%%%
%
%
\noindent and $|\eta_{\nu}|^2$ the half-lives of different pairs of isotopes,
and the lower limit of the half-life of the third as an additional
constraint. They are presented in Figs. \ref{fig:fig2} - \ref{fig:fig4}.
In Fig. \ref{fig:figTeC} we show the solutions for
$|\eta_\nu|^2$ and $|\eta_R|^2$
for $T^{0\nu}_{1/2}(^{130}Te)= 3\times 10^{24}$ y and
$T^{0\nu}_{1/2}(^{100}Mo)= (2.42 - 2.63)\times 10^{24}$ y.
In contrast to the cases illustrated in 
Figs. \ref{fig:fig1} - \ref{fig:fig4},
most of the solution values of $|\eta_\nu|^2$ and $|\eta_R|^2$ 
are excluded by taking into account the lower bound
on $T^{0\nu}_{1/2}(^{76}Ge)$ given in eq. (\ref{limit}).

\begin{figure}[htbp]
\centering%
\subfigure
   {\includegraphics[width=7cm]{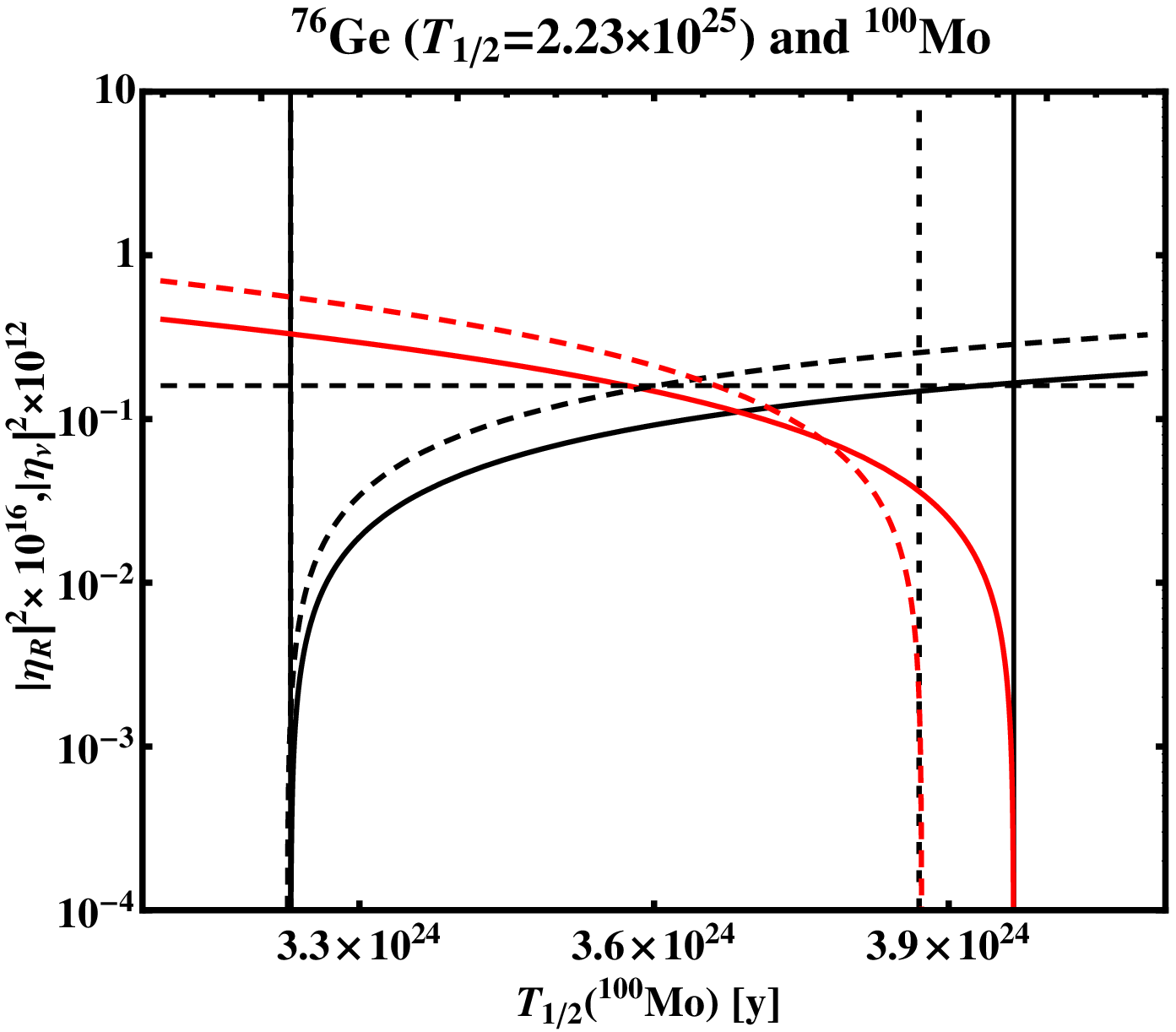}}
 \vspace{2mm}
 \subfigure
   {\includegraphics[width=7cm]{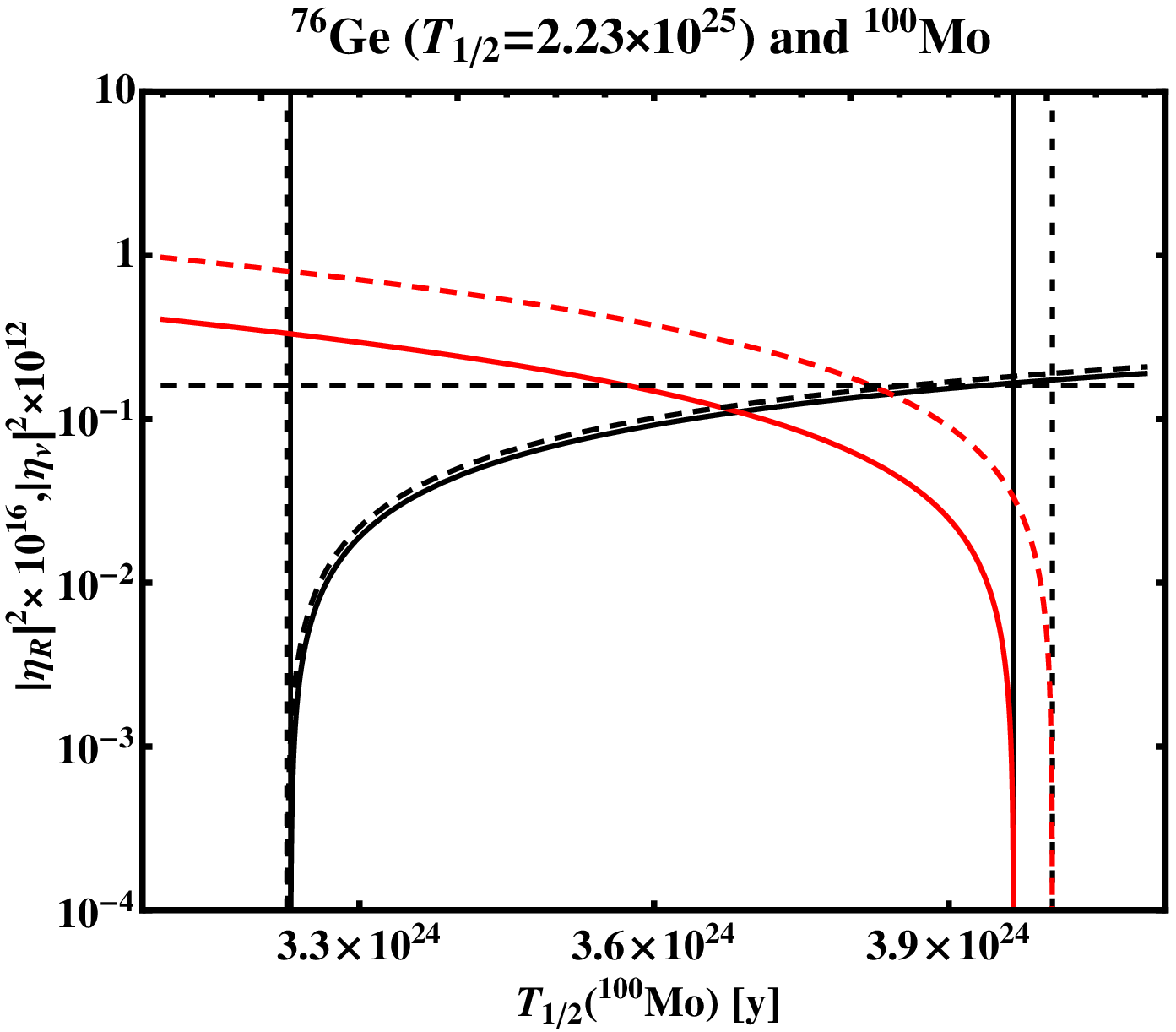}}
\vspace{2mm}
\subfigure
   {\includegraphics[width=7cm]{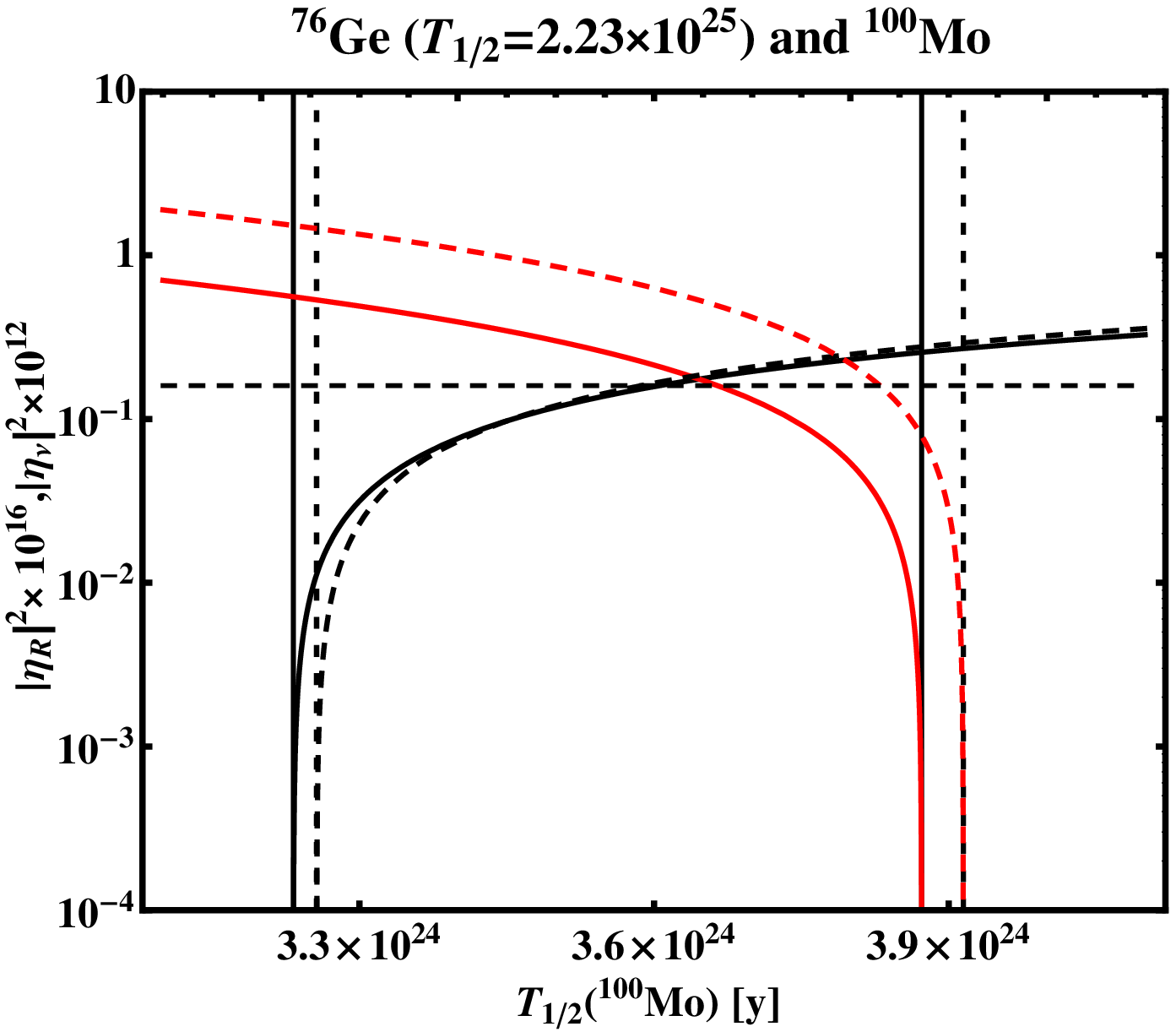}}
 \vspace{2mm}
 \subfigure
   {\includegraphics[width=7cm]{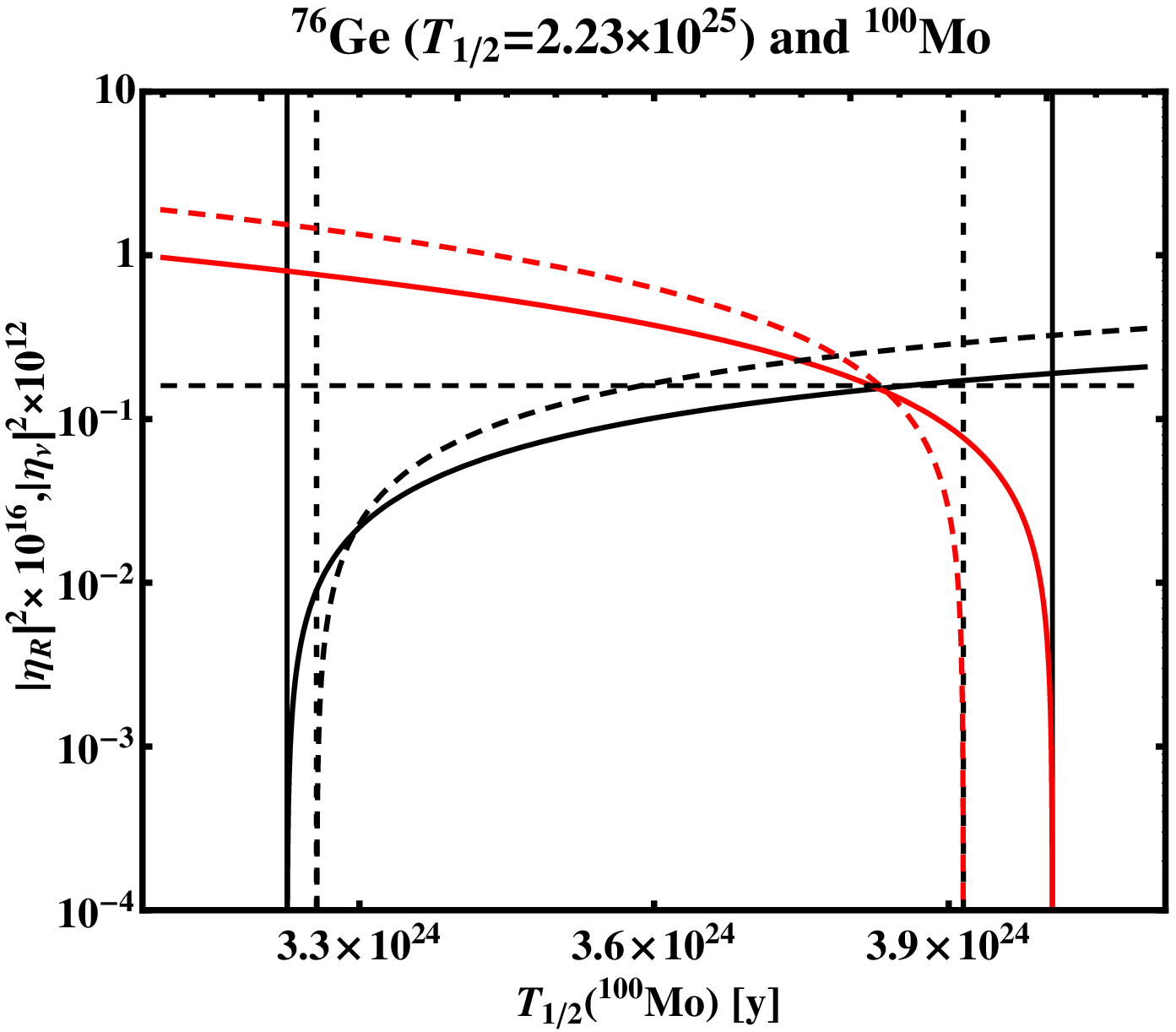}}
\vspace{-0.8cm}
\caption{\label{fig:comparison1} 
Solutions for $|\eta_\nu|^2$ (black lines) 
and $|\eta_R|^2$ (red lines), obtained by
fixing $T_1 = T^{0\nu}_{1/2}(^{76}Ge)=2.23 \times 10^{25}$ yr
and $T_2 = T^{0\nu}_{1/2}(^{100}Mo)$
and using the sets of NMEs
calculated using the ``large basis'' and 
i) CD-Bonn potential, $g_A= 1.25$ (solid lines) 
and $g_A=1$ (dashed lines) (upper left panel);
ii) CD-Bonn (solid lines) and Argonne (dashed lines) 
potentials with $g_A=1.25$ (upper right panel);
iii) CD-Bonn (solid lines) and Argonne (dashed lines) 
potentials with $g_A=1.0$ (lower left panel);
iv) Argonne potential with $g_A= 1.25$ (solid lines) and 
$g_A=1$ (dashed lines) (lower right panel).
The physical (positive) solutions for
$|\eta_\nu|^2$ and $|\eta_R|^2$ 
shown with solid (dashed) lines 
are delimited by two vertical solid (dashed) lines. The horizontal dashed 
line corresponds to the prospective upper 
limit \cite{MainzKATRIN} \meff < 0.2 eV. 
}
\end{figure}

\begin{figure}[htbp]
\centering%
\subfigure
   {\includegraphics[width=7cm]{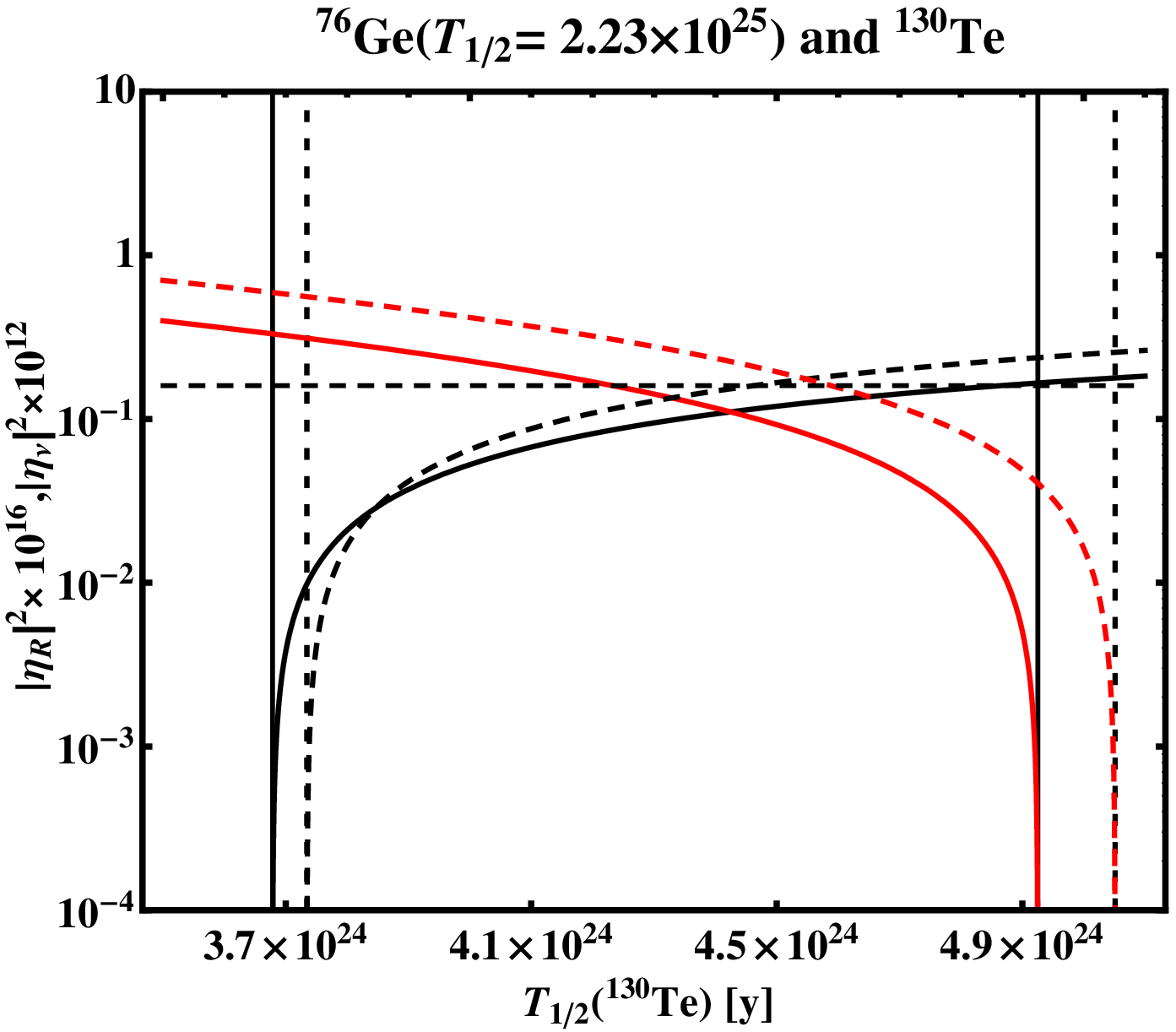}}
 \vspace{2mm}
 \subfigure
   {\includegraphics[width=7cm]{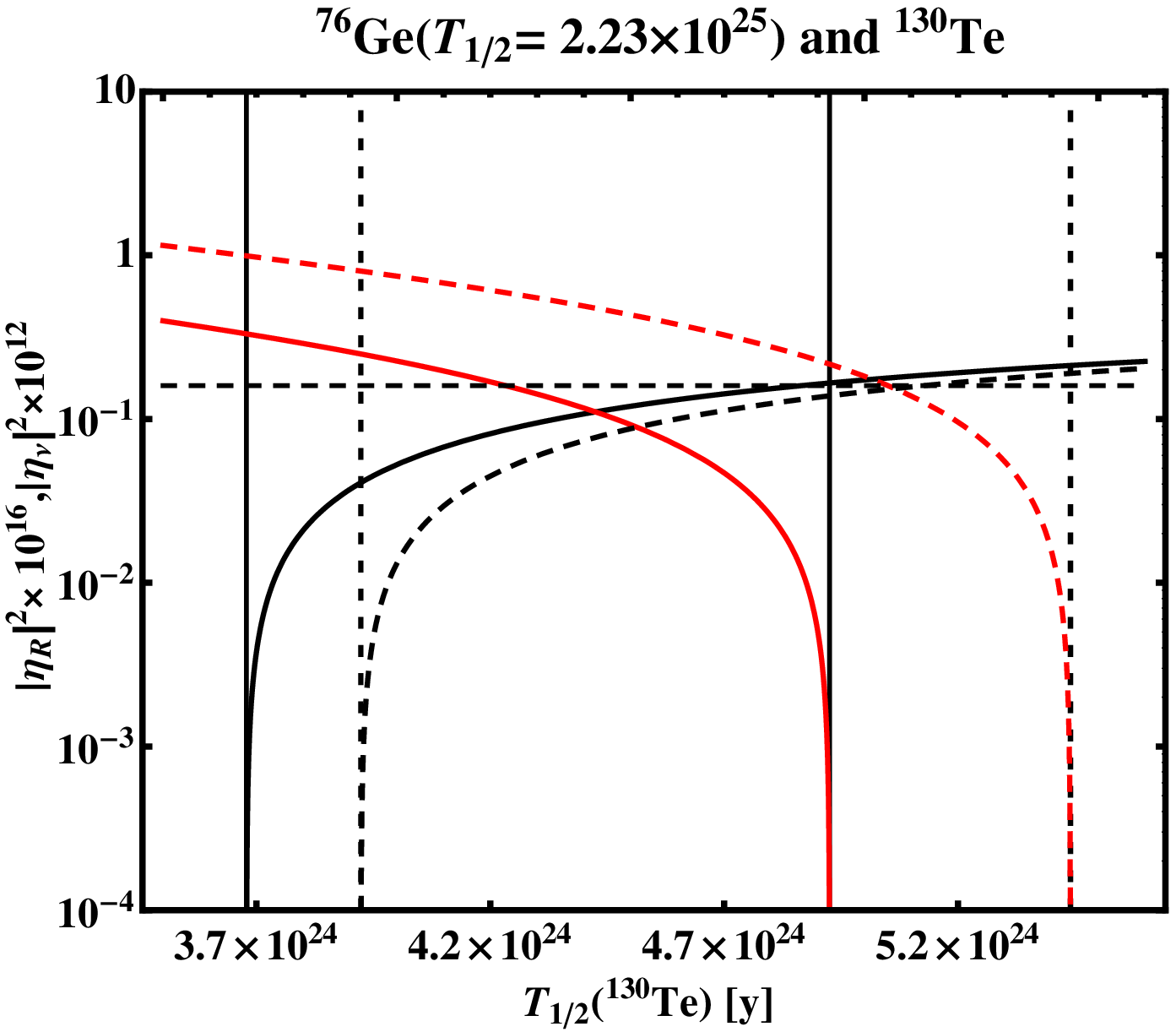}}
\vspace{2mm}
\subfigure
   {\includegraphics[width=7cm]{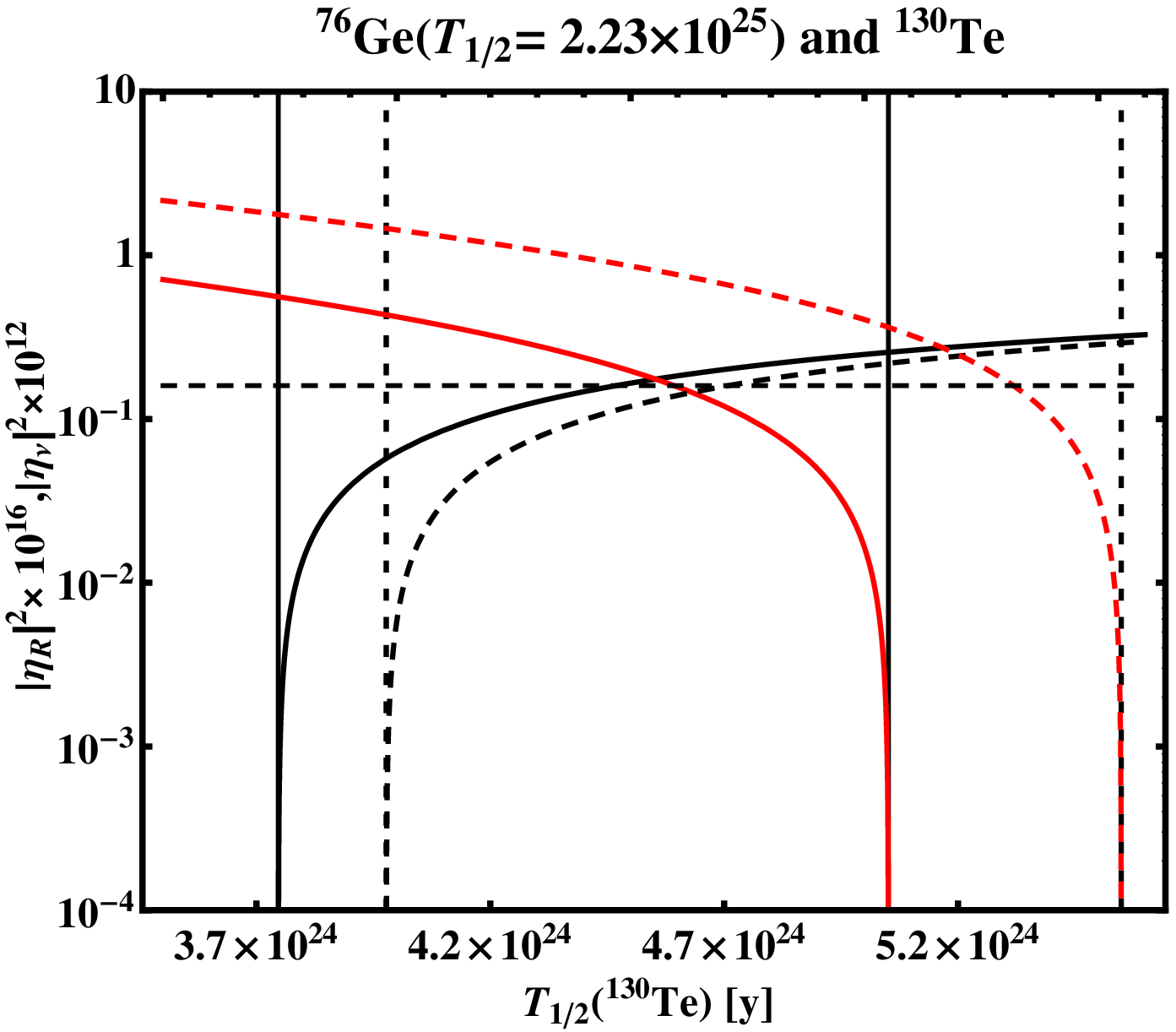}}
 \vspace{2mm}
 \subfigure
   {\includegraphics[width=7cm]{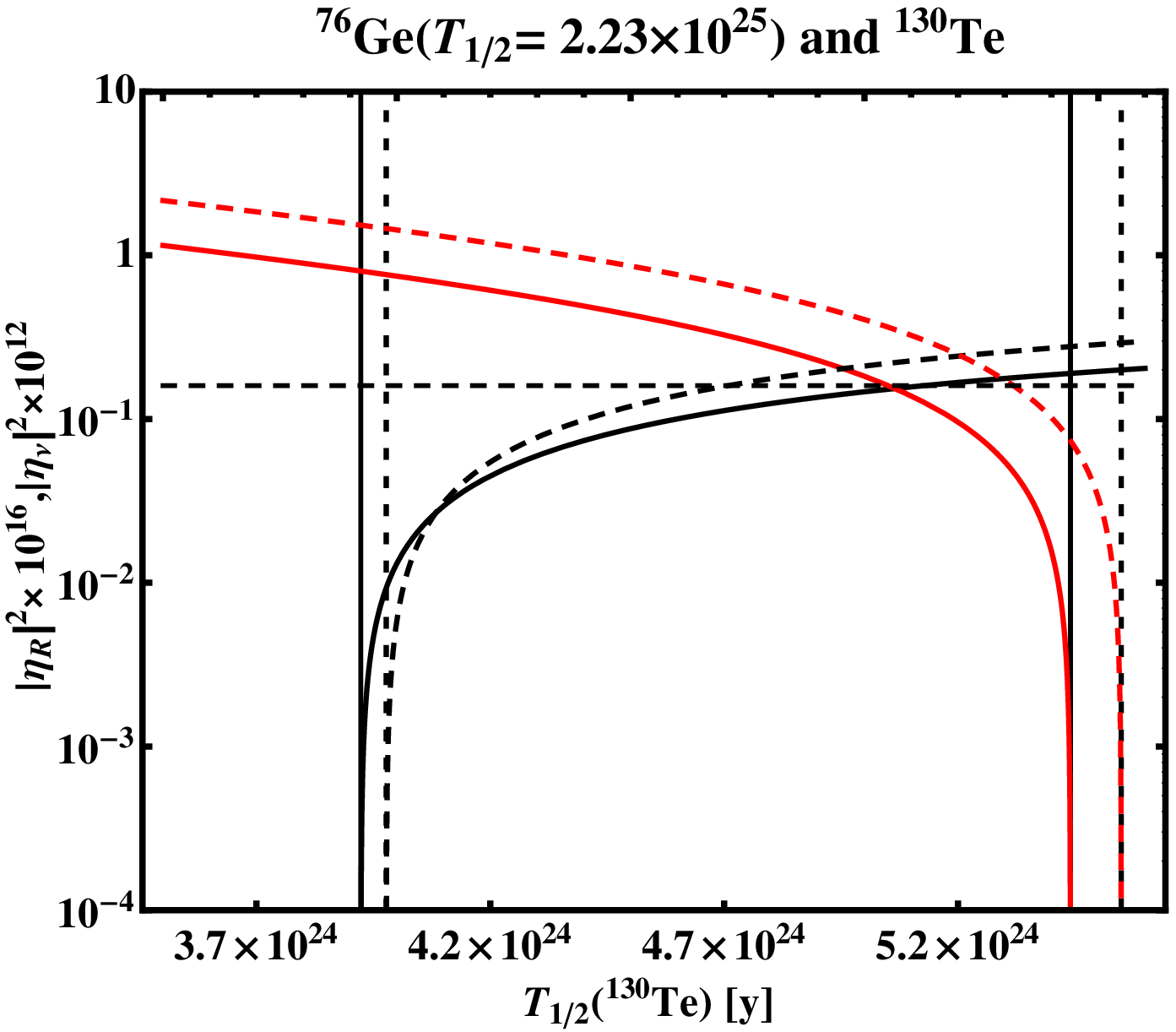}}
\vspace{-0.8cm}
\caption{\label{fig:comparison2} 
The same as in Fig. \ref{fig:comparison1}, but for 
$T_1 = T^{0\nu}_{1/2}(^{76}Ge)=2.23 \times 10^{25}$ yr
and $T_2 =T^{0\nu}_{1/2}(^{130}Te)$. 
% Solutions for $|\eta_\nu|^2$ (black line) 
% and $|\eta_R|^2$ (red line) in the case of two not-interfering 
% mechanisms using Cd-Bonn and Argonne potentials with different $g_A$ values, 
% fixing $T^{0\nu}_{1/2}(^{76}Ge)=2.23 \times 10^{25}$ yr and studying the 
% solutions for $T^{0\nu}_{1/2}(^{130}Te)$. 
% Upper left panel: CD-Bonn potential with 
% $g_A= 1.25$ (thick lines) and $g_A=1$ (dashed lines). 
% Upper right panel: $g_A=1.25$ assuming CD-Bonn 
% potential (thick lines) and Argonne potential (dashed lines). 
% Lower left panel: $g_A=1$ assuming CD-Bonn 
% potential (thick lines) and Argonne potential (dashed lines).
% Lower right panel: Argonne potential with 
% $g_A= 1.25$ (thick lines) and $g_A=1$ (dashed lines). 
% The physical (positive) solutions
% are delimited in the former cases by the 
% two vertical thick lines and in the latter by two dashed vertical lines.
}
\end{figure}

 We have studied also the dependence of the results
discussed above on the value of $g_A$ and the  
NMEs used. This was done using the ``large basis''
NMEs, obtained with the CD-Bonn and Argonne potentials
for the two values of $g_A=1.25;~1.0$.
Some of the results of this study 
are presented graphically in Figs. \ref{fig:comparison1} and 
\ref{fig:comparison2}.
The  horizontal dashed line in these two figures 
corresponds again to the prospective limit 
from the upcoming KATRIN experiment 
\cite{MainzKATRIN}.  
We note that in the cases studied by us, 
changing the value of $g_A$ from 1.25 to 1.0 
for a given potential (CD-Bonn or Argonne) 
% \vspace{-0.8cm}

%%%%%%%%%%%%%%%%%%%%%%%%%%%%%%%%%%%%%%%%%
%%%%%%%%%%%%%%%%%%%%%%%%%%%%%%%%%%%%%
% \vspace{-0.8cm}
\noindent does not lead to a significant change of 
the solutions for $|\eta_\nu|^2$ and $|\eta_R|^2$:
the change is smaller 
 than approximately 10\%.
The solutions exhibit a larger variation 
when for a given $g_A$ and basis, 
the NMEs calculated with the CD-Bonn 
potential are replaced by the NME's obtained 
with the Argonne potential (Fig. \ref{fig:comparison2}, 
upper right and lower left panels). In this case,
as we have mentioned earlier, given $T_1$, 
the interval of allowed values 
of the half-life of the second nucleus $T_2$  
changes somewhat. In the specific cases shown, 
e.g., in Fig.  \ref{fig:comparison2} (upper right and 
lower left panels), $T_1 \equiv 
T^{0\nu}_{1/2}(^{76}Ge) = 2.23\times 10^{25}$ y and
the interval of interest of values of 
$T_2 \equiv T^{0\nu}_{1/2}(^{100}Mo)$ 
shifts to larger values. Obviously, 
the solution values of the parameters 
$|\eta_\nu|^2$  and $|\eta_R|^2$,
obtained with the two different 
sets of the NMEs, can differ drastically
in the vicinity of the maximum 
and minimum values of $T_2$, 
as is also seen in Figs. \ref{fig:comparison1} and 
\ref{fig:comparison2}.
If a given extreme value of $T_2$, say ${\rm max}(T_2)$,
obtained with one set of NMEs, belongs to the 
interval of allowed values of $T_2$, found with 
a second set of NMEs, 
one of the fundamental parameters, calculated 
at ${\rm max}(T_2)$ with the first set of NMEs
can be zero, and can have a relatively large 
nonzero value when calculated with the second 
set of NMEs. Moreover, 
there are narrow intervals of values 
of $T_2$ for which there exist physical 
solutions for 
$|\eta_\nu|^2$  and $|\eta_R|^2$ if one uses 
the NMEs obtained with the CD-Bonn potential 
and there are no physical solutions 
for the NMEs derived with the Argonne potential.
If the measured value of $T_2$ falls in such an interval,
this can imply that either the two mechanisms considered are 
not at work in $\betabeta$-decay, or one of the two sets 
of NMEs does not describe correctly the nuclear transitions.
As Figs. \ref{fig:comparison1} and \ref{fig:comparison2} 
indicate, the data from the KATRIN experiment can help 
limit further the solutions for $|\eta_\nu|^2$, obtained 
with NMEs calculated with the CD-Bonn potential and 
$g_A=1.0$ or with the Argonne potential.

 Let us note finally that Figs. \ref{fig:comparison1} and 
\ref{fig:comparison2} were obtained using hypothetical half-lives 
of $^{76}Ge$ and $^{100}Mo$. We get similar results 
if we use as input hypothetical half-lives of a different pair 
of nuclei, $^{76}Ge$ and $^{130}Te$,
$^{130}Te$ and $^{100}Mo$, etc.

% \vspace{-0.8cm}
%%%%%%%%%%%%%%%%%%%%%%%%%%%%%%%%%%%%%%%%%%%%%%%%%
%%%%%%%%%%%%%%%%%%%%%%%%%%%%%%%%%%%%%%%%%%%%%%%%%
%%%%%%%%%%%%%%%%%%%%%%%%%%%%%%%%%%%%%%%%%%%%%%%%%
% \vspace{-0.8cm}
%%%%%%%%%%%%%%%%%%%%%%%%%%%%%%%%%%%%%%%%%%%%%%%%%
%%%%%%%%%%%%%%%%%%%%%%%%%%%%%%%%%%%%%%%%%%%%%%%%%

% \newpage

%%%%%%%%%%%%%%%%%%%%%%%%%%%%%%
%
\section{\label{sec:analII}   Two ``Interfering'' Mechanisms}
%
%%%%%%%%%%%%%%%%%%%%%%%%%%%%

 Neutrinoless double beta decay can be triggered by two competitive
mechanisms whose interference contribution to the
$\betabeta$-decay rates is non-negligible. 
In this Section we analyze the case
of light Majorana neutrino exchange and gluino exchange.
From equation (\ref{hlint}) it is possible to extract the values
of $|\eta_\nu|^2$, $|\eta_{\lambda'}|^2$ and $\cos\alpha$ setting up a
system of three equation with these three unknowns using as input
the ``data'' on the half-lives of three different nuclei.
The solutions are given by:
%%%%%%%%%%%%%%%%%%%%%%%%%%%%%%%%%
\be
|\eta_\nu|^2= \frac{D_1}{D}\,,~~ |\eta_{\lambda'}|^2 = \frac{D_2}{D}\,,~~
z \equiv 2\cos\alpha|\eta_\nu||\eta_{\lambda'}| =  \frac{D_3}{D}\,,
\label{intsol1}
\ee
%%%%%%%%%%%%%%%%%%%%%%%%%%%%%%%
%
where $D$, $D_1$, $D_2$ and $D_3$ are the following determinants
%%%%%%%%%%%%%%%%%%%%%%%%%%%%%%%%%%%%%%%%%
\be
D = \left| \begin{array}{ccc}
 ({M'}^{0\nu}_{1,\nu })^2 & ({M'}^{0\nu }_{1,\lambda' })^2 & {M'}^{0\nu }_{1,\lambda' } {M'}^{0\nu}_{1,\nu } \\
 ({M'}^{0\nu}_{2,\nu })^2 & ({M'}^{0\nu }_{2,\lambda' })^2 & {M'}^{0\nu }_{2,\lambda' } {M'}^{0\nu}_{2,\nu }\\
 ({M'}^{0\nu}_{3,\nu })^2 & ({M'}^{0\nu }_{3,\lambda '})^2 & {M'}^{0\nu }_{3,\lambda' } {M'}^{0\nu}_{3,\nu }
\end{array}
\right|\,,~~~
D_1 = \left| \begin{array}{ccc}
1/ T_1 G_1 & ({M'}^{0\nu }_{1,\lambda' })^2 & {M'}^{0\nu }_{1,\lambda' } {M'}^{0\nu}_{1,\nu } \\
1/ T_2 G_2 & ({M'}^{0\nu }_{2,\lambda' })^2 & {M'}^{0\nu }_{2,\lambda' } {M'}^{0\nu}_{2,\nu } \\
1/ T_3 G_3 & ({M'}^{0\nu }_{3,\lambda' })^2 & {M'}^{0\nu }_{3,\lambda' } {M'}^{0\nu}_{3,\nu }
\end{array} \right|\,,
\label{DD1}
\ee
%%%%%%%%%%%%%%%%%%%%%%%%%%%%%%%%%%%%%%%
%%%%%%%%%%%%%%%%%%%%%%%%%%%%%%%%%%%%%%%
\be
D_2 =  \left| \begin{array}{ccc}
({M'}^{0\nu}_{1,\nu })^2 & 1/ T_1 G_1 &  {M'}^{0\nu }_{1,\lambda' } {M'}^{0\nu}_{1,\nu } \\
({M'}^{0\nu}_{2,\nu })^2 & 1/ T_2 G_2 &  {M'}^{0\nu }_{2,\lambda' } {M'}^{0\nu}_{2,\nu } \\
({M'}^{0\nu}_{3,\nu })^2 & 1/ T_3 G_3 &  {M'}^{0\nu }_{3,\lambda' } {M'}^{0\nu}_{3,\nu }
\end{array}
\right|\,,~~~
D_3 = \left| \begin{array}{ccc}
  ({M'}^{0\nu}_{1,\nu })^2& ({M'}^{0\nu }_{1,\lambda' })^2 & 1/ T_1 G_1 \\
  ({M'}^{0\nu}_{2,\nu })^2& ({M'}^{0\nu }_{2,\lambda' })^2 & 1/ T_2 G_2\\
  ({M'}^{0\nu}_{3,\nu })^2& ({M'}^{0\nu }_{3,\lambda' })^2 & 1/ T_3 G_3
\end{array} \right|\,.
\label{D2D3}
\ee
%%%%%%%%%%%%%%%%%%%%%%%%%%%%%%%%%%%
%
We must require that $|\eta_\nu|^2$
and $|\eta_{\lambda'}|^2$ be non-negative and that
the factor
$2\cos\alpha|\eta_\nu||\eta_{\lambda'}|$
in the interference term satisfies:
%%%%%%%%%%%%%%%%%%%%%%%%%%%%%%%%%%%
\be
-2|\eta_\nu||\eta_{\lambda'}| \leq
2\cos\alpha|\eta_\nu||\eta_{\lambda'}| \leq2|\eta_\nu||\eta_{\lambda'}|.
\label{fase}
\ee
%%%%%%%%%%%%%%%%%%%%%
%
 If we fix (i.e. have data on)
the half-lives of two of the
nuclei and combine these with the
condition in eq. (\ref{fase}), we can obtain
the  interval of values of the half-life of
the third nucleus, which is compatible with
the data on the half-lives of the two other
nuclei and the mechanisms considered.
The minimal (maximal) value of this
interval of half-lives of the third nucleus
is obtained for $\cos\alpha = +1$
($\cos\alpha = -1$).
Examples of the
intervals of half-life values of the third nucleus
obtained using the half-life values of the
other two nuclei
\footnote{Technically this is done in the following way.
Fixing the half-lives of two isotopes, $T_1$ and $T_2$,
and varying the half-life of the third isotope   $T_3$
in a certain interval,
we obtained  $|\eta_\nu|^2$, $|\eta_{\lambda'}|^2$ and
$z= 2\cos\alpha|\eta_\nu||\eta_{\lambda'}|$ as a function
of $T_3$. Requiring that $|\eta_\nu|^2 > 0$, $|\eta_{\lambda'}|^2 >0$
and that $-2\eta_\nu||\eta_{\lambda'}| \leq z \leq 2\eta_\nu||\eta_{\lambda'}|$
determines the interval of physically allowed values of
$T_3$ (given $T_1$, $T_2$ and the mechanisms of
$\betabeta$-decay considered). This interval 
of physically allowed values of $T_3$  
is shown in Table \ref{tab:table3}.}
for the $\betabeta$-decay mechanisms
discussed are listed in Table \ref{tab:table3}.
The results reported in Table \ref{tab:table3} are 
obtained with NMEs corresponding to the CD-Bonn potential,
the ``large basis'' and $g_A=1.25$.
%%%%%%%%%%%%%%%%%%%%%%%%%%%%%%%%%%%%%%%%%%%
\begin{table}[h!]
\centering \caption{
\label{tab:table3} Ranges of 
half-lives of $T_3$ in the case of two interfering 
mechanisms: the light Majorana neutrino exchange and 
gluino exchange dominance.}

\renewcommand{\arraystretch}{0.8}
{\footnotesize\tt
\begin{tabular}{|l|l|c|}
\hline \hline
 T$^{0\nu}_{1/2}$[y](fixed) &  T$^{0\nu}_{1/2}$[y](fixed) & Allowed  \\
\hline
 T$(Ge)= 2.23\cdot10^{25}$  &  T$(Mo)=5.8\cdot10^{24}$   &    $ 5.99\cdot10^{24}\leq T(Te) \leq 7.35 \cdot10^{24}$\\
  T$(Ge)=2.23\cdot10^{25}$  &  T$(Te)=3\cdot10^{24}$     &    $2.46 \cdot10^{24} \leq  T(Mo) \leq 2.82 \cdot10^{24}$\\
 T$(Ge)= 10^{26}$           &  T$(Mo)=5.8\cdot10^{24}$   & $6.30 \cdot10^{24}\leq T(Te) \leq 6.94 \cdot10^{24}$\\
          T$(Ge)= 10^{26}$  &  T$(Te)=3\cdot10^{24}$     & $ 2.55  \cdot10^{24} \leq T (Mo) \leq 2.72 \cdot10^{24}$\\
 T$(Ge)= 2.23\cdot10^{25}$  &  T$(Te)=3\cdot10^{25}$     & $ 2.14  \cdot10^{25} \leq T (Mo) \leq 3.31 \cdot10^{25}$\\
   T$(Ge)= 10^{26}$         &  T$(Te)=3\cdot10^{25}$     & $2.38 \cdot10^{25} \leq T (Mo) \leq 2.92 \cdot10^{25}$\\
\hline\hline
\end{tabular}}
\end{table}
%%%%%%%%%%%%%%%%%%%%%%%%%%%%%%%%%%%%%%%%
%

% \newpage

\begin{figure}[h!]
  \begin{center}
 \subfigure
{\includegraphics[width=7cm]{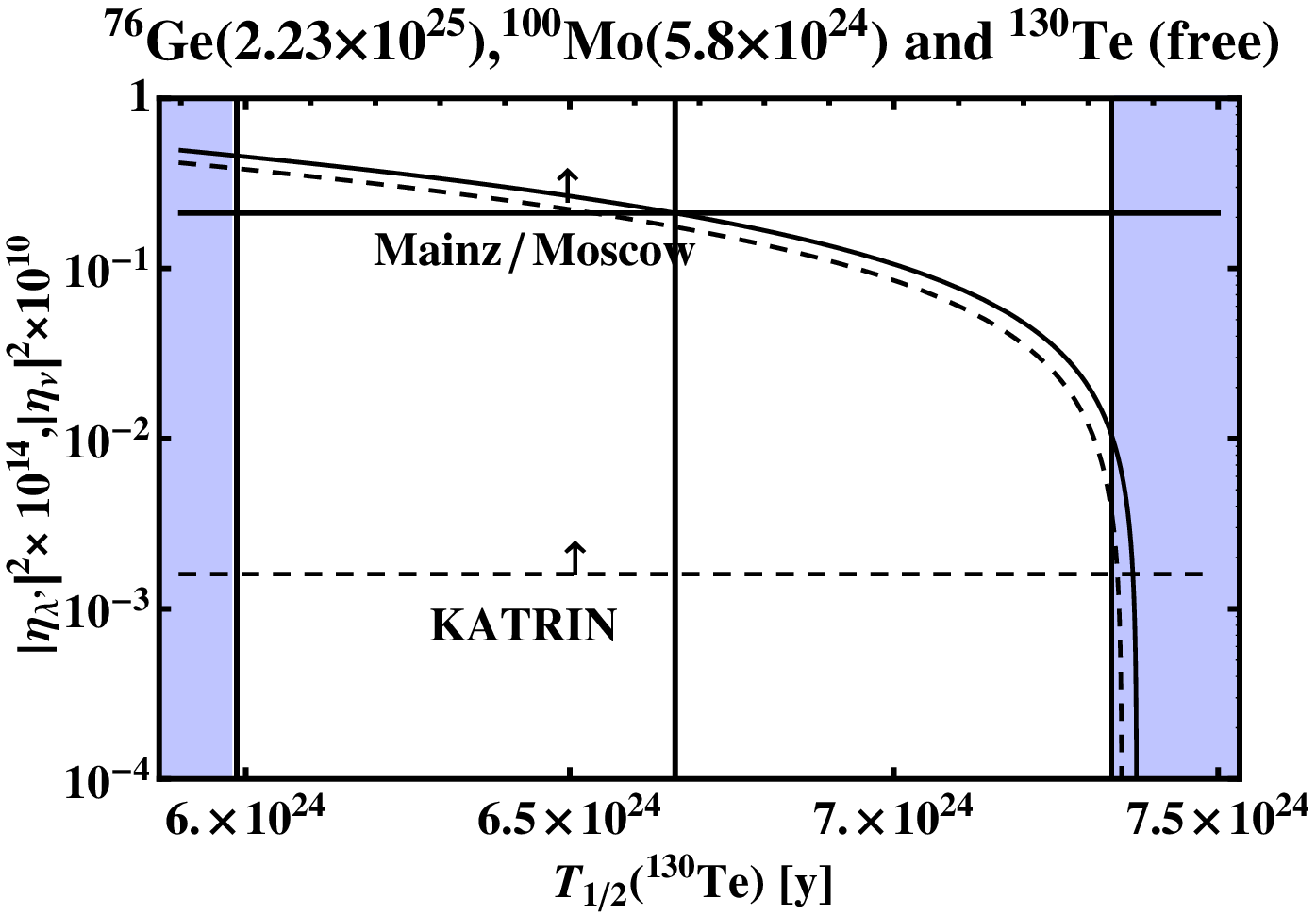}}
% \vspace{5mm}
 \subfigure
 {\includegraphics[width=7cm]{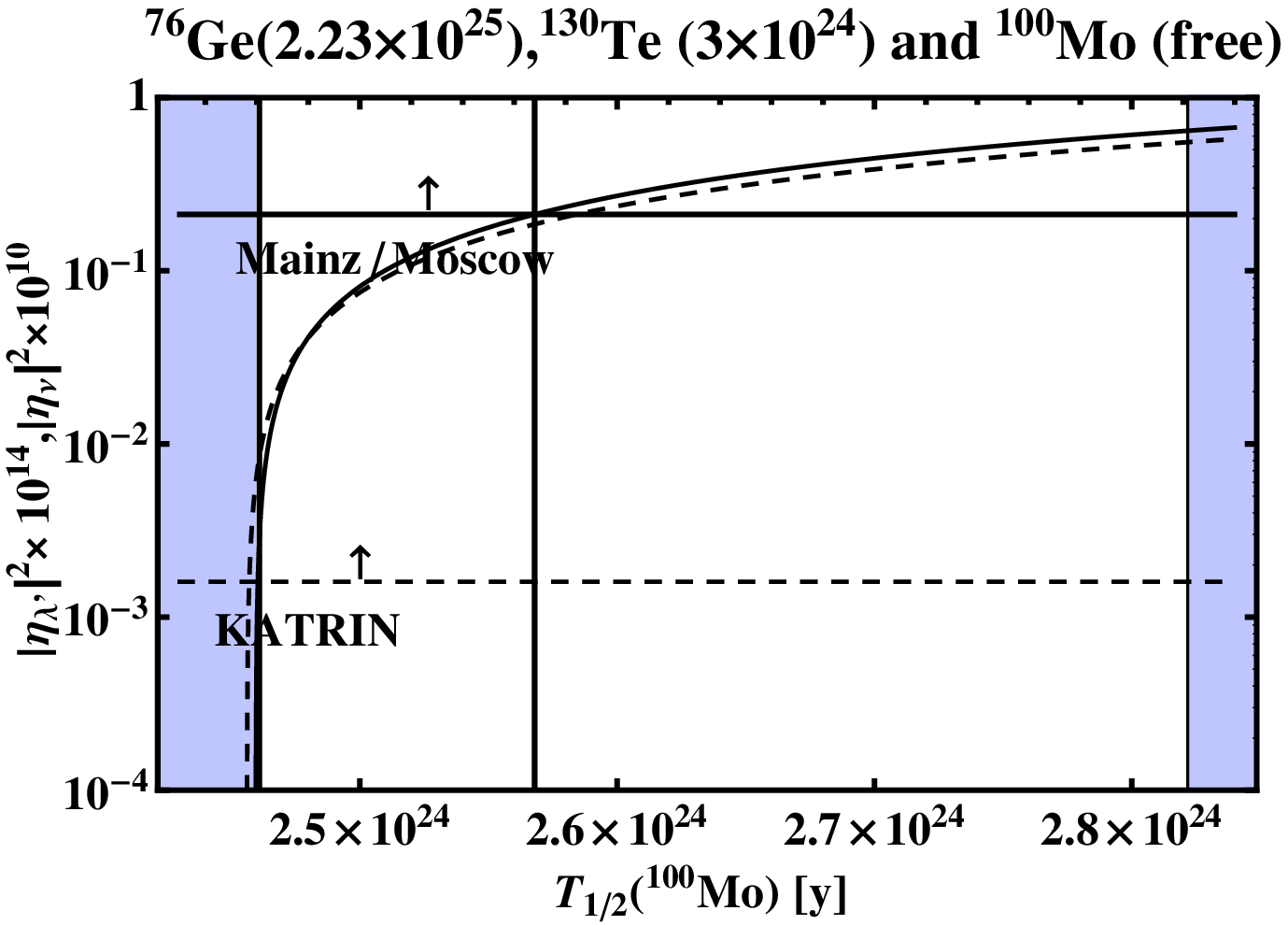}}
  \end{center}
\vspace{-1.0cm}
    \caption{\label{fig:fig5}
The values of the rescaled parameters
$|\eta_\nu|^2$ (thick line) and $|\eta_{\lambda'}|^2$ (dashed lined),
obtained as solutions of the system of  equations 
(\ref{hlint}) for fixed values of 
$T^{0\nu}_{1/2}(^{76}Ge)$ and $T^{0\nu}_{1/2}(^{100}Mo)$ 
and values of $T^{0\nu}_{1/2}(^{130}Te)$ 
lying in a specific interval. The physical
allowed regions correspond to the
areas shown in white, while the areas
shown in gray are excluded. The horizontal solid (dashed) 
line corresponds to the upper limit  \cite{MoscowH3,MainzKATRIN} 
\meff < 2.3 eV (prospective upper 
limit \cite{MainzKATRIN} \meff < 0.2 eV). See text for details.
}
\end{figure}

\begin{figure}[h!]
  \begin{center}
 \subfigure
 {\includegraphics[width=7cm]{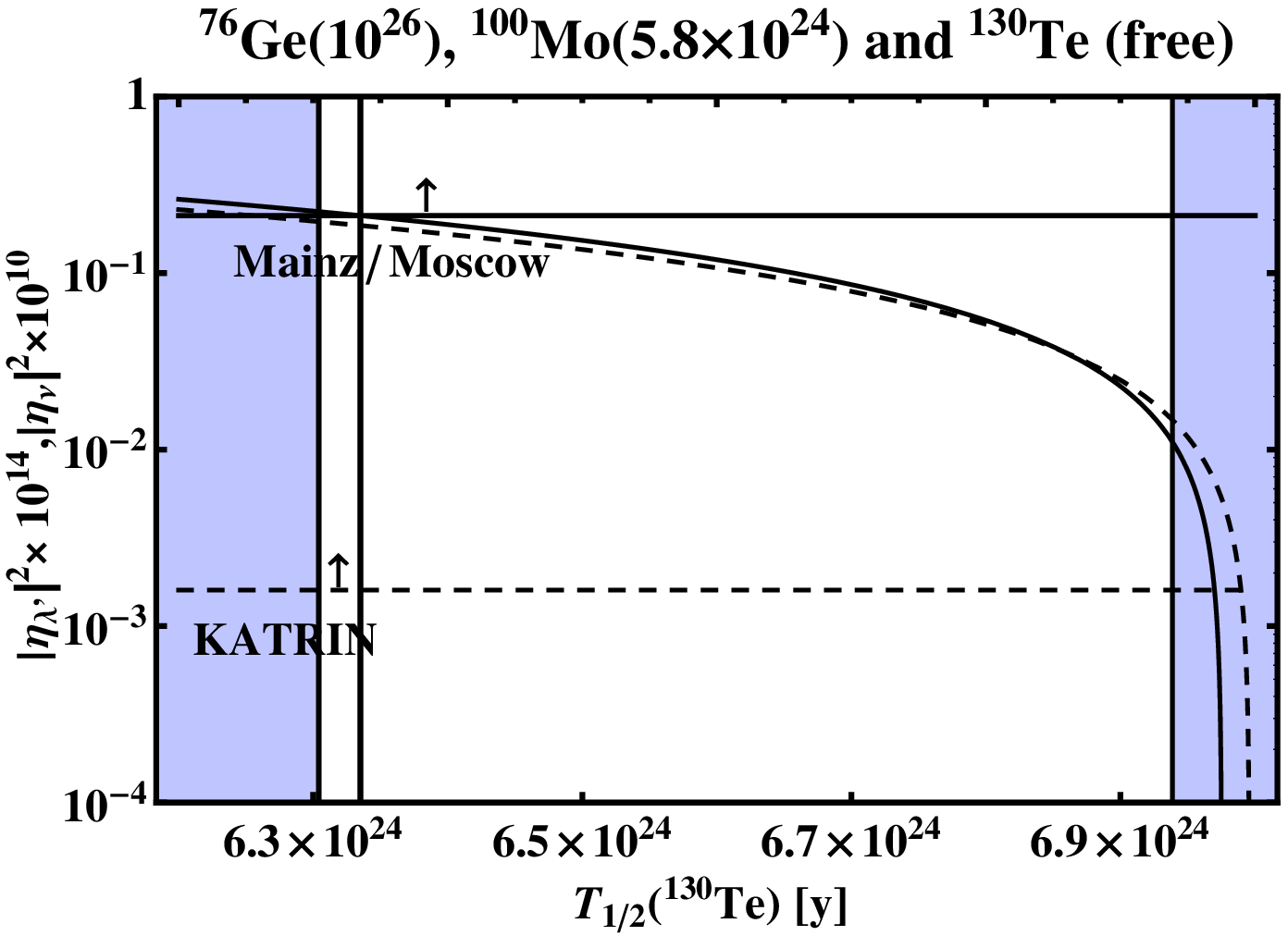}}
 \vspace{5mm}
 \subfigure
  {\includegraphics[width=7cm]{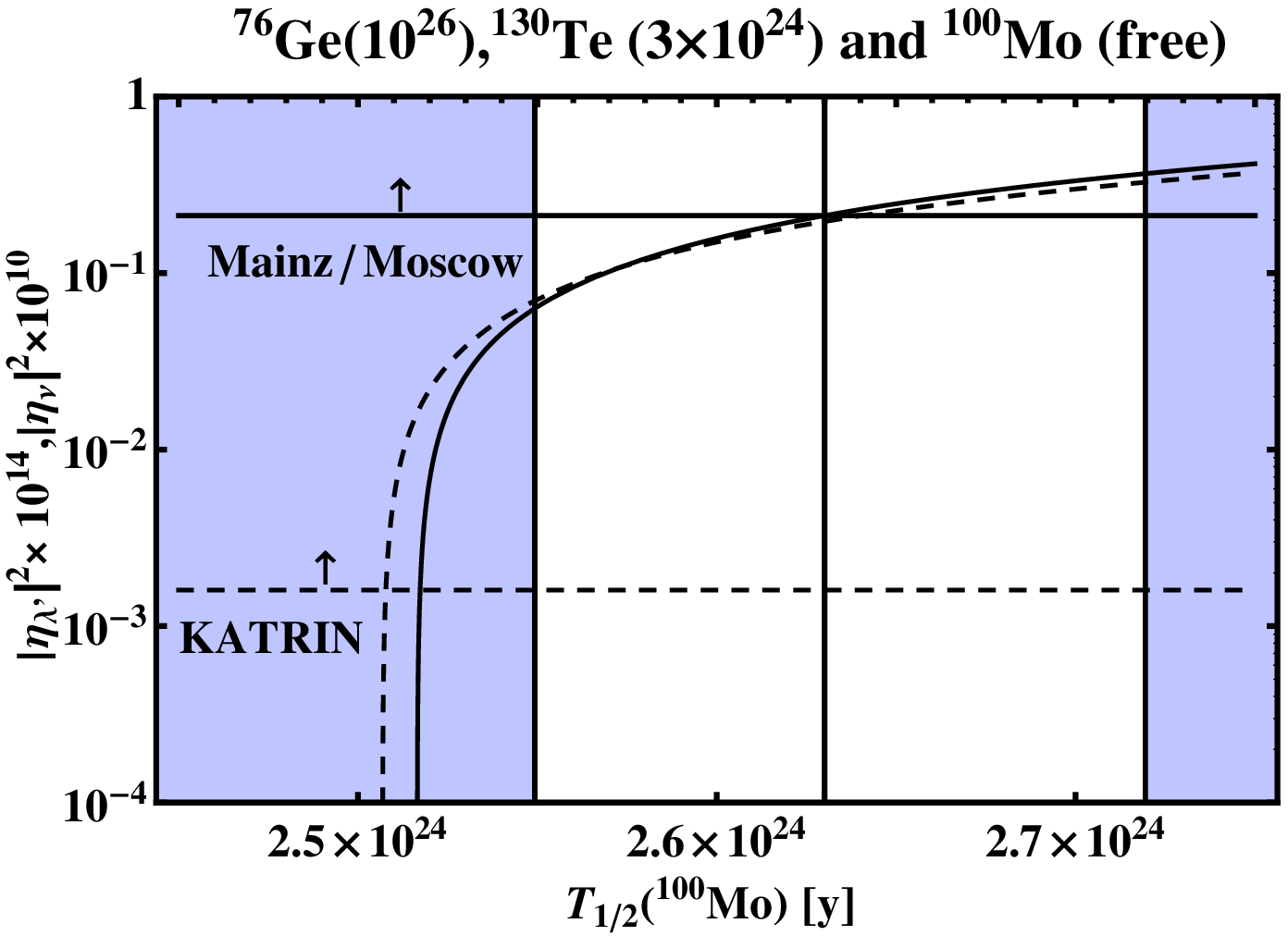}}
     \end{center}
\vspace{-1.0cm} 
   \caption{\label{fig:fig6}
The same as in Fig. \ref{fig:fig5} for a different
set of values of the three half-lives used as input
in the analysis.
}
\end{figure}

\begin{figure}[h!]
  \begin{center}
 \subfigure
  %  {\includegraphics[scale=0.44]{GeminTemaxfixed.eps}}
 {\includegraphics[width=7cm]{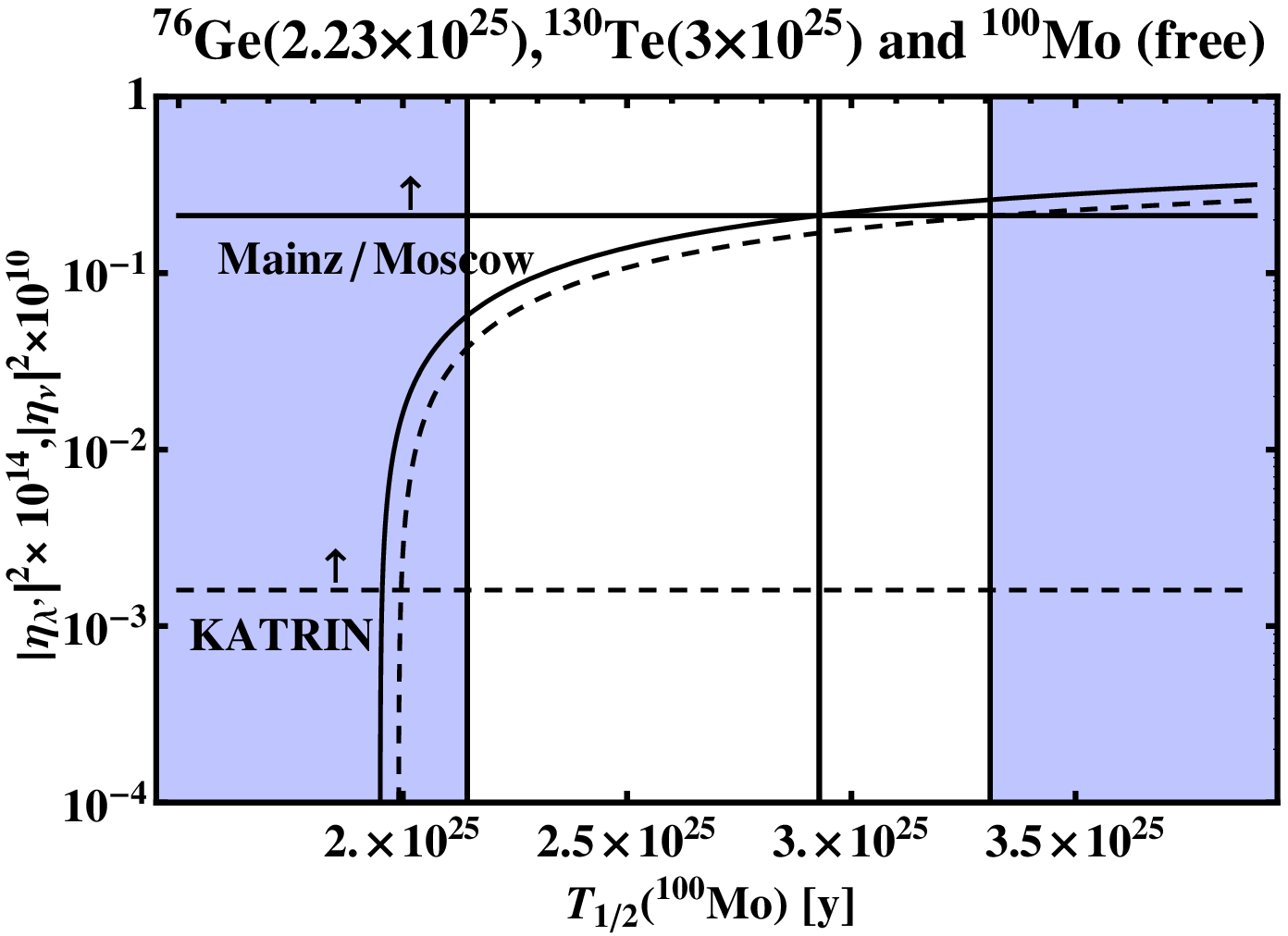}}
 \vspace{5mm}
 \subfigure
   {\includegraphics[width=7cm]{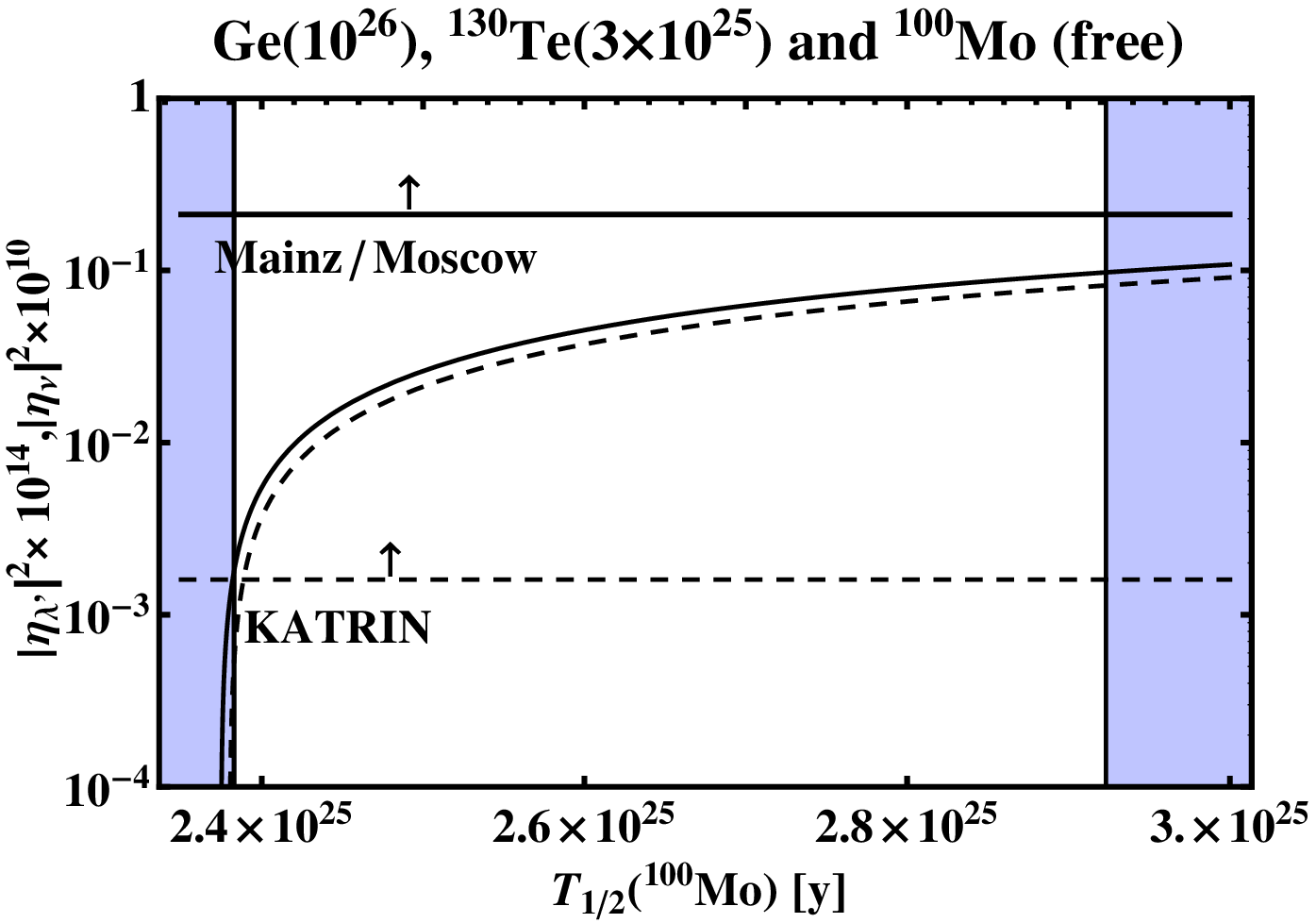}}
     \end{center}
\vspace{-1.0cm}
    \caption{\label{fig:fig7}
The same as in Figs. \ref{fig:fig5} and \ref{fig:fig6}
for a different set of values of the three half-lives
used as input in the analysis.
% Light Majorana neutrino exchange (thick line)
% and gluino exchange (dashed line)
}
\end{figure}

\begin{figure}[h!]
  \begin{center}
 \subfigure
   {\includegraphics[width=7cm]{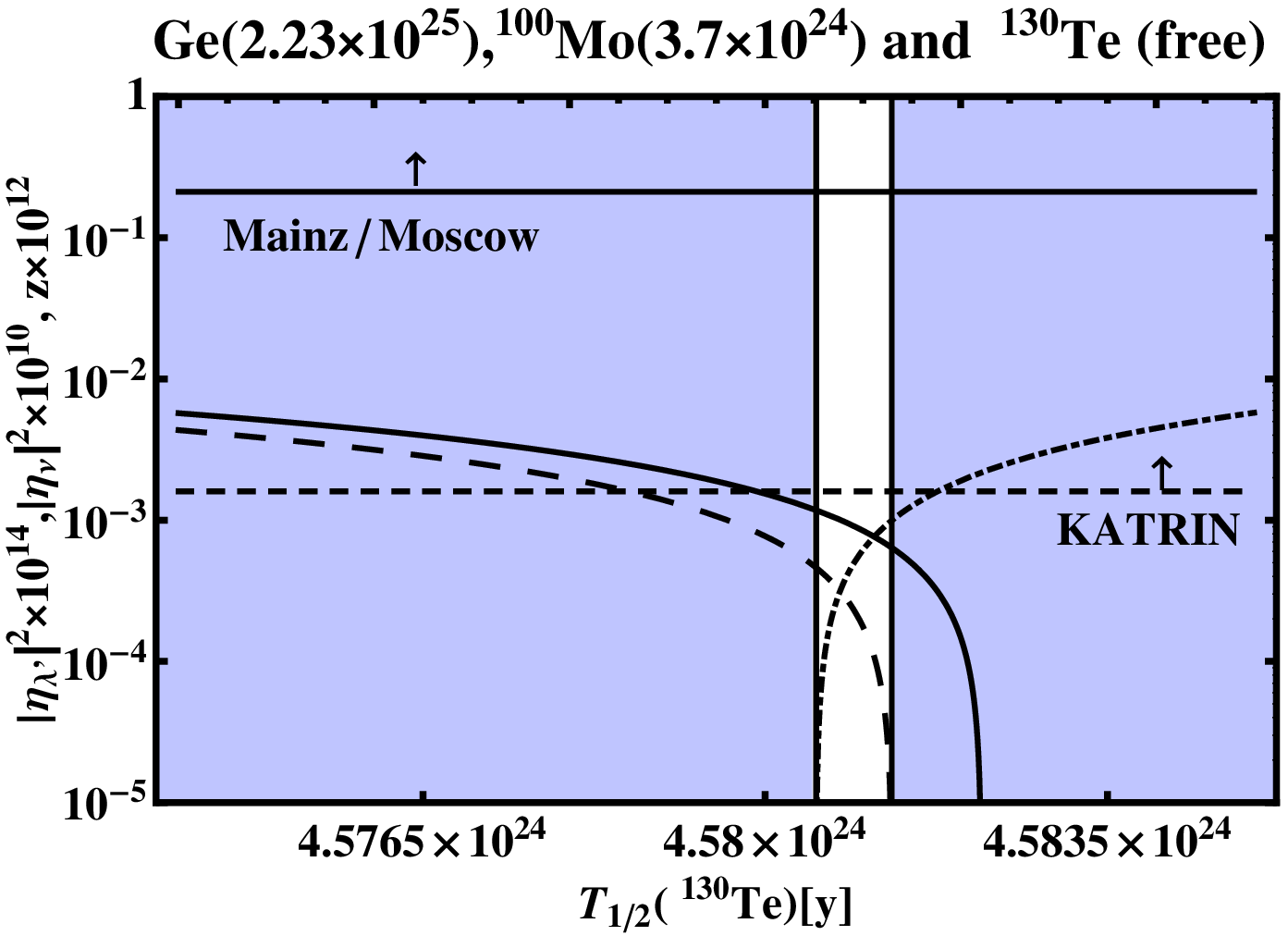}}
 \vspace{5mm}
 \subfigure
   {\includegraphics[width=6.7cm]{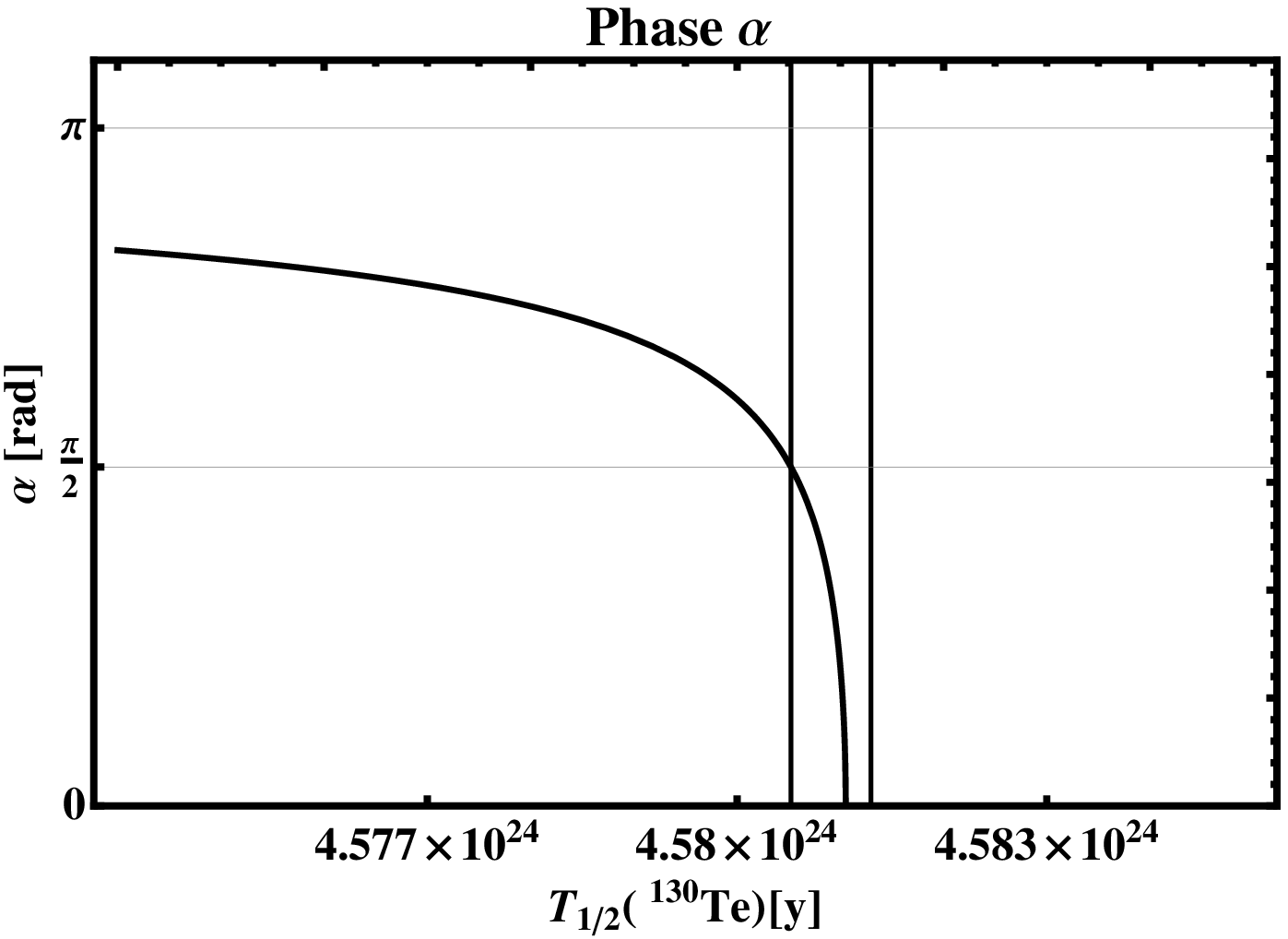}}
     \end{center}
    \caption{\label{fig:fig8}
Left Panel: the values of
$|\eta_\nu|^2\times 10^{10}$ (thick line),
$|\eta_{\lambda'}|^2\times 10^{14}$ (dashed line) and
$z= 2\cos\alpha|\eta_\nu||\eta_{\lambda'}|\times 10^{12}$
(dot-dashed line) corresponding
to the half-lives of $^{76}Ge$,
$^{100}Mo$ and $^{130}Te$
indicated on the figure.
The interval of values of
$T^{0\nu}_{1/2}(^{130}Te)$
between the two vertical lines
corresponds to physical
(positive) solutions
for $|\eta_\nu|^2$ and
$|\eta_{\lambda'}|^2$ as well as to
a positive $z$
(i.e. to a constructive interference
between the contributions due to the two
mechanisms). The horizontal solid (dashed) 
line corresponds to the upper limit  \cite{MoscowH3} 
\cite{MainzKATRIN} \meff<2.3 eV (prospective upper 
limit \cite{MainzKATRIN} \meff < 0.2 eV). See text for details.
% Notice the range in which
% all the three functions are positive.
Right Panel: the phase $\alpha$.}
\end{figure}

We show in few illustrative figures
(Figs. \ref{fig:fig5} - \ref{fig:fig8}) the results of
the determination of $|\eta_\nu|^2$,
$|\eta_{\lambda'}|^2$ and $\cos\alpha$
using different values of half-lives 
of the three nuclei
$^{76}$Ge, $^{100}$Mo and $^{130}$Te
from the intervals given in eq. (\ref{limit}).
The lower bounds of the half-lives
quoted in eq. (\ref{limit}) are
taken into account.
In these figures the physical
allowed regions correspond to the
areas shown in white, while the areas
shown in gray are excluded.
The allowed interval of values of the half-life
of the 3rd nucleus, corresponding to
the white areas, are listed in the 3rd column
of Table \ref{tab:table3}.
The results presented in
Figs.  \ref{fig:fig5} - \ref{fig:fig8}
are derived using the NMEs, calculated with 
the CD-Bonn potential, 
the ``large basis'' and $g_A=1.25$.

  It is interesting to note that for the
two fixed half-life values
used to obtain Figs.
\ref{fig:fig5}, \ref{fig:fig6} and
\ref{fig:fig7}, the interference
between the contributions of
the two mechanisms considered is destructive:
one finds using these values that
for most of the physical (positive) solutions
for $|\eta_\nu|^2$ and
$|\eta_{\lambda'}|^2$,  $\cos\alpha$ is
negative. Moreover, the rescaled
parameters  $|\eta_\nu|^2\times 10^{10}$ and
$|\eta_{\lambda'}|^2\times 10^{14}$ in most of
the solution regions have very close values.
This is due to the fact that
for most of the physically allowed values of
$|\eta_\nu|^2$ and $|\eta_{\lambda'}|^2$,
each of the two terms including
$|\eta_\nu|^2$ or $|\eta_{\lambda'}|^2$
as a factor in the right hand side of eq. (\ref{hlint}) is
much larger than the free term in the left hand side
of eq. (\ref{hlint}). As a consequence, in order to explain  the
``data'' (i.e. the chosen values of the half-lives of the
three isotopes) there should be a strong
mutual compensation between the
contributions due to the two mechanisms.
This is possible only if
$|\eta_\nu|^2({M'}^{0\nu}_{i, \nu})^2$ and
$|\eta_{\lambda'}|^2({M'}^{0\nu}_{i,\lambda'})^2$ have
close values and $\cos\alpha \cong -1$.

  In the case of destructive interference between the two 
contributions, $\meff$ can have values which exceed 
the limit on the absolute scale of neutrino masses 
set by the $^3H$ $\beta$-decay experiments 
\cite{MoscowH3,MainzKATRIN},
eq. (\ref{etanuH3MM}).
This limit is indicated as a horizontal solid line 
in Figs.  \ref{fig:fig5} - \ref{fig:fig7}. 
It leads to further constraints on the physical 
solution for $|\eta_\nu|^2$, and thus for 
$|\eta_{\lambda'}|^2$. 

 As we have already indicated, 
a more stringent limit on the absolute neutrino mass scale
and therefore on $\meff$ is planned to be obtained in the 
KATRIN experiment \cite{MainzKATRIN}: it is given 
in eq. (\ref{etanuKatrin}).
The KATRIN prospective upper bound is shown as a 
horizontal dashed line in 
Figs. \ref{fig:fig5} - \ref{fig:fig7}.
As the results presented in 
Figs.  \ref{fig:fig5} - \ref{fig:fig7} indicate,
if the limit of 0.2 eV will be reached in 
KATRIN experiment, if will lead to severe 
constraints on the solutions for $|\eta_\nu|^2$ 
obtained in the cases we have considered, 
strongly disfavoring (if not ruling out) 
essentially all of them.

  In Fig.  \ref{fig:fig8} we illustrate the possibility 
of constructive interference between the light neutrino 
and the gluino exchange contributions.
The solutions shown in Fig.  \ref{fig:fig8} are not 
constrained by the limits  obtained 
in the $^3H$ $\beta$-decay experiments 
\cite{MoscowH3,MainzKATRIN};
they also satisfy the prospective upper limit 
from KATRIN experiment.
% \newpage
%%%%%%%%%%%%%%%%%%%%%%%%%%%%%%%
%%%%%%%%%%%%%%%%%%%%%%%%%%%%%%%%%%%
% \vspace{-0.8cm}
%%%%%%%%%%%%%%%%%%%%%%%%%%%%%%%%%%
%%%%%%%%%%%%%%%%%%%%%%%%%%%%%%%%%%
%%%%%%%%%%%%%%%%%%%%%%%%%%%%%%%%%
%%%%%%%%%%%%%%%%%%%%%%%%%%%%%%%%%%%%
% \vspace{-0.8cm}
%%%%%%%%%%%%%%%%%%%%%%%%%%%
%%%%%%%%%%%%%%%%%%%%%%%%%%%%%%%%

% \newpage

It is not difficult to
derive from eqs. (\ref{intsol1}) - (\ref{D2D3})
the general conditions under
which $|\eta_\nu|^2$ and $|\eta_{\lambda'}|^2$
are positive and the
interference between the light neutrino
and gluino exchange contributions
is constructive (destructive), i.e.
$\cos\alpha$ (or $z$) is positive (negative).
We will illustrate them later  
using again the NMEs calculated with 
the CD-Bonn potential, the ``large basis'' 
and $g_A=1.25$. 

  Consider first the conditions 
for constructive interference.
We will introduce a somewhat simplified 
notations in this part of the article:
$T_i$, $G_i$, $M_i$ and $\Lambda_i$ 
for $i = 1,2,3$ will denote respectively 
the half-life, phase space factor,
light neutrino and dominant gluino 
exchange NMEs for $^{76}$Ge ($i=1$),
$^{100}$Mo ($i=2$) and $^{130}$Te ($i=3$).
The first thing to notice is that 
it follows from Table \ref{table.1}
that the ratios of NMEs 
$M_i/\Lambda_i$ satisfy the inequalities:
%%%%%%%%%%%%%%%%%%%%%%%%%%%%%%%%%%%%%%%%
\be 
\frac{M_{1}}{\Lambda_{1}} > \frac{M_{2}}{\Lambda_{2}}> 
\frac{M_{3}}{\Lambda_{3}}\,.
\label{MiLambdai}
\ee
%%%%%%%%%%%%%%%%%%%%%%%%%%%%%%%
%
This implies that the determinant $D$, defined in eq. (\ref{DD1}),
is negative:
%%%%%%%%%%%%%%%%%%%%%%%%%%%%%%%%%%%%%
\be 
D = \Ml{1}^2\Ml{2}^2\Ml{3}^2(\frac{\Mn{2}}{\Ml{2}}-
\frac{\Mn{1}}{\Ml{1}})
(\frac{\Mn{3}}{\Ml{3}}-\frac{\Mn{1}}{\Ml{1}})
(\frac{\Mn{3}}{\Ml{3}}-\frac{\Mn{2}}{\Ml{2}}),
\label{D}
\ee 
%%%%%%%%%%%%%%%%%%%%%%%%%%%%%%%%%%
%
Consequently, in order to have 
$|\eta_\nu|^2 > $, $|\eta_{\lambda'}|^2> 0$
and constructive interference between the two 
contributions, i.e. $z > 0$, 
all three determinants $D_1$, $D_2$ and $D_3$, 
defined in eqs. (\ref{DD1}) and (\ref{D2D3}),
have to be negative: $D_a < 0$, $a=1,2,3$.
Given the half-life $T_1$ and the NMEs 
$M_i$ and $\Lambda_i$, these three conditions are 
satisfied if each of the two half-lives 
$T_2$ and $T_3$ lies in specific intervals 
\footnote{The quoted solutions are valid, as can be shown, 
provided $M_{3}/\Lambda_{3} < M_{2}/\Lambda_{2} < 
 0.5(1 + \sqrt{5}) M_{3}/\Lambda_{3}$, which is fulfilled 
for the NMEs given in Table \ref{table.1}.}:
%%%%%%%%%%%%%%%%%%%%%%%%%%%%%%%%%%%%%%%%%%%%%%%%%%%%%
\ba A) &&
\begin{cases}
\frac{\Ml{1}^2}{\Ml{2}^2}\,\frac{G_1}{G_2}\,T_1 < T_2 
\leq\frac{\Mn{1}\Ml{1}}{\Mn{2}\Ml{2}}\,\frac{G_1}{G_2}\,T_1\,,   \\
\vspace{0.5cm}
\frac{(\Mn{2}^2\Ml{1}^2-\Mn{1}^2\Ml{2}^2)\,\frac{G_2}{G_3}\,T_2}
{(\Mn{3}^2\Ml{1}^2 - \Mn{1}^2\Ml{3}^2) + 
\frac{T_2G_2}{T_1 G_1} (\Mn{2}^2\Ml{3}^2-\Mn{3}^2\Ml{2}^2)}
<  T_3   < 
\frac{(\Mn{2}\Ml{2}\Ml{1}^2-\Mn{1}\Ml{1}\Ml{2}^2)\,\frac{G_2}{G_3}\,T_2}
{  (\Mn{3}\Ml{3}\Ml{1}^2-\Mn{1}\Ml{1}\Ml{3}^2)+  \frac{T_2G_2}{T_1 G_1} (\Mn{2}\Ml{2}\Ml{3}^2-\Mn{3}\Ml{3}\Ml{2}^2)}\,;
\end{cases}\nonumber\\
\vspace{0.8cm}
B)&&
\begin{cases} 
\frac{\Mn{1}\Ml{1}}{\Mn{2}\Ml{2}}\,\frac{G_1}{G_2}\,T_1 <T_2
<\frac{\Mn{1}^2}{\Mn{2}^2}\frac{G_1}{G_2}\, T_1\,,  \\
\vspace{0.8cm}
\frac{(\Mn{2}^2\Ml{1}^2-\Mn{1}^2\Ml{2}^2)\,\frac{G_2}{G_3}\, T_2}
{  (\Mn{3}^2\Ml{1}^2-\Mn{1}^2\Ml{3}^2)+  
\frac{T_2G_2}{T_1 G_1} (\Mn{2}^2\Ml{3}^2-\Mn{3}^2\Ml{2}^2) }
<  T_3   <
\frac{(\Mn{2}\Ml{2}\Mn{1}^2-\Mn{1}\Ml{1}\Mn{2}^2)\, 
\frac{G_2}{G_3}\, T_2}
{ (\Mn{3}\Ml{3}\Mn{1}^2-\Mn{1}\Ml{1}\Mn{3}^2) 
+  \frac{T_2G_2}{T_1 G_1} (\Mn{2}\Ml{2}\Mn{3}^2-\Mn{3}\Ml{3}\Mn{2}^2)}\,.
\end{cases}
\label{zpositive}
\ea
%%%%%%%%%%%%%%%%%%%%%%%%%%%%%%%%%%%%%%%%%%%%%%
%
  For the NMEs calculated with 
the CD-Bonn potential, 
the ``large basis'' and $g_A=1.25$
and given $T_1 \neq 0$, the conditions
for constructive interference, $z > 0$, read:
%%%%%%%%%%%%%%%%%%%%%%%%%%%%%%%%%%%%%%%
\be
 z > 0\,:
% T_1>0,
\begin{cases}
% T_2 = 0.14 T_1 & T_3 =\dfrac{4.44\times 10^{14} T_1 T_2}
% {3.74\times 10^{14}T_1-9.32\times 10^{13} T_2}\\
0.14\, T_1 < T_2 \leq 0.16\, T_1\,, & \dfrac{4.44\, T_1\,T_2}
{3.74\, T_1 - 0.93\,T_2}
\leq T_3\leq \dfrac{2.10\, T_1\, T_2}{1.78\, T_1 - 0.47\, T_2}\,; \\
0.16\, T_1 < T_2 < 0.18\, T_1\,, &\dfrac{4.44\, T_1\,
T_2}{3.74\, T_1 -0.93\, T_2}\leq T_3\leq
\dfrac{4.10\, T_1\, T_2}{3.44\, T_1 - 0.81\,T_2}\,.
% T_2 = 0.18 T_1 & T3 = \dfrac{4.44\times10^{14} T_1 T_2}{3.74
% \times10^{14} T_1 - 9.32 \times10^{13} T_2}
\end{cases}
\label{constrint}
\ee
%%%%%%%%%%%%%%%%%%%%%%%%%%%%%%%%%%%%%%%%%%%%%%%
%
These conditions imply that given $T_1$, a constructive 
interference is possible only if $T_2$ lies in a relatively 
narrow interval and $T_3$ has a value in extremely 
narrow intervals, the interval for $T_2$ 
being determined by the value of $T_1$, while that for 
$T_3$ -  by $T_1$ and the interval for $T_2$.
The fact that both the intervals for $T_2$ and $T_3$ 
are so narrow is a consequence of the values 
% of the phase space factors $G_i$ and
of the NMEs used, more precisely, 
of the fact that, for each of the 
two mechanisms discussed,
the NMEs for the three nuclei considered differ 
relatively little: we have 
$|M_i - M_j| << M_i,M_j$, 
$|\Lambda_i - \Lambda_j| << \Lambda_i,\Lambda_j$,
$i \neq j = 1,2,3$, and typically 
$|M_i - M_j|/(0.5 (M_i + M_j)) \sim 10^{-1}$,
$|\Lambda_i - \Lambda_j|/(0.5(\Lambda_i + \Lambda_j)) 
\sim (10^{-2} - 10^{-1})$.
We get similar results for the 
other sets of NMEs, 
quoted in Table \ref{table.1} and 
calculated with the ``large basis''. 
In order to have a constructive interference 
in a much wider interval of values of $T_2$, i.e.,  
to have the minimal value of $T_2$ much smaller 
than the maximal value of $T_2$ in case A) 
in eq. (\ref{zpositive}), 
for instance, the following inequality has to be satisfied:
$\Ml{1}/\Ml{2} << \Mn{1}/\Mn{2}$.
An inspection of Table \ref{table.1} shows that 
this inequality is not satisfied by any of the 
relevant sets of NMEs. 
Numerically,  the intervals of values of $T_2$ and $T_3$  
given  in eq. (\ref{constrint}), for which $z > 0$, 
are very similar to those quoted in eq. (\ref{CDBonn125}). 
 
 For the value of T($^{76}$Ge)= 2.23$\cdot$10$^{25}$ y,
for instance, the conditions for a
constructive interference are given by:
%%%%%%%%%%%%%%%%%%%%%%%%%%%%%%%%%
\be
\begin{split}
% 3.18 \cdot 10^{24}~{\rm y} <T_2 \leq 3.55 \cdot 10^{24}~{\rm y},&\quad
% \frac{ 3.96 T_2}{3.34 - 3.73\cdot 10^{-26}~{\rm y^{-1}} T_2}<T_3<
% \frac{1.88 T_2}{1.58 - 1.87\cdot 10^{-26}{\rm y^{-1}} T_2}\\
% 3.55\cdot 10^{24}~{\rm y} <T_2<3.97\cdot 10^{24}~{\rm y},&\quad
% \frac{3.96 T_2}{3.34 - 3.73\cdot 10^{-26}{\rm y^{-1}} T_2}<T_3<
% \frac{3.65 T_2}{3.07 - 3.25 \cdot 10^{-26}{\rm y^{-1}} T_2}.
%
3.18 \cdot 10^{24}~{\rm y} <T_2 \leq 3.55 \cdot 10^{24}~{\rm y},&\quad
\frac{ 1.19 T_2}{1.00 - 1.12\cdot 10^{-26}~{\rm y^{-1}} T_2}<T_3<
\frac{1.19 T_2}{1.00 - 1.18\cdot 10^{-26}{\rm y^{-1}} T_2}\\
3.55\cdot 10^{24}~{\rm y} <T_2<3.97\cdot 10^{24}~{\rm y},&\quad
\frac{1.186 T_2}{1.00 - 1.117\cdot 10^{-26}{\rm y^{-1}} T_2}<T_3<
\frac{1.189 T_2}{1.00 - 1.059 \cdot 10^{-26}{\rm y^{-1}} T_2}\,.
\label{constructint1}
\end{split}
\ee
%%%%%%%%%%%%%%%%%%%%%%%%%%%%%%
%
Given the fact that $3.18 \cdot 10^{24}~{\rm y} <T_2 \leq 
3.97 \cdot 10^{24}~{\rm y} $ 
and that $T_2$ enters in the denominators of the 
limiting values of $T_3$ multiplied by $10^{-26}~{\rm y^{-1}}$,
the interval of values of $T_3$ of interest is 
extremely narrow. We have  $z > 0$ for, e.g.,
T($^{76}$Ge)= 2.23$\cdot$10$^{25}$ y,
T($^{100}$Mo)= 3.7$\cdot$10$^{24}$ y and T($^{130}$Te)=
4.58$\cdot$10$^{24}$ y, as is also illustrated 
in Fig. \ref{fig:fig8}.

  There are cases in which one has
$|\eta_\nu|^2 = 0$ or $|\eta_{\lambda'}|^2=0$.
The general conditions for having 
$|\eta_\nu|^2 = 0$ or $|\eta_{\lambda'}|^2=0$ 
can be derived from eqs. (\ref{intsol1}) - (\ref{D2D3}) 
and read:
%%%%%%%%%%%%%%%%%%%%%%%%%%%%%%%%%%%%%%%%%%%%
\ba 
|\eta_\nu|^2 = 0, |\eta_{\lambda'}|^2 \neq 0, &&
\begin{cases}
T2=\frac{\Ml{1}^2}{\Ml{2}^2}\frac{G_1}{G_2}\, T_1 \\
\vspace{0.5cm}
T_3 =   
\frac{(\Mn{2}\Ml{2}\Ml{1}^2-\Mn{1}\Ml{1}\Ml{2}^2)\,\frac{G_2}{G_3}\,T_2}
{  (\Mn{3}\Ml{3}\Ml{1}^2-\Mn{1}\Ml{1}\Ml{3}^2)+  \frac{T_2G_2}{T_1 G_1} (\Mn{2}\Ml{2}\Ml{3}^2-\Mn{3}\Ml{3}\Ml{2}^2) }\,;
\end{cases}\nonumber\\
\vspace{0.8cm}
|\eta_{\lambda'}|^2 = 0, |\eta_\nu|^2 \neq 0  &&
\begin{cases} 
T_2=
\frac{\Mn{1}^2}{\Mn{2}^2}\frac{G_1}{G_2}\, T_1\,,  \\
\vspace{0.8cm}
T_3   =
\frac{(\Mn{2}\Ml{2}\Mn{1}^2-\Mn{1}\Ml{1}\Mn{2}^2)\, 
\frac{G_2}{G_3}\, T_2}
{ (\Mn{3}\Ml{3}\Mn{1}^2-\Mn{1}\Ml{1}\Mn{3}^2) +
\frac{T_2G_2}{T_1 G_1} (\Mn{2}\Ml{2}\Mn{3}^2-\Mn{3}\Ml{3}\Mn{2}^2) }\,.
\end{cases}
\label{etazeq0}
\ea
%%%%%%%%%%%%%%%%%%%%%%%%%%%%%%%%%%%%%%%%%%%%
%
They correspond to some of the limiting values 
of $T_2$ and $T_3$ in eq. (\ref{zpositive}). 
We will illustrate them below numerically
using the NMEs calculated with 
the CD-Bonn potential, 
the ``large basis'' and $g_A=1.25$. If, for instance,
one fixes $T_1 \equiv$ T($^{76}$Ge)= 2.23$\cdot$10$^{25}$,
we have i) $|\eta_\nu|^2 = 0$ (and zero interference term)
for $T_2=3.18\cdot 10^{24}$ y and
$T_3= 3.91 \cdot 10^{24}$ y;
ii) $|\eta_{\lambda'}|^2 = 0$ (and zero interference term)
for $T_2 = 3.97\cdot 10^{24}$ y and
$T_3 = 4.93\cdot 10^{24}$ y,
where $T_2$ and $T_3$ denote
the half-lives of  $^{100}$Mo and $^{130}$Te,
respectively. In general, given $T_1$
we have $|\eta_\nu|^2 = 0$,  $|\eta_{\lambda'}|^2 \neq 0$ if
%%%%%%%%%%%%%%%%%%%%%%%%%%%%%%%%%%%%%%%%
\be
T_2 = 0.14 T_1\,,~~~
T_3 =\dfrac{2.10\, T_1\, T_2}{1.78\, T_1 - 0.47\, T_2}\, \cong 0.18\,T_1\,,
\label{etanu0}
\ee
%%%%%%%%%%%%%%%%%%%%%%%%%%%%%%%%%%%%%%%
%
and  $|\eta_{\lambda'}|^2 = 0$, $|\eta_\nu|^2 \neq 0$ provided
%%%%%%%%%%%%%%%%%%%%%%%%%%%%%%%%%%%%%%
\be
T_2 = 0.18 T_1\,,~~~ T_3 = \dfrac{4.10\, T_1 T_2}
{3.44\, T_1 - 0.81\, T_2}
\cong 0.22\, T_1\,.
\label{etalam0}
\ee
%%%%%%%%%%%%%
%
The conditions for having zero inteference term, $z = 0$, but 
$|\eta_\nu|^2 \neq 0$ or $|\eta_{\lambda'}|^2 \neq 0$, read:
%%%%%%%%%%%%%%%%%%%%%%%%%%%%%%%%%%%%%
\be
\begin{cases}
\frac{\Ml{1}^2}{\Ml{2}^2}\,\frac{G_1}{G_2}\,T_1 < T_2
<\frac{\Mn{1}^2}{\Mn{2}^2}\frac{G_1}{G_2}\, T_1\,, \\
\vspace{0.8cm}
T_3=\frac{(\Mn{2}^2\Ml{1}^2-\Mn{1}^2\Ml{2}^2)\,\frac{G_2}{G_3}\,T_2}
{(\Mn{3}^2\Ml{1}^2 - \Mn{1}^2\Ml{3}^2) + \frac{T_2G_2}{T_1
G_1} (\Mn{2}^2\Ml{3}^2-\Mn{3}^2\Ml{2}^2)} \,.
\end{cases}
\ee
%%%%%%%%%%%%%%%%%%%%%%%%%%%%%%%%%%%%%%%%%
%
Given $T_1$, the general conditions for destructive 
interference, i.e. for $z < 0$, can be derived in a similar way.
They read:
%%%%%%%%%%%%%%%%%%%%%%%%%%%%%%%%%%%%
\ba A) &&
\begin{cases}
0 < T_2 \leq \frac{\Ml{1}^2}{\Ml{2}^2}\,\frac{G_1}{G_2}\,T_1,\\
0<  T_3   < 
\frac{(\Mn{2}\Ml{2}\Ml{1}^2-\Mn{1}\Ml{1}\Ml{2}^2)\,\frac{G_2}{G_3}\,T_2}
{   (\Mn{3}\Ml{3}\Ml{1}^2-\Mn{1}\Ml{1}\Ml{3}^2)+  \frac{T_2G_2}{T_1 G_1} (\Mn{2}\Ml{2}\Ml{3}^2-\Mn{3}\Ml{3}\Ml{2}^2) }\,;
\end{cases} \label{znegative1} \\ % \nonumber\\
\vspace{0.8cm}
B)&&
\begin{cases} 
\frac{\Ml{1}^2}{\Ml{2}^2}\,\frac{G_1}{G_2}\,T_1 <T_2
\leq \frac{\Mn{1}^2}{\Mn{2}^2}\,\frac{G_1}{G_2}\,T_1\,,  \\
\vspace{0.8cm}
0<  T_3   <
\frac{(\Mn{2}^2\Ml{1}^2-\Mn{1}^2\Ml{2}^2)\,\frac{G_2}{G_3}\,T_2}
{(\Mn{3}^2\Ml{1}^2 - \Mn{1}^2\Ml{3}^2) + 
\frac{T_2G_2}{T_1 G_1} (\Mn{2}^2\Ml{3}^2-\Mn{3}^2\Ml{2}^2) }\,;
\end{cases}\label{znegative2} \\  % \nonumber\\
\vspace{0.8cm}
C)&&
\begin{cases} 
\frac{\Mn{1}^2}{\Mn{2}^2}\,\frac{G_1}{G_2}\,T_1 <T_2
<\frac{\Mn{1}\Ml{1}\Mn{3}-\Ml{3}\Mn{1}^2}{\Mn{2}\Ml{2}\Mn{3}-\Ml{3}\Mn{2}^2}\,\frac{G_1}{G_2}\,T_1\,,  \\
\vspace{0.8cm}
0<  T_3 <  \frac{(\Mn{2}\Ml{2}\Mn{1}^2-\Mn{1}\Ml{1}\Mn{2}^2)\,\frac{G_2}{G_3}\,T_2}
{(\Mn{3}\Ml{3}\Mn{1}^2-\Mn{1}\Ml{1}\Mn{3}^2)+  \frac{T_2G_2}{T_1 G_1} (\Mn{2}\Ml{2}\Mn{3}^2-\Mn{3}\Ml{3}\Mn{2}^2) }\,;
\end{cases}\label{znegative3} \\ % \nonumber\\
\vspace{0.8cm}
D)&&
\begin{cases} 
T_2\geq
\frac{\Mn{1}\Ml{1}\Mn{3}-\Ml{3}\Mn{1}^2}{\Mn{2}\Ml{2}\Mn{3}-\Ml{3}\Mn{2}^2}\,\frac{G_1}{G_2}\,T_1\,,  \\
\vspace{0.8cm}
 T_3   >0.
\end{cases}
\label{znegative4}
\ea
%%%%%%%%%%%%%%%%%%%%%%%%%%%%%%%%%%%%%%
%
Obviously, one has to take into account 
the existing experimental lower 
limits on $T_2$ and $T_3$ in 
eqs. (\ref{znegative1}) - (\ref{znegative4}).
We will give next the  ``numerical''  equivalent of the conditions  (\ref{znegative1}) - (\ref{znegative4}), 
obtained with NMEs calculated with the CD-Bonn potential, 
the ``large basis'' and $g_A=1.25$:
%%%%%%%%%%%%%%%%%%%%%%%%%%%%%%%%%%%%%%%%%%
\be z < 0\,:
% \qquad T_1>0,
\begin{cases}
T_2\leq 0.14\, T_1\,, & T_3\leq \dfrac{2.10\,T_1\, T_2}{1.78 T_1-0.47 T_2}\,;\\
0.14\, T_1<T_2\leq 0.18\, T_1\,, & T_3\leq\dfrac{4.44\, T_1\, T_2}
{3.74\,T_1-0.93\, T_2}\,;\\
0.18\,T_1 < T_2< 4.23\, T_1\,,  & T_3\leq \dfrac{4.10\, T_1\, T_2}
{3.44\, T_1 - 0.81\, T_2}\,;\\
T_2 \geq 4.23\,T_1 & T_3 > 0\,.
\end{cases}
\label{destructint}
\ee
%%%%%%%%%%%%%%%%%%%%%%%%%%%%%%
%
The intervals  of values of $T_2$ and $T_3$ 
in eqs. (\ref{znegative1}) - (\ref{destructint})
are very different from those corresponding 
to the case of two ``non-interfering''  $\betabeta$-decay 
mechanisms given in eq. (\ref{CDBonn125}),  
the only exception being  the second set of 
intervals in eq. (\ref{destructint}), 
which partially overlap with those in eq. (\ref{CDBonn125}). 
This difference can allow  to discriminate
experimentally between the two possibilities of  
$\betabeta$-decay being triggered by two ``non-interfering'' 
mechanisms or by two ``destructively interfering'' 
mechanisms. 
We have check how the intervals of values of the half-life $T_3$ 
given in Table \ref{tab:table3}, corresponding to NMEs 
derived with the CD-Bonn potential, $g_A=1.25$ 
and the ``large basis'', change when one uses the NMEs 
obtained with the same potential and basis, but using $g_A=1.0$,
as well as the NMEs found with the Argonne potential for $g_A=1.25;~1.0$ 
and the ``large basis''. The results are shown in Tables 
\ref{tab:table4} - \ref{tab:table6}. We see that for certain 
values of the hypothetical half-lives of the two nuclei,
the interval of allowed values of 
the half-life of the third nucleus 
becomes noticeably larger when calculated with NMEs, 
corresponding to  $g_A=1.0$ or to the Argonne 
potential. This is due to a relatively deep 
compensation between the three terms
in the $\betabeta$-decay rate 
of the third nucleus in the case of a 
negative interference term 
(destructive interference). 
  
  Similar analysis can be performed for any other pair of 
``interfering'' mechanisms assumed to be operative in 
$\betabeta$-decay.
We note also that the extension 
of the analysis to more than two mechanisms generating the 
$\betabeta$-decay is rather straightforward.

% \clearpage

%%%%%%%%%%%%%%%%%%%%%%%%%%%%%%%%%%%%%%%%
\begin{table}[h!]
\centering 
\caption{\label{tab:table4}
CD-Bonn potential and $g_A=1$}
\renewcommand{\arraystretch}{0.8}
{\footnotesize\tt
\begin{tabular}{|l|l|c|}
\hline \hline
 T$^{0\nu}_{1/2}$[y](fixed) &  T$^{0\nu}_{1/2}$[y](fixed) & Allowed  \\
\hline
T$(Ge)= 2.23\cdot10^{25}$  &  T$(Mo)=5.8\cdot10^{24}$   &    $ 3\cdot10^{24}\leq T(Te) \leq 8.62 \cdot10^{24}$\\
T$(Ge)=2.23\cdot10^{25}$   &  T$(Te)=3\cdot10^{24}$     &    $2.55\cdot10^{24} \leq  T(Mo) \leq 6.18\cdot10^{24}$\\
T$(Ge)= 2.23\cdot10^{25}$  &  T$(Te)=3\cdot10^{25}$     & $ 1.33  \cdot10^{25} \leq T (Mo) \leq 3.88 \cdot10^{26}$\\
T$(Ge)= 10^{26}$           &  T$(Mo)=5.8\cdot10^{24}$   & $3.62\cdot10^{24}\leq T(Te) \leq 6.04 \cdot10^{24}$\\
T$(Ge)= 10^{26}$  	   &  T$(Te)=3\cdot10^{24}$     & $ 3.11 \cdot10^{24} \leq T (Mo) \leq 4.70 \cdot10^{24}$\\
T$(Ge)= 10^{26}$           &  T$(Te)=3\cdot10^{25}$     & $2.15 \cdot10^{25} \leq T (Mo) \leq 8.29 \cdot10^{25}$\\
\hline\hline
\end{tabular}}
\end{table}
%%%%%%%%%%%%%%%%%%%%%%%%%%%%%%%%%%%%%%%%

%%%%%%%%%%%%%%%%%%%%%%%%%%%%%%%%%%%%%%%%
\begin{table}[h!]
\centering \caption{
\label{tab:table5}
Argonne potential and $g_A=1.25$}
\renewcommand{\arraystretch}{0.8}
{\footnotesize\tt
\begin{tabular}{|l|l|c|}
\hline \hline
 T$^{0\nu}_{1/2}$[y](fixed) &  T$^{0\nu}_{1/2}$[y](fixed) & Allowed  \\
\hline
T$(Ge)= 2.23\cdot10^{25}$  &  T$(Mo)=5.8\cdot10^{24}$   &    $ 3\cdot10^{24}\leq T(Te) \leq 9.22 \cdot10^{24}$\\
T$(Ge)=2.23\cdot10^{25}$   &  T$(Te)=3\cdot10^{24}$     &    $2.55\cdot10^{24} \leq  T(Mo) \leq 7.92\cdot10^{24}$\\
T$(Ge)= 2.23\cdot10^{25}$  &  T$(Te)=3\cdot10^{25}$     & $1.19  \cdot10^{25} \leq T (Mo) \leq 2.55 \cdot10^{27}$\\
T$(Ge)= 10^{26}$           &  T$(Mo)=5.8\cdot10^{24}$   & $3.15\cdot10^{24}\leq T(Te) \leq5.85 \cdot10^{24}$\\
T$(Ge)= 10^{26}$  	   &  T$(Te)=3\cdot10^{24}$     & $ 3.25\cdot10^{24} \leq T (Mo) \leq 5.49 \cdot10^{24}$\\
T$(Ge)= 10^{26}$           &  T$(Te)=3\cdot10^{25}$     & $2.08 \cdot10^{25} \leq T (Mo) \leq 1.20 \cdot10^{26}$\\
\hline\hline
\end{tabular}}
\end{table}
%%%%%%%%%%%%%%%%%%%%%%%%%%%%%%%%%%%%%%%%

%%%%%%%%%%%%%%%%%%%%%%%%%%%%%%%%%%%%%%%%
\begin{table}[h!]
\centering \caption{
\label{tab:table6}
Argonne Potential and $g_A=1$}
\renewcommand{\arraystretch}{0.8}
{\footnotesize\tt
\begin{tabular}{|l|l|c|}
\hline \hline
 T$^{0\nu}_{1/2}$[y](fixed) &  T$^{0\nu}_{1/2}$[y](fixed) & Allowed  \\
\hline
T$(Ge)= 2.23\cdot10^{25}$  &  T$(Mo)=5.8\cdot10^{24}$   &    $ 3\cdot10^{24}\leq T(Te) \leq 1.11 \cdot10^{25}$\\
T$(Ge)=2.23\cdot10^{25}$   &  T$(Te)=3\cdot10^{24}$     &    $2.63\cdot10^{24} \leq  T(Mo) \leq 2.04\cdot10^{25}$\\
T$(Ge)= 2.23\cdot10^{25}$  &  T$(Te)=3\cdot10^{25}$     & $9.19  \cdot10^{24} \leq T (Mo) \leq 2.36 \cdot10^{26}$\\
T$(Ge)= 10^{26}$           &  T$(Mo)=5.8\cdot10^{24}$   & $3\cdot10^{24}\leq T(Te) \leq5.07 \cdot10^{24}$\\
T$(Ge)= 10^{26}$  	   &  T$(Te)=3\cdot10^{24}$     & $ 3.82\cdot10^{24} \leq T (Mo) \leq 9.44 \cdot10^{24}$\\
T$(Ge)= 10^{26}$           &  T$(Te)=3\cdot10^{25}$     & $1.96 \cdot10^{25} \leq T (Mo) \leq 6.54 \cdot10^{26}$\\
\hline\hline
\end{tabular}}
\end{table}
%%%%%%%%%%%%%%%%%%%%%%%%%%%%%%%%%%%%%%%%
%
\newpage

 Finally, we would like to  point out to one additional consequence of 
the ``positivity'' conditions and the condition the interference term 
should satisfy when two ``interfering'' mechanisms are responsible 
for the $\betabeta$-decay. Let us denote the two 
fundamental parameters characterising the two 
mechanisms by $\eta_{\beta}$ and $\eta_{\kappa}$.
Then, given the half-life of one isotope, say of $^{76}Ge$ ($T_1$),
and an experimental lower bound on the half-life of a second isotope, 
e.g., of $^{130}Te$ ($T_3$), the conditions 
 $|\eta_{\beta}|^2 > 0$, $|\eta_{\kappa}|^2 > 0$ 
and  $-|\eta_{\beta}||\eta_{\kappa}|\leq |\eta_{\beta}||\eta_{\kappa}| \cos \alpha_{\beta\kappa} \leq |\eta_{\beta}||\eta_{\kappa}|$, 
imply a constraint on the half-life of any third isotope, 
say of $^{100}Mo$ ($T_2$).
This latter constraint depends noticeably on the type of 
the two ``interfering'' mechanisms generating the $\betabeta$-decay 
and can be used, in principle,  to discriminate between the different 
possible pairs of ``interfering'' mechanisms. 
Below we illustrate this result by deriving the constraint 
one obtains on the half-life of $^{100}Mo$, $T_2$, 
assuming that the half-life of  $^{76}Ge$ is 
$T_1 =  2.23\times 10^{25}$ y and taking into account
the current  experimental lower bound on the half-life 
of  $^{130}Te$,  $T_3 > 3.0\times 10^{24}$ y.
Using these ``data'' as input, 
the NMEs calculated with the CD-Bonn and Argonne potentials, 
the ``large basis'' and $g_A=1.25$,
we get the following constraint 
on $T_2$ for the different pairs of ``interfering'' 
mechanisms discussed by us 
(the numbers in brackets are obtained with the NMEs
corresponding to the Argonne potential, unless otherwise 
indicated).
\\

\noindent {\bf Light Neutrino and gluino exchange mechanisms:}
%%%%%%%%%%%%%%%%%
\be  
T_2 \equiv T^{0\nu}_{1/2}(^{100}Mo) > 2.46~(2.47)\times 10^{24}~{\rm y}.
% T_2 \equiv T^{0\nu}_{1/2}(^{100}Mo) > 2.461~(2.472)\times 10^{24}~{\rm y}. 
% Argonne: 2.47164\times 10^{24}~{\rm y}.
\ee
%%%%%%%%%%%%%%%%%%%%%%
%
Increasing the value of $T^{0\nu}_{1/2}(^{76}Ge)$ leads to the increasing of 
the value of the lower limit.\\

\noindent {\bf Light Neutrino and LH Heavy neutrino exchange mechanisms:}
%%%%%%%%%%%%%%%%%%%%%%%%%
\be 
T^{0\nu}_{1/2}(^{100}Mo) >2.78~(2.68)\times 10^{24}~{\rm y}.
% CD-Bonn: T_2 \equiv T^{0\nu}_{1/2}(^{100}Mo) > 2.784\times 10^{24}~{\rm y}.
% Argonne: T^{0\nu}_{1/2}(^{100}Mo) > 2.682\times 10^{24}~{\rm y}. 
\ee
%%%%%%%%%%%%%%%%%%%%%%%%
%
The value of the lower limit increases with 
the increasing of the value of the half-life of $^{76}Ge$.\\

\noindent {\bf LH Heavy neutrino and gluino exchange mechanisms:}
%%%%%%%%%%%%%%%%%%%%%%%%
 \be 
1.36\times 10^{24}~{\rm y} < T^{0\nu}_{1/2}(^{100}Mo) < 
3.42\times 10^{24}~{\rm y}\,.
% 1.358\times 10^{24}~{\rm y} < T^{0\nu}_{1/2}(^{100}Mo) < 
% 3.416\times 10^{24}~{\rm y}\,.
% Argonne: (T^{0\nu}_{1/2}(^{100}Mo) >5.966\times 10^{23}~{\rm y}).
\ee
%%%%%%%%%%%%%%%%%%%%%%%%%%%%%%%
%
Increasing the value of $T^{0\nu}_{1/2}(^{76}Ge)$ leads to a shift 
of the interval  to larger values,
and for a sufficiently large $T^{0\nu}_{1/2}(^{76}Ge) > 10^{26}$ y -
 even just to a lower bound on $T^{0\nu}_{1/2}(^{100}Mo)$.
For $T_1 = 10^{26}$ y, for instance, we find  
$4.19\times 10^{24}~{\rm y} < T^{0\nu}_{1/2}(^{100}Mo) < 
3.39\times 10^{25}~{\rm y}$. 
% from 4.18487*10^{24} to 3.38552*10^{25}

 Using the NMEs derived with the Argonne potential we find a very different 
result - only a lower bound:
$T^{0\nu}_{1/2}(^{100}Mo) > 5.97\times 10^{23}~{\rm y}$.
% 5.966
The difference between the results obtained with 
the two sets of NMEs can be traced to fact that the 
determinant $D$ in eqs. (\ref{intsol1}) 
and  (\ref{DD1}), calculated with the 
second set of NMEs, has opposite sign to that, 
calculated with the first set of NMEs. 
As a consequence, the dependence of the physical 
solutions for $|\eta^L_N|^2$ and $|\eta_{\lambda'}|^2$ 
on  $T_1$, $T_2$ and $T_3$ in the two cases of 
NMEs is very different.
\\

\noindent {\bf Squarks-neutrino and gluino exchange mechanisms:}
%%%%%%%%%%%%%%%%%%%%%%%%%
\be 
T^{0\nu}_{1/2}(^{100}Mo) > 7.92~(22.1)\times 10^{23}~{\rm y}. 
% T^{0\nu}_{1/2}(^{100}Mo) > 7.916~(22.12)\times 10^{23}~{\rm y}. 
% Argonne: T^{0\nu}_{1/2}(^{100}Mo) >2.212\times 10^{24}~{\rm y}.
\ee
%%%%%%%%%%%%%%%%%%%%%%%%
%
For larger values of  $T^{0\nu}_{1/2}(^{76}Ge)$, 
this lower bound assumes larger values.

 We see that the two sets of NMEs lead to 
quite different results in the cases of the 
LH heavy neutrino and gluino exchange and
squarks-neutrino and gluino exchange mechanisms.
Nevertheless, the constraints thus obtained can be used, 
e.g., to exclude some of the possible cases of 
two ``interfering'' mechanisms inducing the 
$\betabeta$-decay.
Indeed, if, for instance, it is confirmed that 
 $T^{0\nu}_{1/2}(^{76}Ge) =  2.23\times 10^{25}$ y, 
and in addition it is established, taking all 
relevant uncertainties into account, 
that $T^{0\nu}_{1/2}(^{100}Mo) \leq 10^{24}~{\rm y}$,
that combined with the experimental 
lower limit on $T^{0\nu}_{1/2}(^{130}Te)$ 
would rule out 
i) the light neutrino and gluino exchanges, and
ii) the light neutrino and LH heavy neutrino exchanges,
% and iii) the LH heavy neutrino and gluino exchanges, 
as possible mechanisms generating the $\betabeta$-decay.

%%%%%%%%%%%%%%%%%%%%%%%%%%%%
%
\section{Summary and Conclusions}
%
%%%%%%%%%%%%%%%%%%%%%%%%%%

 In the present article we have 
considered the possibility of several different 
mechanisms contributing to the
$\betabeta$-decay amplitude in the general case
of CP nonconservation. The mechanisms discussed 
are light Majorana neutrino exchange, 
exchange of  heavy  Majorana neutrinos 
coupled to (V-A) currents, exchange of heavy Majorana neutrinos 
coupled to (V+A) currents, 
lepton charge non-conserving couplings in SUSY theories
with $R$-parity breaking. Of the latter we have 
concentrated on the so-called ``dominant gluino exchange'' 
mechanism. Each of these mechanisms is characterised 
by a specific fundamental lepton number violating (LNV) 
parameter. The latter are defined in 
Section 2. We have investigated in detail 
the cases of two ``non-interfering''
and two ``interfering'' mechanisms, generating 
the $\betabeta$-decay. In the analysis we 
have performed, we have used 
hypothetical $\betabeta$-decay 
half-lives of the following three 
isotopes: $^{76}$Ge, $^{100}$Mo and $^{130}$Te.
They are denoted as $T_1$, $T_2$ and $T_3$, 
respectively.
Four sets of nuclear matrix elements (NMEs) 
of the decays of these three nuclei were 
utilised:  they were obtained with two different 
nucleon-nucleon potentials (CD-Bonn and Argonne) 
and two different values  of the axial 
coupling constant $g_A=1.25;~1.0$ 
(see Table \ref{table.1}).

   If the $\betabeta$-decay is induced by  
two ``non-interfering'' mechanisms, 
which for concreteness we have 
considered to be the light LH Majorana neutrino 
exchange and the heavy RH Majorana neutrino
exchange with $(V+A)$ currents, 
one can determine the squares of the 
absolute values of the two LNV parameters, 
characterising these mechanisms, 
$|\eta_\nu|^2$ and $|\eta_R|^2$,
from data on the half-lives 
of two nuclear isotopes. 
We have done that using as input 
all three possible pairs of 
half-lives of  $^{76}$Ge, 
$^{100}$Mo and $^{130}$Te, 
chosen from the intervals 
given in eq. (\ref{limit})
and satisfying the existing 
experimental lower limits, as well as 
the  half-life of the $\betabeta$-decay 
of $^{76}Ge$, claimed to be 
observed in \cite{Klap04}:
$T^{0\nu}_{1/2}(^{76}Ge)= 
2.23^{+0.44}_{-0.31}\times 10^{25}$ y.
We find that if the half-life  of one of  the 
three nuclei is measured, 
the requirement that
$|\eta_\nu|^2 \geq 0$ and $|\eta_R|^2 \geq 0$
(``positivity condition'')
constrains  the other two half-lives 
(and the $\betabeta$-decay  half-life of any other 
$\betabeta$-decaying isotope for that matter)
to lie in specific  intervals, 
determined by the measured 
half-life and the relevant NMEs
(see eqs. (\ref{PosC}) - (\ref{Argonne12510})).
This  feature is common to all cases 
of  two ``non-interfering'' mechanisms 
generating the $\betabeta$-decay.
The indicated  specific  
half-life intervals for the various isotopes, 
are stable with respect to the 
change of the NMEs  (within the sets of 
NMEs considered by us)
used to derive  them. The intervals depend, in general, 
on the type of the two 
``non-interfering'' mechanisms. 
However, these differences in the cases of the 
$\betabeta$-decay triggered by 
the exchange of heavy Majorana neutrinos 
coupled to (V+A) currents  and
i) the light  Majorana neutrino exchange, or
 ii) the gluino 
exchange mechanism, or i ii)  the squark-neutrino exchange 
mechanism,  are extremely  small. One of the consequences of 
this feature of the different pairs of ``non-interfering'' 
mechanisms considered  by us is that if it will be possible 
to rule out one of them as the cause of  $\betabeta$-decay, 
most likely one will  be able to rule out all three of them. 
Using the indicated difference 
to get information about 
the specific pair of  ``non-interfering''  mechanisms 
possibly operative in $\betabeta$-decay
requires, in the cases considered by us, an 
extremely high precision in the measurement of the 
$\betabeta$-decay half-lives of the 
isotopes considered.
 The levels of precision required 
 seem impossible to achieve in 
 the foreseeable future.
 If it is experimentally established that
 any of the indicated intervals of  
half-lives  lies outside the interval of physical 
solutions  of $|\eta_\nu|^2$ and  $|\eta_R|^2$, obtained 
taking into account all relevant uncertainties, 
one would be led to conclude that the 
$\betabeta$-decay is not generated by the 
two mechanisms  considered. 
The constraints  under discussion will not be 
valid, in general,  if the  $\betabeta$-decay is 
triggered by two  ``interfering''  mechanisms 
with a non-negligible (destructive)  interference term, or 
by more than  two mechanisms none  of which 
plays a subdominat role in $\betabeta$-decay.

  We have studied also the dependence of the 
physical solutions for 
$|\eta_\nu|^2$ and  $|\eta_R|^2$ obtained  
on the NMEs used. 
Some of the results of this study 
are presented graphically in Figs. \ref{fig:comparison1} and 
\ref{fig:comparison2}.
We found that the solutions
can exhibit a significant variation 
with the NMEs used.
Given the half-life $T_1$, 
the interval of allowed values 
of the half-life of the second nucleus $T_2$, 
determined from the ``positivity conditions'',
$|\eta_\nu|^2 \geq 0$,  $|\eta_R|^2 \geq 0$,
changes somewhat with the change 
of the NMEs. 
The solution values of the parameters 
$|\eta_\nu|^2$  and $|\eta_R|^2$,
obtained with the two different 
sets of the NMEs, can differ drastically
in the vicinity of the maximum 
and minimum values of $T_2$
(Figs. \ref{fig:comparison1} and 
\ref{fig:comparison2}).
If a given extreme value of $T_2$, say ${\rm max}(T_2)$,
obtained with one set of NMEs, belongs to the 
interval of allowed values of $T_2$, found with 
a second set of NMEs, 
one of the fundamental parameters, calculated 
at ${\rm max}(T_2)$ with the first set of NMEs
can be zero, and can have a relatively large 
nonzero value when calculated with the second 
set of NMEs. Moreover, 
there are narrow intervals of values 
of $T_2$ for which there exist physical 
solutions for 
$|\eta_\nu|^2$  and $|\eta_R|^2$ if one uses 
the NMEs obtained with the CD-Bonn potential 
and there are no physical solutions 
for the NMEs derived with the Argonne potential.
If the measured value of $T_2$ falls in such an interval,
this can imply that either the two mechanisms considered are 
not at work in $\betabeta$-decay, or one of the two sets 
of NMEs does not describe correctly the nuclear transitions.

 Neutrinoless double beta decay can be generated
by two competitive
mechanisms whose interference contribution to the
$\betabeta$-decay rates is non-negligible. 
 In the case when two
``interfering'' mechanisms are responsible
for the $\betabeta$-decay, the squares of the 
absolute values of the two relevant parameters and the 
interference term parameter, which involves 
the cosine of an unknown relative phase 
of the two fundamental parameters, 
can be uniquely determined, 
in principle, from data on the half-lives 
of three nuclei. We have analyzed in detail the case
of light Majorana neutrino exchange 
and gluino exchange. In this case the 
parameters which are determined 
from data on the half-lives 
are $|\eta_\nu|^2$, $|\eta_{\lambda'}|^2$
and $z = 2\cos\alpha\,|\eta_\nu||\eta_{\lambda'}|$. 
The physical solutions for these parameters
have to satisfy the conditions 
$|\eta_\nu|^2 \geq 0$, $|\eta_{\lambda'}|^2\geq 0$ 
and $-\,2|\eta_\nu||\eta_{\lambda'}|\leq z \leq 
2|\eta_\nu||\eta_{\lambda'}|$.
The latter condition implies that 
given the half-lives of two isotopes, $T_1$ and $T_2$,
the half-life of any third isotope $T_3$
is constrained to lie is a specific interval, 
if the mechanisms considered are indeed   
generating the $\betabeta$-decay.
If further the half-life of one isotope $T_1$  is known, 
for the interference to be constructive (destructive) 
the half-lives of any other pair of 
isotopes $T_2$ and $T_3$, 
should belong to specific intervals. 
These intervals depend on whether 
the interference between the two contributions in the 
$\betabeta$-decay rate is constructive or destructive.
We have derived in analytic form 
the general  conditions 
for i) constructive interference ($z > 0$),
ii) destructive interference ($z< 0$), 
iii)   $|\eta_\nu|^2 = 0$, $|\eta_{\lambda'}|^2 \neq 0$,
iv)  $|\eta_\nu|^2 \neq 0$,  $|\eta_{\lambda'}|^2 = 0$ and 
v) $z = 0$,  $|\eta_\nu|^2 \neq 0$, $|\eta_{\lambda'}|^2 \neq 0$.

We have found that,  given $T_1$, a constructive 
interference is possible only if $T_2$ lies in a relatively 
narrow interval and $T_3$ has a value in extremely 
narrow intervals, the interval for $T_2$ 
being determined by the value of $T_1$, while that for 
$T_3$ -  by $T_1$ and the interval for $T_2$.
The fact that both the intervals for $T_2$ and $T_3$ 
are so narrow is a consequence of the fact that, 
for each of the  two mechanisms discussed,
the NMEs for the three nuclei considered differ 
relatively little: the relative difference between 
the nuclear matrix elements of any two nuclei 
does not exceed 10\%.

 The intervals  of values of $T_2$ and $T_3$ 
corresponding to destructive interference 
 ( eqs. (\ref{znegative1}) - (\ref{destructint}))
are very different from those corresponding 
to the cases of constructive interference and 
of the  two ``non-interfering''  $\betabeta$-decay 
mechanisms we have considered
(eq.(\ref{CDBonn125})).  
Within the set of $\betabeta$-decay mechanisms studied by us,
this difference can allow  to discriminate
experimentally between the  possibilities of the   
$\betabeta$-decay being triggered by two `` destructively interfering'' 
mechanisms or  by two ``constructively  interfering''  
or by two ``non- interfering''  mechanisms. 

  We have shown also that further significant constraints on 
the physical solutions for  the fundamental parameter 
$|\eta_\nu|^2$  in the case of the light Majorana 
neutrino exchange mechanism  and the gluino 
exchange (or any other  ``interfering'')  mechanism
 can be obtained by using 
the current and the prospective 
upper bounds on the absolute scale 
of neutrino masses from the past  \cite{MoscowH3,MainzKATRIN} 
and the upcoming KATRIN  \cite{MainzKATRIN} 
$^3H$ $\beta$-decay experiments of 
2.3 eV and 0.2 eV,  respectively.
Our results show that the KATRIN prospective 
upper bound of 0.2 eV, if reached, 
 could imply particularly  stringent 
constraints in the cases of   ``destructively  interfering'' 
 mechanisms  one of which is the light neutrino exchange, 
 to the point of strongly disfavoring 
 (or even excluding)  some of them.
The KATRIN prospective upper bound could be used 
to constrain also the fundamental parameters  
of  two ``non-interfering'' mechanisms, 
one of which is the light Majorana neutrino exchange.
This bound could eliminate,  in particular, 
some parts of the  half-life solution intervals where 
there is a significant dependence of the values of
$|\eta_\nu|^2$ obtained on the 
set of NMEs used.
 
  The measurements of the 
half-lives with rather high precision and the 
knowledge of the relevant nuclear matrix elements with 
relatively small uncertainties is crucial for establishing 
whether more than one mechanisms are operative in 
$\betabeta$-decay. The method considered by us
can be generalised to the case of more than two
$\betabeta$-decay mechanisms.
It allows to treat the cases of 
CP conserving and CP nonconserving 
couplings generating the $\betabeta$-decay
in a unique way.

{\bf Acknowledgements.} 
This work was supported in part by the INFN program 
on ``Astroparticle Physics'', by the Italian MIUR program 
on ``Neutrinos, Dark Matter and  Dark Energy in the Era of LHC'' 
(A.M. and S.T.P.) and by the World Premier International 
Research Center Initiative (WPI Initiative), 
MEXT,  Japan  (S.T.P.). A.F. and F. \v S.  acknowledge support of the
Deutsche Forschungsgemeinschaft within the project
436 SLK 17/298. The work of F. \v S. was also
partially supported by the VEGA Grant agency of the
Slovak Republic under contract 1/063/09.

{\bf Note Added.}   The possibility of several mechanisms being 
operative in $\betabeta$-decay is also discussed 
in  another recent preprint \cite{ELisietalMM11},  where 
the sets  of nuclear matrix elements 
given in Table  \ref{table.1}  are  also used.
However,   the aspects of the  problem of multiple 
mechanisms generating the $\betabeta$-decay 
investigated in  the present article and in 
the preprint \cite{ELisietalMM11}
are  very different and, apart from 
the description of the nuclear matrix elements, 
the two studies practically do not overlap.

\end{document}